\documentclass[USenglish,oneside,twocolumn]{article}

\usepackage[dvipsnames]{xcolor} % More color names.

\usepackage[utf8]{inputenc}%(only for the pdftex engine)
\usepackage[big]{dgruyter_NEW}
 
\DOI{foobar}

\cclogo{\includegraphics{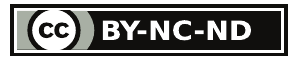}}

\def\ForceBleck

\usepackage[T1]{fontenc}
\usepackage{type1ec}

\usepackage{mathtools, dsfont}
\usepackage{amssymb}
\usepackage{nicefrac}
\usepackage{amsmath}
\usepackage{amsthm}
\usepackage{thmtools}
\usepackage{amsfonts, pifont}
\usepackage[dvipsnames]{xcolor} % More color names.
\usepackage{colortbl}
\usepackage{pdfpages} % For including whole PDF's.
\usepackage{multicol}
\usepackage{multirow} % Multiple rows on table.
\usepackage{enumerate} % Format enumerate.
\usepackage[breakable,skins]{tcolorbox} % Allows for colored boxes.

\usepackage{float}

\usepackage{listings}
\usepackage{cprotect}

\usepackage{graphicx,caption,subcaption}
\usepackage{tabularx} % Used to set max width of a table.
\usepackage{soul} % For highlighting text.
\usepackage{soulutf8} % To allow accents within higlighted text.
\usepackage{twoopt} % Allows for two optional parameters in one command.
\usepackage{varwidth}  % Provides varwidth environment.
\usepackage{lipsum} % Allows for automatically generating long text.
\usepackage{xspace} % Allows for smart spacing after macros via \xspace.
\def\ReviewsOn{} % Comment out to enable review mode, in which comments are turned off, but only text within a command \review{} is highlighted. This mode is only allowed if Bleck is switched off, because Bleck mode has preference over everything else.

\def\CommentsOn{} % Comment out to hide all mark-up comments from the text. This mode is only allowed if Bleck is switched off, because Bleck mode has preference over everything else.
\ifdefined\ForceBleck % If commandline forces Bleck mode. 
	\def\Bleck{}
	\undef\CommentsOn{}
	\undef\ReviewsOn{} 
\else 
	\ifdefined\ForceDraft % If commandline forces Draft mode.
		\undef\Bleck{} 
		\def\CommentsOn{}
		\def\ReviewsOn{} 
	\else
		\ifdefined\ForceReview % If commandline forces Review mode.
			\undef\Bleck{} 
			\undef\CommentsOn{}
			\def\ReviewsOn{} 
		\fi	
	\fi
\fi

\ifdefined\Bleck % Bleck forces all comments and reviews to be switched off.
	\newcommand{\DraftOnly}[1]{} % 
	\undef\CommentsOn{} % This forces comments to be switched off.
	\undef\ReviewsOn{} % This forces review highlights to be switched off.
\else
	\ifdefined\ForceReview % \ForceReview forces comments to be switched off, but reviews to be switched on. 
		\newcommand{\DraftOnly}[1]{} % 
		\def\ReviewsOn{} % This forces review highlights to be switched off.
	\else 
		\newcommand{\DraftOnly}[1]{#1}
	\fi
\fi

\ifdefined\ReviewsOn % In case reviews are to be shown.
	\newcommand{\review}[1]{% Highlights reviews in color.
		{{\leavevmode\color{magenta}{#1}}}\xspace
	}
	\newcommand{\reviewfootnote}[1]{\footnote{\review{\textbf{Nota de revisão:} #1}}} % Shows review note.
\else
	\newcommand{\review}[1]{#1} % Shows reviews in original color.
	\newcommand{\reviewfootnote}[1]{} % Hides review note.
\fi

\ifdefined\CommentsOn % In case comments are to be shown.

\else % In case comments are to be hidden.

\fi % End if draftbox.

\ifdefined\CommentsOn % In case comments are to be shown.
	\newcommand{\draftbox}[2]{ % Box for stuff that only appears on a draft.
		\begin{center}
			\begin{tcolorbox}[
				enhanced jigsaw,
				breakable,
				left=2pt,
				right=2pt,
				top=2pt,
				bottom=2pt,
				bicolor,
				colframe = black!70,
				colbacklower = black!70,
				colback  = purple!5!white,
				coltitle = white,
				title = {Begin draft \texttt{$\blacktriangleright$#1}},
				halign lower = right,
				]
				\textcolor{black}{\texttt{#2}}
				\tcblower
				\textcolor{white}{\texttt{End draft}}
			\end{tcolorbox}					
		\end{center} 
	}
\else % In case comments are to be hidden.
	\newcommand{\draftbox}[1]{} % Hides draft boxes.
\fi % End if draftbox.

\newtcolorbox{todobox}[3][]
{
	enhanced jigsaw,
	enforce breakable,
	hbox,
	left=1pt,
	right=1pt,
	top=1pt,
	bottom=1pt,
	colframe = #2!20,
	colback  = #2!100,
	coltitle = #2!20!black,  
	title    = {#3},
	#1,
}

\newtcolorbox{donebox}[3][]
{
	enhanced jigsaw,
	breakable,
	hbox,
	left=1pt,
	right=1pt,
	top=1pt,
	bottom=1pt,
	colframe = #2!50,
	borderline={0.5mm}{0mm}{black,dashed},
	colback  = #2!50,
	coltitle = #2!20!black,  
	title    = {#3},
	#1,
}

\ifdefined\CommentsOn % In case comments are to be shown.
	
	\def \msboxcolor {Cerulean}
	\def \mstextinboxcolor {white}
	\def \ms{\ctext{\msboxcolor}{\mstextinboxcolor}{\mbox{[\msname]}}}
	\newcommandtwoopt{\mstodo}[2][to do][]{
		\begin{todobox}{\msboxcolor}{}
			\begin{varwidth}{\linewidth}
				\textcolor{\mstextinboxcolor}{\texttt{#2$\blacktriangleright$\ms: #1}}
			\end{varwidth}
		\end{todobox}
	}
	\newcommandtwoopt{\msdone}[2][to do][]{
		\begin{donebox}{black}{}
			\begin{varwidth}{\linewidth}
				\textcolor{white}{\texttt{#2$\blacktriangleright$\ms: #1}}
			\end{varwidth}
		\end{donebox}
	}
	\newcommand{\msdraft}[1]{
		\draftbox{\ms}{#1}
	}
\else % In case comments are to be hidden.
	
	\def \msboxcolor {}
	\def \mstextinboxcolor {}
	\def \ms{}
	\newcommandtwoopt{\mstodo}[2][to do][]{}
	\newcommandtwoopt{\msdone}[2][to do][]{}
	\newcommand{\msdraft}[1]{}
\fi % End if for Mário.

\ifdefined\CommentsOn % In case comments are to be shown.
	
	\def \ghnboxcolor {PineGreen}
	\def \ghntextinboxcolor {white}
	\def \ghn{\ctext{\ghnboxcolor}{\ghntextinboxcolor}{\mbox{[\ghnname]}}}
	\newcommandtwoopt{\ghntodo}[2][to do][]{
		\begin{todobox}{\ghnboxcolor}{}
			\begin{varwidth}{\linewidth}
				\textcolor{\ghntextinboxcolor}{\texttt{#2$\blacktriangleright$\ghn: #1}}
			\end{varwidth}
		\end{todobox}
	}
	\newcommandtwoopt{\ghndone}[2][to do][]{
		\begin{donebox}{black}{}
			\begin{varwidth}{\linewidth}
				\textcolor{white}{\texttt{#2$\blacktriangleright$\ghn: #1}}
			\end{varwidth}
		\end{donebox}
	}
	\newcommand{\ghndraft}[1]{
		\draftbox{\ghn}{#1}
	}
\else % In case comments are to be hidden.
	
	\def \ghnboxcolor {}
	\def \ghntextinboxcolor {}
	\def \ghn{}
	\newcommandtwoopt{\ghntodo}[2][to do][]{}
	\newcommandtwoopt{\ghndone}[2][to do][]{}
	\newcommand{\ghndraft}[1]{}	
\fi % End if for Gabriel.

\ifdefined\CommentsOn % In case comments are to be shown 

\def \abboxcolor {OrangeRed}
\def \abtextinboxcolor {white}
\def \ab{\ctext{\abboxcolor}{\abtextinboxcolor}{\mbox{[\abname]}}}
\newcommandtwoopt{\abtodo}[2][to do][]{
	\begin{todobox}{\abboxcolor}{}
		\begin{varwidth}{\linewidth}
			\textcolor{\abtextinboxcolor}{\texttt{#2$\blacktriangleright$\ab: #1}}
		\end{varwidth}
	\end{todobox}
}
\newcommandtwoopt{\abdone}[2][to do][]{
	\begin{donebox}{black}{}
		\begin{varwidth}{\linewidth}
			\textcolor{white}{\texttt{#2$\blacktriangleright$\ab: #1}}
		\end{varwidth}
	\end{donebox}
}
\newcommand{\abdraft}[1]{
	\draftbox{\ab}{#1}
}
\else % In case comments are to be hidden 

\def \abboxcolor {}
\def \abtextinboxcolor {}
\def \ab{}
\newcommandtwoopt{\abtodo}[2][to do][]{}
\newcommandtwoopt{\abdone}[2][to do][]{}
\newcommand{\abdraft}[1]{}
\fi % End if for Annabelle

\ifdefined\CommentsOn % In case comments are to be shown 

\def \tashboxcolor {YellowOrange}
\def \tashtextinboxcolor {black}
\def \tash{\ctext{\tashboxcolor}{\tashtextinboxcolor}{\mbox{[\tashname]}}}
\newcommandtwoopt{\tashtodo}[2][to do][]{
	\begin{todobox}{\tashboxcolor}{}
		\begin{varwidth}{\linewidth}
			\textcolor{\tashtextinboxcolor}{\texttt{#2$\blacktriangleright$\tash: #1}}
		\end{varwidth}
	\end{todobox}
}
\newcommandtwoopt{\tashdone}[2][to do][]{
	\begin{donebox}{black}{}
		\begin{varwidth}{\linewidth}
			\textcolor{white}{\texttt{#2$\blacktriangleright$\tash: #1}}
		\end{varwidth}
	\end{donebox}
}
\newcommand{\tashdraft}[1]{
	\draftbox{\tash}{#1}
}
\else % In case comments are to be hidden 

\def \tashboxcolor {}
\def \tashtextinboxcolor {}
\def \tash{}
\newcommandtwoopt{\tashtodo}[2][to do][]{}
\newcommandtwoopt{\tashdone}[2][to do][]{}
\newcommand{\tashdraft}[1]{}		
\fi % End if for Natasha

\ifdefined\CommentsOn % In case comments are to be shown 

\def \cmboxcolor {Purple}
\def \cmtextinboxcolor {white}
\def \cm{\ctext{\cmboxcolor}{\cmtextinboxcolor}{\mbox{[\cmname]}}}
\newcommandtwoopt{\cmtodo}[2][to do][]{
	\begin{todobox}{\cmboxcolor}{}
		\begin{varwidth}{\linewidth}
			\textcolor{\cmtextinboxcolor}{\texttt{#2$\blacktriangleright$\cm: #1}}
		\end{varwidth}
	\end{todobox}
}
\newcommandtwoopt{\cmdone}[2][to do][]{
	\begin{donebox}{black}{}
		\begin{varwidth}{\linewidth}
			\textcolor{white}{\texttt{#2$\blacktriangleright$\cm: #1}}
		\end{varwidth}
	\end{donebox}
}
\newcommand{\cmdraft}[1]{
	\draftbox{\cm}{#1}
}
\else % In case comments are to be hidden 

\def \cmboxcolor {}
\def \cmtextinboxcolor {}
\def \cm{}
\newcommandtwoopt{\cmtodo}[2][to do][]{}
\newcommandtwoopt{\cmdone}[2][to do][]{}
\newcommand{\cmdraft}[1]{}
\fi % End if for Carroll

\ifdefined\CommentsOn % In case comments are to be shown 

\def \allboxcolor {yellow}
\def \alltextinboxcolor {black}
\def \all{\ctext{\allboxcolor}{\alltextinboxcolor}{\mbox{[\allname]}}}
\newcommandtwoopt{\alltodo}[2][to do][]{
	\begin{todobox}{\allboxcolor}{}
		\begin{varwidth}{\linewidth}
			\textcolor{\alltextinboxcolor}{\texttt{#2$\blacktriangleright$\all: #1}}
		\end{varwidth}
	\end{todobox}
}
\newcommandtwoopt{\alldone}[2][to do][]{
	\begin{donebox}{black}{}
		\begin{varwidth}{\linewidth}
			\textcolor{white}{\texttt{#2$\blacktriangleright$\all: #1}}
		\end{varwidth}
	\end{donebox}
}
\newcommand{\alldraft}[1]{
	\draftbox{\ab}{#1}
}
\newcommand\See[1]{\marginpar{\tiny#1$\downarrow$}}
\newcommand\Here[1]{\marginpar{\tiny#1$\uparrow$}}
\else % In case comments are to be hidden 

\def \allboxcolor {}
\def \alltextinboxcolor {}
\def \all{}
\newcommandtwoopt{\alltodo}[2][to do][]{}
\newcommandtwoopt{\alldone}[2][to do][]{}
\newcommand{\alldraft}[1]{}
\newcommand\See[1]{}
\newcommand\Here[1]{}
\fi % End if for All

\newtheorem{theorem}{Theorem}

\newtheorem{example}[theorem]{Example}

\newcommand{\cala}{\mathcal{A}}

\newcommand{\cald}{\mathcal{D}}

\newcommand{\calx}{\ensuremath{\mathcal{X}}}
\newcommand{\caly}{\ensuremath{\mathcal{Y}}}

\newcommand\NF[2] {\ensuremath{\nicefrac{#1}{#2}}}

\newcommand{\Set}[1]{\{#1\}}

\newcommand{\attrset}{\cala}

\newcommand{\longdb}{\mathcal{{L}_{D}}}
\newcommand{\longaggr}[1]{\textit{agreg}(#1)}
\newcommand{\idattr}{a_{\textit{id}}}

\newcommand{\domain}[1]{\textit{dom}(#1)}

\newcommand\QIF {\textit{QIF}}
\newcommand\ARX {\textit{ARX}}
\newcommand\INEP {\textsc{inep}}
\newcommand\INEPallcaps {\textsc{INEP}}
\newcommand\GDPR {\textsc{gdpr}}
\newcommand\LAI {\textsc{lai}}
\newcommand\LGPD {\textsc{lgpd}}
\newcommand\CIPSEA {\textsc{cipsea}}
\newcommand\MMM {\textsf{M}}
\newcommand\RRR {\textsf{R}}
\newcommand\AAA {\textsf{A}}
\newcommand\SSS {\textsf{S}}
\newcommand\LLL {\textsf{L}}
\newcommand\III {\textsf{I}}
\newcommand\CCC {\textsf{C}}

\newcommand\Joint[2]		{#1{\kern-0.6pt\smalltriangleright\kern.3pt}#2} % needs hack (below) to get \smalltriangleright from mathabx without loading it.

\newcommand\Table[1] {Tbl.~\ref{#1}}
\newcommand{\id}{\textit{id}}
\newcommand{\age}{\textit{age}}
\newcommand{\gender}{\textit{gender}}
\newcommand{\grade}{\textit{grade}}
\newcommand{\disability}{\textit{disability}}

\newcommand{\male}{\texttt{M}}
\newcommand{\female}{\texttt{F}}
\newcommand{\five}{\texttt{E}}
\newcommand{\four}{\texttt{D}}
\newcommand{\three}{\texttt{C}}
\newcommand{\two}{\texttt{B}}
\newcommand{\one}{\texttt{A}}
\newcommand{\yes}{\texttt{yes}}
\newcommand{\no}{\texttt{no}}
\newcommand{\IMS}{\textsf{IMS}} % Individual-target Membership Single dataset
\newcommand{\CMS}{\textsf{CMS}} % Collective-target Membership Single dataset
\newcommand{\IRS}{\textsf{IRS}} % Individual-target Re-identification Single dataset
\newcommand{\CRS}{\textsf{CRS}} % Collective-target Re-identification Single dataset
\newcommand{\IAS}{\textsf{IAS}} % Individual-target Attribute-inference Single dataset
\newcommand{\CAS}{\textsf{CAS}} % Collective-target Attribute-inference Single dataset
\newcommand{\IML}{\textsf{IML}} % Individual-target Membership Longitudinal dataset
\newcommand{\CML}{\textsf{CML}} % Collective-target Membership Longitudinal dataset
\newcommand{\IRL}{\textsf{IRL}} % Individual-target Re-identification Longitudinal dataset
\newcommand{\CRL}{\textsf{CRL}} % Collective-target Re-identification Longitudinal dataset
\newcommand{\IAL}{\textsf{IAL}} % Individual-target Attribute-inference Longitudinal dataset
\newcommand{\CAL}{\textsf{CAL}} % Collective-target Attribute-inference Longitudinal dataset

\newcommand{\pg}[1]{A{#1}} % General premisses (for both longitudinal and single-dataset attacks). Used in unified presentation for USENIX paper.
\def\ojoin{\setbox0=\hbox{$\Join$}%
	\rule[0.05ex]{.27em}{.4pt}\llap{\rule[1.1ex]{.27em}{.4pt}}}
\def\leftouterjoin{\mathbin{\ojoin\mkern-7.8mu\Join}}
\newcommand{\countquery}[1]{\texttt{count}_{q}}

\DeclareFontFamily{U}{mathb}{\hyphenchar\font45}
\DeclareFontShape{U}{mathb}{m}{n}{
	<5> <6> <7> <8> <9> <10> gen * mathb
	<10.95> mathb10 <12> <14.4> <17.28> <20.74> <24.88> mathb12
}{}
\DeclareSymbolFont{mathb}{U}{mathb}{m}{n}
\DeclareMathSymbol{\smalltriangleright}{2}{mathb}{"9B}% name to be checked.
\newcommand\Sec[1] {Sec.\ \ref{#1}}

\newcommand{\qm}[1]{``#1''} % Quotation marks

\definecolor{sand}{rgb}{0.839,0.824,0.769}

\begin{document}

%\author[0]{Anonymized for the review process}
%\affil[0]{-}

\author[1]{M\'ario S.\ Alvim}

\author[2]{Natasha Fernandes}

\author[3]{Annabelle McIver}

\author[4]{Carroll Morgan}

\author[5]{Gabriel H.\ Nunes}

\affil[1]{UFMG, Brazil, e-mail: msalvim@dcc.ufmg.br}

\affil[2]{Macquarie University, Australia, e-mail: natasha.fernandes@mq.edu.au}

\affil[3]{Macquarie University, Australia, e-mail: annabelle.mciver@mq.edu.au}

\affil[4]{UNSW and Trustworthy Systems, Australia, e-mail: carroll.morgan@unsw.edu.au}

\affil[5]{UFMG, Brazil, e-mail: ghn@nunesgh.com}

% V 1032 Carroll 211128
%\title{\huge A refactorisation of privacy analyses, its advantage and justification,
%	and its application to very large public databases}
%\runningtitle{Refactorisation of privacy analyses of public databases}
%\title{\huge Privacy assessment and management for very large datasets: Challenges, advances, and the Brazilian Educational Censuses}
\title{\huge Flexible and scalable privacy assessment for very large datasets\review{,} with an application to official governmental microdata
}
\runningtitle{Flexible and scalable privacy assessment for very large datasets}
\begin{abstract}{%
%\colorO{ % Why is the first { necessary?
%Privacy analyses are typically described with words such as ``prosecutor'', ``journalist'', and ``marketer'' that are informally descriptive of real-world practices and roles. Here we describe privacy instead with mathematical concepts based on recent work in \emph{Quantitative Information Flow}, \QIF, with an immediately accruing advantage that those separate concepts are simple, and that they combine together in a regular way. Furthermore, with them we can not only reconstruct the informally classified practices (above), but systematically uncover new privacy-threatening roles that might not yet have been exploited (but assuredly will be, eventually). In that latter sense, the \QIF\ approach might allow us to get ``ahead of the game'' instead of remaining in the more usual situation of ``catch-up and patch''.\\[1ex]
%%
%We apply our new approach to the Educational Census(es) of Brazil, comprising over 90 attributes of approximately 50 million individuals released (longitudinally) every year  since 2007, and we highlight three significant advantages so enabled: the approach is \emph{flexible}, dealing with attacks both known and novel; that flexibility allows its results to be \emph{explained} both to politicians and to the public in a way they understand; and it is \emph{computationally tractable} for very large datasets.
%\colorM{
We present a systematic refactoring of the conventional treatment of privacy analyses, basing it on mathematical
concepts from the framework of 
\emph{Quantitative Information Flow} ({\QIF}).
%Those separate concepts are simple and combine together in a regular way, and with them we can not only reconstruct
%well-known attack models, but systematically uncover new privacy-threatening roles that might not yet have been exploited. % (but assuredly will be, eventually).	
%This allows us not only to recover well-studied attack models, but systematically uncover new privacy-threatening roles that might not yet have been exploited (but assuredly will be, eventually).
The approach we suggest brings three principal advantages: it is \emph{flexible}, allowing for precise quantification and comparison of privacy risks for attacks both known and novel; 
%it is \emph{computationally tractable} for very large datasets;
\review{it can be \emph{computationally tractable} for very large, longitudinal datasets;}
and its results are \emph{explainable} both to politicians and to the general public.
We apply our approach to a very large case study: the Educational Censuses of Brazil, curated by the governmental agency {\INEP}, which comprise over 90 attributes of approximately 50 million individuals released longitudinally every year  since 2007.
These datasets have only very recently (2018--2021) attracted legislation to regulate their privacy --- while at the same time continuing to maintain the openness that \review{had been sought} in Brazilian society. 
\INEP's reaction to that legislation was the genesis of our project with them.
In our conclusions here we share the scientific, technical, and communication lessons we learned in the process.
%}}
}\end{abstract}

%  \classification[PACS]{}
%  \communicated{...}
%  \dedication{...}

  \journalname{Proceedings on Privacy Enhancing Technologies}
\DOI{Editor to enter DOI}
  \startpage{1}
  \received{..}
  \revised{..}
  \accepted{..}

  \journalyear{..}
  \journalvolume{..}
  \journalissue{..}

\maketitle
%
%
%\DraftOnly{
  %=============================
  % GENERAL INFO
  %=============================
%  \input{instructions.tex}

  %\input{new-structure.tex}
  %-----------------------------
%}
%
%
%=============================
% SECTION
%=============================
\section{Introduction}\label{s102818}

{%\color{orange} %1119

\msdraft{General instructions for reviewed version:
	\begin{enumerate}
		\item When introducing any new text to the main body of the paper, do it via the command \review{$\backslash\texttt{review}\{\text{new text}\}$}, so it will show up in a different color (\review{magenta}) when we submit the new version.
		\item PETS papers given a decision of Major Revision
		% or Accept with Shepherding (aka Minor Revision) 
		may revise their paper \textbf{with 16 main-body pages}, excluding bibliography and clearly-marked appendices, \textbf{and 21 pages total}.
		\item Major Revision notes are here: \url{https://docs.google.com/document/d/1ZBBTQEOupsy8fVaJZRDrH1I27pRNv-G5iAL5QESSfUA/edit}
	\end{enumerate}
}

\tashdone[Fulfill these promises:
\begin{enumerate}
	\item We'll add a paragraph mentioning existing works using QIF for differential privacy and searchable encryption.
	%plus detailed examples of QIF models for membership inference and re-identification attacks.
	\item QIF's generality is highlighted in previous works applying QIF to differential-privacy on medium-sized datasets (1), linkage/intersection attacks against k-anonymity/differential-privacy (2), and searchable encryption (3). We'll include a paragraph on this.
	\item We use "vulnerability" in its established sense in QIF: a measure of inference risk. We'll make it clear upfront.
	\item QIF isn't limited to average-case measures: it's also applied to the max-case setting generally (4) and differential privacy specifically (4,5). QIF's primary limitation is computational tractability of algebraic reasoning; however a central contribution of our paper is to demonstrate its feasibility even so.
\end{enumerate}
][\ms]
\msdone[I have put 1-2 in the related work in Section 8, 3 as a footnote in Section 4.1 and 4 is included in the footnote at the end of Section 3.][\tash]

\ghndone[Fulfill these promises:
\begin{enumerate}
	\item We'll enhance discussion on relevant legislation and specify that INEP’s datasets usages include demographic research, policy-making, and governmental planning.
\end{enumerate}
][\ms]

\cmdone[Fulfill these promises:
\begin{enumerate}
	\item Provide generic guidance on the usage of QIF for explainability purposes.	
\end{enumerate}
][\ms]

\msdone[Fulfill these promises:
\begin{enumerate}
	\item Write response letter.
\end{enumerate}
]

% --- 1121
%\colorM{
Privacy preservation in the release of governmental data about individuals has led recently to legislation in many contexts.
\review{Notable examples include the European \emph{General Data Protection Regulation} (\GDPR)% 
%from 2016
~\cite{EU16}, 
the United States' \emph{Confidential Information Protection and Statistical Efficiency Act} (\CIPSEA)% from  2002 
~\cite{US02}, and the Australian review % 
%in 2015 
of its \emph{Privacy Act}%
%from 1988 
~\cite{AU1988}.}
There are, however, three principal problems concerning this kind of legislation. 
%}

%\colorM{
%A first problem is that, since it is formulated at the level of governments or higher, the data affected can be huge.
%\alltodo[Is it intentional that we have 1 problem in one paragraph, and then the 2 other problems in the following paragraph?][\ms]
One problem is that legislation usually addresses \emph{known} privacy issues (since they are what brought the issues to the public eye), but when new ways of violating privacy are found (which can happen overnight), the original legislation must still apply (since changing legislation is difficult and time-consuming). 
\review{A second problem is that, since such legislation is formulated at the level of governments or higher, the data affected can be huge and longitudinal.}
And thirdly, the legislation must be couched in terms that politicians and the public understand, even though 
achieving compliance 
to it is (eventually) a question of mathematics and computer code.  It is crucial\review{,} therefore\review{,} to 
have a link between those two worlds, one that identifies 
%link these two worlds, identifying
meaningful threats 
while minimizing possible waste of resources
on non-threats.
%}

%\colorM{
In this paper we consider all three issues, grounding our approach on 
%relatively recent 
\review{decision- and information-theoretic 
principles 
%known collectively as 
of} 
\emph{Quantitative Information Flow}
\review{(\QIF)}~\cite{Clark:01:ENTCS,Smith09,McIver:10,Alvim20:Book}.
\review{\QIF\ has been successfully applied to a variety of privacy and security analyses, including searchable encryption%
% on medium-sized datasets
~\cite{jurado2021formal}, 
%modeling 
intersection and linkage attacks against $k$-anonymity~\cite{fernandes2018processing}, and differential privacy~\cite{DBLP:conf/csfw/0001FP19}.}
\alltodo[I brought \tash's reference to previous uses of QIF in security from related work to here, as I think it puts it more in evidence.][\ms]
% BANDICOOT!!! --- Can something of Malacaria et al be cited here, without comment?
%\alltodo[Regarding BANDICOOT: Done!][\ms]
%QIF's importance to the security community has been 
%recognized in various forms: its seminal paper received the \emph{ETAPS 2020 Test of Time Award},\footnote{\url{https://etaps.org/2020/test-of-time-award}} and a later work (co-authored by the proponent of this project) was awarded the \emph{NSA's 3rd Annual Best Scientific Cybersecurity Paper}.~\footnote{\url{https://cps-vo.org/node/21539\#winning}}
%stakeholders and the public.}
%
\review{In the context of the present work, we} name the three challenges introduced above
\emph{flexibility}, 
\emph{scalability}, and
\emph{explainability}, and now consider each one in turn.
% ^ 1121 Carroll 211128 
\review{We then 
%introduce a case study that puts 
put} 
our approach to a real-world test: a thorough formal analysis of privacy
issues in the official Educational Censuses of Brazil,
the very large \INEP~\footnote{The An\'{i}sio Teixeira National Institute of Educational Studies and Research: \url{https://www.gov.br/INEP}} datasets.
%}

\subsection[What are the challenges?]{The challenge of flexibility}\label{s191650}

%\cmdraft{
%One challenge is that the current tools are not set up to be able to use for two reasons:  they are unable to provide a uniform basis for the attacks AND they cannot deal with longitudinal collections nor very large datasets.
%}

The first challenge is to ensure that all meaningful threats 
are recognised, whilst minimizing resources wasted on non-threats. 
And the problem here is that current attack practices are either 
\emph{ad hoc} or constrained to particular scenarios (as discussed ahead in \Sec{s191720}). 

The impact of focussing on known scenarios
is illustrated by the very comprehensive \ARX\ tool~\cite{ARX,Prasser14,Prasser15,Prasser20}:
it supports the analysis of re-identification risk under 
%\colorB{
	El Emam's ``prosecutor'', ``journalist'', and ``marketer'' attack models \cite{ElEmam13:Book,ElEmam13:Guide,Prasser20}.
%}
%
%precisely the attack roles (\Sec{sec:priv_attack_classes}) mentioned above
%\colorM{
\ARX\ has been remarkably successful in many applications --
including e.g. in the MIRACUM %medical data sharing 
network in Germany with data of about 3 million patients with 70 million facts \cite{Prokosch18}, 
and in a Norwegian re-identification analysis of
medical data with
over 5 million records \cite{Ursin17}.
However, 
\ARX\ could not manage \INEP's censuses:
%Despite \ARX's notable successes in many applications,\,%
%\footnote{\ARX\ was employed e.g.
%	by the MIRACUM %medical data sharing 
%	network in Germany
%	on data of about 3 million patients with 70 million facts \cite{Prokosch18}, 
%	and in a Norwegian re-identification analysis of
%	medical data with
%	over 5 million records \cite{Ursin17}.}
%it could not manage our case study 
%(\INEP's censuses):
the tool is limited to datasets of at most $2^{31}{-}1$ cells,\,%
\footnote{Noting this limitation,
	our team contacted \ARX's curators and discussed an update to overcome it~\cite{Ramon:20:GithubArxIssue}. The fix has been submitted and is now under evaluation.
	\msdone[Once paper is accepted, de-anonymize citation~\cite{Ramon:20:GithubArxIssueAnonimized} to
	~\cite{Ramon:20:GithubArxIssue}.][\ms]
}
that is ${\sim}23$ million records of 92 attributes each. That is smaller than {\INEP}'s dataset even for a single year.
%}
Furthermore, \ARX\ provided only a fixed selection of privacy degradation
measures, all of them related to re-identification and not e.g.\ supporting direct assessment of attribute-inference risks.
%\colorB{
But, more importantly, \ARX\ was not designed to support the full expressiveness of \QIF\ analyses, 
\review{including reasoning about longitudinal attacks in which the adversary has uncertainty about the linkage of a particular individual's records across datasets,}
and so it could not be naturally extended to encompass attack models other than those hard-coded in the tool already.
%} % (\Sec{s191632}).}
%And \ARX\ --supporting the ``standard'' attack roles rather than the more basic \QIF\ ``conceptual Lego''--, could not easily be extended for attacks other than those hard-coded 
%in the tool already (\Sec{s191632}).
Other popular tools face similar issues (as discussed in \Sec{s095512}).

\subsection[Where does ``scale'' come from?]{The challenge of
	scalability}\label{s103817}

Our concrete example --and the motivation for this work-- was
\INEP's\,
\review{longitudinal} collection of official educational-statistics datasets for the whole of Brazil.
Updated yearly since 2007, those contain \emph{microdata} (i.e.\ for individuals) 
for (nearly) every student in the country, and at all levels 
%\colorB{
(from elementary to graduate schools).
%}
Once processed, the data are released to the Internet 
%\colorB{
where they are freely available.
%}
%: there, they are available to everybody, friend or foe.
Even just 
%\colorB{
one year's data contain
%} 
about 90 attributes for approximately 50 million students --- around 25\% of the entire Brazilian population.
\review{This collection of official longitudinal microdata 
is conspicuously huge (even on the world stage).
It is used for governmental planning, especially in the 
allocation of the budget of the Ministry of Education's National Fund of Educational Resources,\footnote{\review{\emph{Fundeb}: \url{https://www.fnde.gov.br/financiamento/fundeb}}}
and by civil society both in Brazil and abroad in many ways,
including in demographic research \cite{Carnoy17,Dalmon19}, and policy-making and -monitoring \cite{Costa15,Martins16,Rosa19,Rosa20}.}
%\ghndone[Can you try to recover the Stanford institution that uses our data? Then extend the discussion on how the data is used.][\ms]

However, a new privacy law \cite{BR18} \review{inspired by} the European \GDPR\ came into effect in Brazil in 2021,
and {\INEP} was suddenly forced to perform a thorough
exploration of possible vulnerabilities in their datasets.
Although previous analyses had provided anecdotal examples of re-identification risks \cite{Queiroz15},
still there had been no systematic analysis of how widespread these risks actually were in the \emph{full} dataset collection.
\review{This demands the consideration of
the adversary's confidence in the accuracy 
of her linkage of records across the datasets 
in the longitudinal collection, which directly affects
also the accuracy of her inferences and, consequently, leakage.
This task is relatively easy if individuals' 
identifiers are persistent across datasets,
%every individual has the same
%identifier across all datasets, % in the collection, 
but becomes significantly more challenging otherwise.
Originally, \INEP\ intended to consider the latter case.}	
Thus the challenge of scale was to analyse
\emph{all} the data, including its longitudinal
aspects.
That is why we were contacted by {\INEP}.
\msdone[Once paper is accepted, maybe reference some documents related to \INEP's project and partnership with UFMG?][\ghn]

\subsection{The challenge of explainability}\label{ss1506}
%1728
\review{% 1731
The third challenge is that the university scientists who discover a vulnerability~\review{\footnote{\review{Note that the term \qm{vulnerability} used throughout refers to the \qm{risk} (to the secret); this is the terminology which has been adopted in the \QIF\ literature~\cite{Alvim20:Book}.}}} in \emph{mathematical} terms must be able to explain the threat it actually poses to those affected, and to do that in \emph{everyday} terms they understand. \emph{``There is a potential decrease of conditional Shannon Entropy''}, for example, may not convince government ministers that ``something must be done'' --- but \emph{``This inference attack might cost the data curator \$$N$''} could concentrate their minds wonderfully.%
% See http://www.thisdayinquotes.com/2009/09/hanging-it-concentrates-mind.html
}% 1731

\review{% 1732
Our \QIF\ approach has two conspicuous features addressing that kind of explainability. First (\Sec{s100834}), 
%the 
\QIF\ analyses %--i.e.\ the numbers produced-- model features that 
relate directly to specific adversarial attacks: % on the dataset: 
what is observed, what it might cost (the adversary) to attempt those observations, and what she might gain if she succeeds. If, e.g., %the 
government data scientists are concerned about re-identification, in \QIF\ terms that can be expressed as an adversary whose intent is to identify anyone at all. 
That is in contrast to threat measures that are based on traditional information theory (e.g.\ Shannon) and whose definitions were designed for quite different purposes (i.e.\ efficiency of encodings).
}% 1732

\review{% 1733
Second (\Sec{s201148}), once preliminary results are delivered, the government might then be able to clarify their concerns, to make them more precise based on what they have just learned: having seen the general risk posed by re-identification, they might \emph{then} be able to see that the concern is not just an adversary who could identify ``anyone'' but, rather, that it is a specific minority group that is now possibly at risk. The modular way in which \QIF\ describes threats allows computer programs built on \QIF\ principles to be re-run immediately with different parameters, instead of having to be re-coded, re-tested and only then re-deployed: for \QIF-structured tools a single change in the ``intent parameter'' might be enough, and the response to the more specific question could be very quickly given.
Quick responses to new questions suggested by earlier answers is a key factor in the explainability of anything.
}% 1733

\subsection{Overview of the {\INEPallcaps} case-study}

Here we look briefly at the genesis of our project \cite{InstitutoNacionalDeEstudosEPesquisasEducacionaisAnisioTeixeira2022,Alvim2022a}. 
More technical detail is given in \Sec{sec:inep-results-synopsis}.%\review{\footnote{\review{In case of acceptance of this manuscript, a link to the full official documentation of the project will be provided here.}}}
\ghndone[If the paper is accepted, fill in the link in this footnote.][\ms]
\mstodo[I have added as references \cite{InstitutoNacionalDeEstudosEPesquisasEducacionaisAnisioTeixeira2022,Alvim2022a}.][\ghn]

%\colorB{
In Brazil, the issues of transparency and of privacy
%preservation 
in the governmental release of data about individuals 
are regulated through two complementary laws%
\review{, further detailed in Appendix~\ref{s2202280000}, 
but whose essence is as follows.}
%} 
On the one hand, a \emph{transparency law} from 2011, known as {\LAI}~\cite{BR11},
adopts a philosophy of \qm{transparency by default}
and requires that information
be publicly available on the Internet: any exceptions 
must be properly justified.
On the other hand, \review{a} new \emph{privacy law}, {\LGPD}~\cite{BR18}, 
restricts the release of data on individuals, prescribing sanctions in the case of non-compliance.

In this context, we were contacted by \INEP\ to search for privacy vulnerabilities in their \emph{already published} datasets:
a longitudinal collection of ${\sim}50$ million records per year, each with ${\sim}90$ attributes. Their current measures
had focussed only on \emph{de-identification}, a known problem,
but the legislation itself did not limit the kind of leaks that might be exploited in the future. And here is where the issues of \emph{flexibility} and \emph{scalability} arise.

For example, it 
%\colorB{
was
%} 
known from the literature that
when non-unique attributes 
are released unaltered (e.g.\ date of birth, 
city of residency, gender),
then those attributes can act as \emph{quasi-identifiers} (QIDs), that is, in combination they can effect a de-anonymization~\cite{Dalenius86,Samarati98,Sweeney00,Narayanan08}.
As mentioned, anecdotal evidence of such risks
had already been identified in {\INEP}'s datasets~\cite{Queiroz15}, but
they were unquantified and narrow in scope.

%\mstodo[In the next paragraph say that, although we have not done it now, we might have to do it in the future. Talk about the law. Talk about Inep's change of policy. Reasonable attack models.][\ms]

More significantly, though, was the possibility of other attacks 
not considered by {\INEP} even anecdotally,
e.g.\ attribute attacks where knowing an individual's city of residence could be used to infer ethnicity. The legislation was broad enough to target those as well --- and of course possibly other attacks that \emph{no-one} had invented yet.
{\INEP} was thus forced %by the broadness of the legislation
to be prepared to look for breaches they had not yet considered%
\review{, and across the longitudinal collection.
That is, whatever we provided to {\INEP} had to be flexible 
enough, and longitudinally scalable, to handle and quantify \emph{future} risks too.}

%\item 
The issue of explainability was also formidable in two ways. First, we had to be able to convince {\INEP} that they were at risk even in cases they thought they were not. That meant putting into everyday terms --and quickly-- a quantified risk that \emph{anyone} could understand (and care about): 
\emph{\qm{Do you know that with 80\% probability we can from the existing data identify who y\underline{our} 
children are, and where they go to school?}} But this had to come from a rigorous mathematical analysis.

The second part of this challenge was that whatever changes {\INEP} was convinced (eventually) to make would \review{likely} face strong resistance 
from the public and lead suddenly to different, new questions --- and so, again, properly
justifying and communicating any change would have to be done carefully and quickly.
%~\footnote{As a related example, the US Census Bureau
%	has faced serious resistance from stakeholders when
%	discussing changes on the current balance between transparency and privacy
%	in their data-publishing methods~\cite{Garfinkel18,Mervis:19:Science}.}
%\colorM{
As a high-profile example, the US Census Bureau
has faced serious resistance from stakeholders when
discussing changes on the current balance between transparency and privacy
in their data-publishing methods~\cite{Garfinkel18,Mervis:19:Science}.
} % 1119

%\alldone[The next section --if the above is OK-- should be (re-)aligned with the new text above.][\ab\ + \cm]

\subsection{Our principal contributions}

The main contributions of this paper are the items below, addressing the challenges we have identified:

\begin{enumerate}
\item We re-factorize attacks %classifications 
along three orthogonal axes: %({\bf \Sec{s191650}}): 
%\begin{enumerate}
%\item 
%\quad 
(i)
\emph{the information sought by the adversary} (mem\-ber\-ship-inference, re-identification, or attribute-inference);
%\item 
%\quad 
(ii)
\emph{the adversary's target} (fixed-individuals, or collective targets); and
%\item 
%\quad 
(iii)
\emph{the adversary's access to datasets} (single datasets, or longitudinal collections).
%\end{enumerate}
As well as comprehensively covering the relevant operational 
scenarios from the literature, this re-factorization identifies some new ones (\Sec{s191720}).
	
%\item \colorB{We use the re-factorization above within a coherent formal framework grounded on {\QIF}, identifying how to instantiate various parameters
%to rigorously capture various attack models of interest	
%	 (\Sec{s100834}).}
\item %\colorB{
	We use the re-factorization above within a coherent formal framework grounded on {\QIF}.
We devise a non-traditional instantiation of the role 
of the adversary's prior knowledge and the channel in 
the \QIF\ model that allows, at the same time, for:
(i) a realistic capture of \INEP's scenario -- in which 
datasets were already of public knowledge even before 
any attack was performed; and
(ii) tractable computations of analyses (\Sec{s100834}).
%}

%\item  We develop algorithms to compute the \QIF\ formulations defined above for very large datasets.

\item We illustrate the flexibility and scalability of our approach with extensive experimental evaluations 
	of both re-identification and attribute-inference attacks
	 in {\INEP}'s extremely large longitudinal collection of Educational Censuses datasets (Sec.~\ref{s091634a} and \ref{s1011aa}).
	To the best of our knowledge, these analyses 
	are the largest and most thorough in scope ever performed 
	on publicly available governmental microdata, and they reveal several insights about the privacy issues of such 
	large releases.
\end{enumerate}	
%\item[] In addition:
Additionally, we provide a free, optimized tool 
%with an optimized implementation 
of our attacks and privacy analyses
(\Sec{s201143}).

\msdone[Point 3. says we develop algorithms but we don't mention them in this paper. I think it should be reworded as a contribution. Also, are we providing an optimised tool? Or a tool only optimised for this dataset?][\tash]
\tashtodo[The algorithms are just optimizations present in the tool. The algorithms are not specific for this dataset, they can work on all datasets in a similar format. I'll ask Gabriel to confirm this.][\ms]
\ghntodo[Can you confirm the above?][\ms]
\alltodo[The tool works on any dataset consisting of microdata, but I am not sure what is meant by \qm{optimized} here. We have managed to compute our results faster with our algorithm than if we had used pandas methods, which in theory are optimized to work with datasets. But when it comes to memory use, I am not sure if we could have done better.][\ghn]

% QUOKKA!!! --- Where do we say how to get the tools?
%\alltodo[Regarding QUOKKA: There is a default section called ``availability'' for USENIX papers, it's at the end of the paper, close to acknowledgments. 
%We must give the link to the tool there, in the final version of the paper (before the final version it may go against anonymity).
%For now we also mention the tool in \Sec{s201143}; we are just waiting for Gabriel to create the link to it (but we also have to keep the link anonymous for review).][\ms]
%\alltodo[Sec.~\ref{s000000}.][\ghn]
	
%\item We compile and present thkey  lessons learned in 
%	the course of our project.
%	For as mathematically sound and experimentally thorough 
%	any formal analysis might been, it can only foster 
%	real change if it persuades decision- and policy- makers 
%	of privacy issues at a concrete level, and then empowers
%	them to make well informed decisions.
%	In this context, we had to identify --and even develop-- ways 
%	to effectively	transmit the results of our formal analyses 
%	to {\INEP}'s agents.

\textbf{Ethics considerations.}
All results in this paper were obtained in a formal 
cooperation with {\INEP}, at their request,
and fully communicated to them.
The agreement permits publication of all 
vulnerabilities found, including all those identified in this paper.
%\colorB{, including all those identified in this paper.
%} 
%The analyzed datasets have been made freely available at the Internet by {\INEP} itself, and 
Following Brazil's transparency law, the datasets and 
all results found are %\colorB{
freely
%} 
available to any citizen.

%\msdone[Just trying to emphasise that everything in this paper is allowed to be published and is done so with the knowledge of INEP.][\tash]

\textbf{Plan of the paper.} 
\Sec{s102913} explains our re-factorization of 
privacy attack models; 
\Sec{s100834} introduces the {\QIF} framework and shows
how it enables the re-factorization; 
%\colorB{
\Sec{s091634a} describes our case study with {\INEP}'s datasets;
% (and their co-operation);
%}
\Sec{s1011aa} explains the vulnerabilities discovered;
\Sec{s091634} presents lessons learned; 
\Sec{s191632} considers prospects;
and
%Finally, \Sec{s191632} presents future work, and
\Sec{s095512} discusses further related work.

%-----------------------------

%=============================
% SECTION
%============================
\section[Rationalizing the privacy landscape]{Rationalizing the landscape of privacy attack models}\label{s102913}

% !TEX root = main.tex

%\cmdraft{
%\begin{itemize}
%	\item The first contribution was to rationalize the privacy vulnerabilities/attacks discovered in the literature (prosecutor etc.)
%	\item Recognize that they can be categorized in a uniform way
%	\item Say what that categorization is
%	\item Notice in particular that all problems have 3 main features: prior knowledge, adversarial intent, information leak model.
%	\item In fact because of that we can express all these privacy scenarios using {\QIF}.
%\end{itemize}
%This Sec.\ \ref{s102913} should finish with a sentence or two that says ``{\QIF}'' is what we will use to do that.
%}

%\msdraft{Old Mario's notes from before the re-organization of the paper on 2021-08-25/26:
%	\begin{itemize}
%		\item 	I think this subsection needs to be thoroughly rethought. 
%		What is important here is only attack models (journalist, marketer, prosecutor, etc.),
%		longitudinal attacks, famous attacks from the literature, and the ARX tool. 
%		\item Perhaps some interesting information (one paragraph or two) on the general legislation about privacy and, in particular, on the Brazilian LGPD that motivates our work.
%		Little else is needed.
%		Concepts such as k-anonymity, l-diversity, t-closeness, dp, etc. should be mentioned only en-passant, and only in reference to existing famous attacks or the ARX tool.
%		Our paper is focused on vulnerabilities, not
%		on mitigation methods.
%	\end{itemize}
%}

There are currently many different approaches to the classification of privacy attacks:
according to the adversary's goals
(e.g.\ membership, identity, or attribute disclosures)~\cite{ARX,Divanis15:Book,Fung10:Book};
according to her target and prior knowledge 
(e.g.\ the prosecutor, journalist, and marketer models)~\cite{Prasser20,ElEmam13:Book,ElEmam13:Guide} etc.
They have been adopted in practice by the popular
ARX data-anonymization tool~\cite{ARX}, among others.
However, their motivation comes from a small number of \emph{concrete} scenarios, rather than
being organized systematically along independent dimensions
and, as a result, the identified attacks might fail to cover the 
threat landscape. (For example,
 attribute-inference attacks on longitudinal collections).
%been studied?)
%\tashtodo[To connect this with the motivation
%of this paper, maybe we can add a sentence saying
%something along the lines that this is a significant
%challenge practitioners may overcome when they
%have to make sure they are providing a thorough
%privacy analysis in real, large-scale scenarios,
%with several potential social repercussions, like
%the one that motivates this paper.][\ms]

%\tashdone[Some extra blurb on ARX was moved to Sargasso]

And so this section rationalizes 
existing attack models into a unified classification which not only covers %\colorB{
various
%} 
attack models already known, but  identifies some new ones. 
We begin by visiting the existing models.

%\subsection{An empirical classification of models}\label{sec:priv_attack_classes}
\subsection{An empirical classification of models}\label{sec:priv_attack_classes}

Some works focus on re-identification of individuals in a microdata release~\cite{Prasser20,ElEmam13:Book,ElEmam13:Guide}. Re-identification attacks are 
(considered to be) of three types depending on the adversary's prior knowledge and target:%
~\footnote{These are described in \cite{ElEmam13:Book} as \review{\qm{risks}}, but here we use the term \review{\qm{models}}.}
\begin{itemize}
	\item The \emph{Prosecutor} attack model: the adversary tries to re-identify a specific individual (target) whose data is known to be in
    the dataset of interest.
%    \alltodo[Why are attack names capitalized (they aren't elsewhere in this paper).][\ms]
	
	\item The \emph{Journalist} attack model: the adversary tries to re-identify a specific individual whose data is \emph{not necessarily} known to be in
    the dataset of interest.%She does it anyway.	
	\item The \emph{Marketer} attack model: the adversary tries to re-identify as many individuals as possible in the dataset of interest.
%\tashdone[I don't \emph{quite} understand the distinctions here. Could we discuss it?][\cm]
% i.e. it measures how many records, on average, could be re-identified.
%	\tashdone[For the prosecutor and journalist models you specified the adversary's target (individual vs.\ collective target) and prior knowledge (target is in the dataset vs.\ isn't), but not the corresponding measure of success. 
%	For the marketer you also specified the measure of success. Should we just say ``the adversary's goal is to re-identify as many individuals as possible, no matter who they are, in the dataset of interest''. 
%	Notice that the classification above is already not orthogonal, because it confounds target with prior knowledge.][\ms]
%	\tashdone[Isn't our \emph{real} view that the adversary's goal is to maximize $V_g$ (minimize $U_\ell$ wrt.\ to the $g$ or $\ell$ by which she's characterized? It's the conversion of the dataset into a channel (which can be done in many ways) and the choice of $g/\ell$ that determines what kind of attack it is. The question is ``Where in this report to we say that?'')][\cm]
%	\tashdone[That is, ``they'' work directly from the data set, and do different things depending on what kind of attack they are modelling or carrying out. \emph{We} on the other hand convert the dataset to a channel first, and choose $g/\ell$ and, from that point on, do the \emph{same} thing no matter what kind of attack it is.][\cm]
%	\cmdone[Is it possible to describe this in \Sec{s100834}?][\tash]
\end{itemize}

% Need to look at the landscape of threats
% Need to say that these come from different groups and can lead to some confusion.
% Book describes prosecutor and journalist as risks, but now they are called models.
% Attacker model
% Can also call the above 3 `scenarios'
% Make clear the new dimensions or why we defined them.
% Where is longitudinal and single dataset?
% Call {\QIF} models vs other models

Yet there are other works that classify according to the type of information sought \cite{ARX,Divanis15:Book,Fung10:Book}:

\begin{itemize}
	\item The \emph{Membership-inference} model: the goal is only to infer whether individuals' data 
	%are present 
	\review{appear}
	in a dataset.
%	\alltodo[Why are attack names capitalized (they aren't elsewhere in this paper).][\ms]
	
   \item The \emph{Re-identification} model: the goal is to link data records to the individuals to whom they refer.
   
   \item The \emph{Attribute-inference}  model: the goal is to infer the value of a sensitive attribute for individuals, 
   %\colorB{
   regardless of whether 
   they were re-identified.
	%}   
   %it was possible to re-identify them.
\end{itemize}

Because the scenarios above pertain to 3 main adversarial features -- her \emph{prior knowledge}, her \emph{targets}, and the \emph{information} she wishes to obtain --  we are now able to suggest a unified classification of attack models.

%We start by proposing a systematic 
%re-factorization of attack classifications
%which systematically covers a variety of 
%relevant operational scenarios from the literature,
%and additionally identify some novel ones.

%As discussed in \Sec{sec:background-attacks}, 

\subsection{An orthogonal classification of models}
\label{s191720}

We re-factorize attacks along three orthogonal axes:

\begin{itemize}
%\begin{itemize}[{Axis} I:]
%\paragraph*{%Axis 1: 
\item \textbf{Axis I:}
\textbf{The information sought by the adversary.} 
We consider
(\MMM) \emph{membership-inference}, %in which the adversary's goal is to infer whether or not individuals are represented in a dataset;
(\RRR) \emph{re-identification}, %in which her goal is to link dataset records to the individuals to whom they refer; 
and
(\AAA) \emph{attribute-inference}.%, in which her goal is to infer the value of a sensitive attribute for individuals, regardless of whether or not it was possible to re-identify these individuals.
%\footnote{Note that re-identification attacks are stronger
%	than both membership and attribute-inference attacks, since success on the
%	former implies success on the latter.}

%\paragraph*{%Axis 2: 
\item \textbf{Axis II:}
\textbf{The adversary's target.}
We consider
 (\III) \emph{individual-targets}, where her goal is to obtain sensitive information on a specific individual; and
 (\CCC) \emph{collective-targets}, where
her goal is to obtain 
sensitive 
information on as many individuals as possible, no matter who they might be.

%\paragraph*{%Axis 3:
\item \textbf{Axis III:} 
\textbf{The adversary's access to datasets.} 
%\colorB{
	We consider
(\SSS) \emph{single-dataset access}, of a single dataset corresponding 
to a specific point in time; and
(\LLL) \emph{longitudinal-dataset access}, where
several versions are accessible, each %corresponding to
for a different time.
%}
\end{itemize}

The above axes yield $3{\times}2{\times}2=12$ 
possible
combinations of attack models, given acronyms in \Table{tab:attack-class}.
%\alldone[Should we mention why some acronyms are
%highlighted in Tab.~\ref{tab:attack-class}, and in two different styles? And do we even need to highlight them at all?][\ms]
%This classification covers the popular attack 
%models mentioned in \Sec{sec:priv_attack_classes}.
Thus  %\Sec{sec:background-attacks}:
the prosecutor model corresponds to \IRS,
the journalist model to \IMS,
and the marketer model to \CRS\ (all bracketed in the table).
But our re-factorization 
covers many other relevant scenarios as well, 
such as e.g.\ attribute-inference attacks on longitudinal
collections (\CAL\ and \IAL, underlined in the table).
%It is important to notice that this categorization captures the popular adversary models mentioned in \Sec{sec:background-attacks}, with the following equivalences: \IRS\ and the prosecutor model, \IMS\ and the journalist model, and \CRS\ and the marketer model.
%\ghndone[Can you make a list of which existing model corresponds to which attack in our Table?
%This is important to highlight why the re-classification is relevant.][\ms]

%In the above classification we did not make use 
%of the adversary's possible \emph{prior knowledge}, which 
%is feature of some of the proposed models. 
In the next section we show how
%prior knowledge, \emph{as well as} 
the above adversarial features are naturally represented in the threat model provided by the \QIF\ framework.
%\tashdone[I removed the comment about the prior not being used -- it is determined by the adversary's access.]

\begin{table}[bt]
	\centering
	\renewcommand{\arraystretch}{1}
	\begin{small}
		$	\begin{array}%{|>{\centering\arraybackslash}m{0.16\columnwidth}||>{\centering\arraybackslash}m{0.12\columnwidth}|>{\centering\arraybackslash}m{0.16\columnwidth}||>{\centering\arraybackslash}m{0.12\columnwidth}|>{\centering\arraybackslash}m{0.16\columnwidth}|}
			{|c||c|c|c|c|}
			\cline{2-5} 
			\multicolumn{1}{%>{\centering\arraybackslash}m{0.16\columnwidth}|
				c|}{} & \multicolumn{2}{%>{\centering\arraybackslash}m{0.33\columnwidth}||
				c|}{\textbf{Single-dataset (\SSS)}} & \multicolumn{2}{%>{\centering\arraybackslash}m{0.33\columnwidth}|
				c|}{\textbf{Longitudinal (\LLL)}} \\ 
			\cline{2-5} %\noalign{\vskip 1mm} \cline{2-5}
			\multicolumn{1}{>{\centering\arraybackslash}m{0.16\columnwidth}|}{} & \textbf{Ind.\ (\III)} & \textbf{Col.\ (\CCC)} & \textbf{Ind.\ (\III)} & \textbf{Col.\ (\CCC)} \\ \cline{2-5} \noalign{\vskip 0.5mm} \hline
			\textbf{Memb.\ (\MMM)}  & \bf[\IMS]  & \CMS & \IML & \CML \\ \hline
			\textbf{Re-id.\ (\RRR)} & \bf[\IRS]  & \bf[\CRS] & \IRL & \CRL \\ \hline
			\textbf{Attr.\ (\AAA)}  & \IAS & \CAS & \underline{\textbf{\IAL}} & \underline{\textbf{\CAL}} \\ \hline
		\end{array}
		$
	\end{small}
	\renewcommand{\arraystretch}{1}
	\caption{Re-factorization of attack models and their acronyms.}
	\label{tab:attack-class}
\end{table}

%Each attack can, then, be formalized in the {\QIF} framework.
%We explain how this can be done the following section.

%-----------------------------

%=============================
% SECTION
%=============================
\section[Quantitative information flow]{%\colorB{
Quantitative information flow:  what it is, and how it induces rationalization%
%}
}\label{s100834}

% !TEX root = main.tex

%\cmtodo[I moved most of the comments a few pages further on, because it was getting too hard to read. (Follow the little marginal links.) I think the main one is Annabelle's 1529, that we are supposed to be linking \QIF's systematic separation and classification of concepts to the unsystematic and overlapping collection of concepts that are ``out there'' and (presumably) were surveyed in an earlier section. That is, I would like to do that in a more structured way than the current version here does. I will go and have a look at her alternative version (in Sargasso) and see whether I can pluck some material from there.]

%=============================
% SECTION
%=============================

%\cmdone[I think the main point of this QIF section is to link it to the previous section ... or at least that is what I was thinking when I suggested this way around. We don't have to say anything about what is relevant in practice, but merely to point out that those axes identified above in the rationalization can all be implemented with QIF. Otherwise this section might seem like it isn't connected?][\ab]

\Sec{s102913} just above surveyed traditional threats to privacy, and in particular their extensive nomenclature (prosecutor, journalist, marketer, etc.).
\emph{Quantitative Information Flow} ({\QIF}) provides a mathematical model in which those varying points of view can almost all be seen as aspects of the same thing, thus streamlining the conceptual approach required: we can therefore focus on the small number of \emph{technical} elements that cause the threats, and treat them in a unified way. \QIF\ can streamline the computations as well, as we see below\review{,} and that helps with scalability.
% KOOKABURRA

\review{\textbf{The philosophy of \QIF.}
The \QIF\ framework's focus is to 
capture the adversary’s knowledge, goals, 
and capabilities, and from that quantify 
the leakage of information caused by a corresponding 
optimal inference attack.
The framework is grounded on sound information-- and 
decision-theoretic principles enabling the rigorous 
assessment of how much information leakage a system allows
\emph{in principle}, and independently from the adversary's computational power~\cite{Alvim20:Book}.
Hence, \QIF\ guarantees hold no matter the particular tactic
or algorithm the adversary employs to execute the attack, 
as what is measured is exactly how much
sensitive information is leaked by the best possible 
such tactic or algorithm. 
}

\review{\textbf{Overview of privacy models in \QIF.}}
{\QIF} (we will see) separates (1) the adversary's \emph{knowledge} from (2) the description of the ``leak'' she is trying to exploit, and that %(2) 
leak description is again separated from (3) her \emph{intentions and capabilities}. The first (1) is  modeled as a probabilistic ``prior''; the second (2) is modeled by Bayesian reasoning to produce a ``hyper-distribution''; and the third (3) is modeled as a 
%``gain'' or ``loss function'',
``gain function''
%
%``vulnerability'' (derived from a ``gain function'') or equivalently an uncertainty (derived from a ``loss function''), both of which are functions from hypers
%%\cmdone[Actually they are functions from \emph{distributions} to reals, and their ``conditional'' versions apply to hypers. Which do we want to emphasize here?]
that gives what could almost 
be regarded as %\$-values. 
monetary values.
We introduce those in turn.
%Later we will write $V_g$ for a vulnerability (function) derived from gain-function $g$, and $U_\ell$ for an uncertainty derived from loss-function $\ell$.
%It is {\QIF}'s smooth treatment of the three together that makes it such a useful conceptual tool.

A \emph{prior} (1) is a 
%\colorB{
probability
%} 
distribution over unknown (but sought after) data, and it models the adversary's knowledge about 
that data even before any leak occurs:
\emph{how likely is it that this person is unmarried?} \emph{How likely is it that this row of the dataset describes Warren Buffett?} A \emph{gain function} (3) for an adversary gives a numerical (expected) valuation of the benefit to her of learning that information: the gain function of an adversary seeking a partner would be high in the first case, but low in the second; 
%\colorB{
	but if she wants to raid a bank account, it would be the opposite. Varying the gain function is how we %will
formalize the attacks from \Table{tab:attack-class}.
%}
%\cmdone[Bill Gates is divorced now. 
%Hence, I think the example isn't quite clear, because he cannot be ruled out as a potential unmarried partner. 
%What if we replace the guy with, say, Mark Zuckerberg?][\ms]

A \emph{hyper-distribution} (2) summarizes mathematically how an adversary uses Bayesian reasoning to exploit an information leak: we write ``hyper'' for short. The specifics of the  information leak are described by a channel: for each possible secret that the adversary wants to learn there is some probability that a particular output (coming from the leak) is observed by the adversary. Combining the channel probabilities with the prior enables posterior reasoning using Bayes' rule so that the adversary is able to revise her knowledge about the potential value of the secret, and better align her intent with what she has just learned. The hyper organizes this reasoning as a marginal probability over observations and, for each observation,  a posterior probability distribution over the secret values. 
% NATASHA-REVIEW: Removed as per reviewers comments. Also this is no longer referenced ahead?
%And --as mentioned at the end of \Sec{s201149}-- a hyper can be a much more compact representation 
% than the conventional conditional-probability matrix, because a \emph{single} posterior can stand for the merging of \emph{many} columns in the matrix: that is how the computations are streamlined.

%is describes mathematically the effect of an information leak exploited by the adversary: it converts a prior into a \emph{distribution} of possible posteriors, each one of which --again via the gain function-- gives an expected valuation to the adversary of using the leaked information. For example a leak of an individual's age would (probably) be used by the adversary to decide whether to go for a partner, or to go for raiding a bank account.
%\cmdone[Can the term `hyper-distribution' be introduced here? It is mentioned later on but not defined (as far as I can see). Also, can we stick just to `channel' rather than abstract channels?][\tash]
%\cmdone[Oh, is the problem his age? Come on, people can get married at any age (above 18)!][\ms]

To be concrete for a moment, we mention that
a popular gain function
is the ``Bayes Vulnerability'' which rewards a correct guess of a secret's value 
%\colorB{
with 
%(so to speak) 
%\$1 
1
if the guess is correct and 
%\$0 
0
otherwise.
%}
It (and other functions like it) was just what was needed by {\INEP} to provide the Brazilian government with hard scientific evidence to estimate the vulnerability of re-identification such as \textit{``There is an 80\% chance that a randomly selected individual can be re-identified in the currently published microdata.''}
%\alltodo[The reader may get confused about the connection between the monetary interpretation of a gain function and the following interpretation in terms of  probabilities.][\ms]
\review{A further benefit of 
	%the approach we are now describing 
	this approach
	is that these \QIF-categorized concepts, 
which can be distilled and explained in terms that {\INEP} care about, can be computed \emph{at scale} if carefully worked out and optimized.}
%That is one way that {\QIF} provides an important bridge, linking the \emph{social} aspects of privacy to the \emph{technical} aspects of privacy.

\review{\textbf{The components of a \QIF\ model, with an example.}}
We now return to the more technical aspects of the {\QIF} model and how it relates to datasets,
how a \emph{secret} is a value of some type \calx, and a secret (data) release, which in \QIF\ is called a \emph{channel}, is a 
%\colorB{
(probabilistic)
%} 
function from \calx\ to some set of observations \caly,
%\,%
%\footnote{More generally, {\QIF} can model probabilistic releases, but we do not (yet) need that feature for the study reported here.}
and how an \emph{adversary} is abstracted to a (gain) function that can be applied to the \review{hyper}, induced by a channel, to determine the advantage accruing to the adversary from using that channel.
%\cmdone[I don't see how to avoid referring to p.\pageref{p1522} here; but I have replaced the parts I used with (not-yet-properly) \LaTeX'd versions (Tab.~\ref{t104949}). I use only its \ref{t104949-a} and \ref{t104949-f}; we could include just those here, and refer to Natasha's calculations for the intermediate details; but also Tab.~\ref{t104949}'s calculations could be given in the Appendix, and might help people to understand Natasha's larger calculations if they are seeking confirmation that they understand..]
%\cmdone[I think it's quite good as it is! The reference forward is quite reasonable.][\ms]

\begin{table}[!tb]
	\begin{subtable}[t]{0.49\linewidth}
		\renewcommand{\arraystretch}{1}
		\centering
		\begin{small}
%			\begin{tabular}{|>{\centering\arraybackslash}m{0.28\linewidth}|
%			>{\centering\arraybackslash}m{0.16\linewidth}
%			>{\centering\arraybackslash}m{0.10\linewidth}
%			>{\centering\arraybackslash}m{0.08\linewidth}|}
%				\hline
%				\raisebox{-0.75mm}{attributes $\blacktriangleright$}
%				& 
%				\multirow{3}{*}{\rotatebox{45}{\small{\textit{lang.}}}} 
%				&			\multirow{3}{*}{\rotatebox{45}{\small{\textit{gender}}}}
%				&			\multirow{3}{*}{\rotatebox{45}{\small{\textit{age}}}}
%				\\[1mm] \cline{1-1}
%				& 
%				& 
%				& 
%				\\[-3mm]
%				\id\ $\blacktriangledown$
%				& 
%				& 
%				& 
%				\\
%				\hline
%				\rowcolor{sand!50!white} 1 & \texttt{Eng.} & \male & ${>}30$ \\
%				2 & \texttt{Port.} & \male & ${\leq}30$ \\ 
%				\rowcolor{sand!50!white} 3 & \texttt{German} & \female & ${\leq}30$ \\
%				4 & \texttt{German} & \male & ${\leq}30$ \\ 
%				\hline
%			\end{tabular}
			$
			\begin{array}{|>{\centering\arraybackslash}m{0.04\linewidth}|>{\centering\arraybackslash}m{0.22\linewidth}|>{\centering\arraybackslash}m{0.15\linewidth}|>{\centering\arraybackslash}m{0.13\linewidth}|}
				\hline
				\id & \textit{lang.} & \textit{gend.} & \textit{age} \\ \hline 
				\rowcolor{sand!50!white} 1 & \texttt{English}  & \male & ${>}30$  \\ 
				2 & \texttt{Port.}  & \male & ${\leq}30$  \\ %\hline
				\rowcolor{sand!50!white} 3 & \texttt{German}  & \female & ${\leq}30$  \\ 
				4 & \texttt{German}  & \male & ${\leq}30$  \\ \hline			
			\end{array}
			$
		\end{small}
		\caption{%Caption for \ref{t104949-a} 
		Original dataset.}
		\label{t104949-a}
	\end{subtable} 
	\hfill
	\begin{subtable}[t]{0.49\linewidth}
		\renewcommand{\arraystretch}{1}
		\centering
		\begin{small}
			$
			\begin{array}{|c|c|}
				\hline
				\textit{language} &
				prior%~$\blacktriangledown$
				\\ \hline
				\rowcolor{sand!50!white} \texttt{English} & \nicefrac{1}{4} \\
				\texttt{Port.} & \nicefrac{1}{4} \\
				\rowcolor{sand!50!white} \texttt{German} & \nicefrac{1}{2} \\
				\hline
			\end{array}
			$
	\end{small}
	\caption{%Caption for \ref{t104949-g} 
	Prior %knowledge 
	on \textit{language}, the adversary's knowledge \textit{before} the leak.}
	\label{t104949-g}
	\end{subtable}

%	\vspace{2mm} %\hfill
	
	\begin{subtable}[b]{\linewidth}%{0.42\linewidth}
	\renewcommand{\arraystretch}{1}
	\centering
	\begin{small}
			$
			\begin{array}{|c|c c c c|}
				\multicolumn{1}{c}{\text{outers~$\blacktriangleright$}}
				& \multicolumn{1}{c}{$\nicefrac{1}{2}$}
				& \multicolumn{1}{c}{$\nicefrac{1}{4}$}
				& \multicolumn{1}{c}{$\nicefrac{1}{4}$}
				& \multicolumn{1}{c}{$0$}
				\\[1mm]
				\hline
				\raisebox{-0.75mm}{\textit{gender, age} $\blacktriangleright$}
				& 
				\multirow{3}{*}{\rotatebox{90}{\small{\male\  $\leq30$~}}} 
				&			\multirow{3}{*}{\rotatebox{90}{\small{\male\  $>30$~}}}
				&			\multirow{3}{*}{\rotatebox{90}{\small{\female\  $\leq30$~}}} 
				&			\multirow{3}{*}{\rotatebox{90}{\small{\female\  $>30$~}}}
				\\[1mm] \cline{1-1}
				& 
				& 
				& 
				& 
				\\[-2mm]
				\raisebox{0.5mm}{\textit{language}~$\blacktriangledown$}
				& 
				& 
				& 
				& 
				\\
				\hline
				\rowcolor{sand!50!white} \texttt{English} & 0 & 1 & 0 & 0 \\
				\texttt{Portuguese} & \nicefrac{1}{2} & 0 & 0 & 0 \\ 
				\rowcolor{sand!50!white} \texttt{German} & \nicefrac{1}{2} & 0 & 1 & 0 \\ \hline
			\end{array}			
			$
		\end{small}
		\caption{%Caption for \ref{t104949-f} 
		Hyper-distribution over \textit{language} given {\gender} and {\age}, modeling the adversary's knowledge \textit{after} the leak -- the outers constitute the probabilities for each observation and the posterior probability distributions in each column summarize what the adversary has learned about the language.}
		\label{t104949-f}
	\end{subtable} 
	
%\bigskip{\it 210916: Still having some trouble with positioning the rotated entries, and the missing right borders\ldots SHAARK}
%\alltodo[Regarding SHAARK: Let me know what rotated entries you want, and I'll do it. Regarding missing right borders: they aren't missing, just not rendered in some pdf viewers at the default zoom; just zoom in and you'll see them.][\ms]
\caption{Summary of the \QIF\ analysis for leaking information about native language from a table of microdata.}
%Caption for Tab. \ref{t104949}}
	\label{t104949} 
\end{table}	

In \Table{t104949-a} we have a 4-row dataset giving for each individual the native language spoken (English, Portuguese, German), the gender (\male, \female) and the age (${\leq,>}30$). The adversary is trying to guess the native language of the person she is about to meet (but has not yet seen), and she assumes the person \review{selected} is equally \review{(i.e.\ uniformly)} likely to be any one of the four in the dataset. We describe her with a gain function yielding \$4 if she guesses right, and \$0 if she guesses wrong. 
The adversary's prior on language (i.e. her knowledge about the sought secret even before meeting the person) is shown in
\Table{t104949-g}, and clearly she will guess German
(the most likely language): 
an expected gain of \$2.

The full procedure for converting the dataset in \Table{t104949-a} into a hyper as shown in \Table{t104949-f} is given in Appendix~\ref{s2110071153} and
used in detail in \Sec{sec:inep-case-study-examples} 
%\colorB{
with a 
%larger, 
more realistic example. We \review{continue} with
%will 
%however 
the small example here to illustrate the systematization that \QIF\ allows.
%}
%\cmdone[Don't you want to mention the prior and the prior vulnerability? Now we have Fig.~\ref{t104949-g} with the prior. If you don't need it, you can comment out the figure.][\ms]

If now our adversary \emph{sees} the person before guessing, the gender and age are leaked. We illustrate the \QIF\ approach by showing that her expected gain increases to \$3. 
%\cmdone[I'm not sure I can parse the above sentence.][\ms]
From Fig\,. \ref{t104949-f} %(final table) 
%we notice that
she sees a ``young'' man with probability \NF{1}{2} and
%that the 
\review{the posterior probabilities for language become \NF{1}{2} for both Portuguese and German:}
%probability that man speaks Portuguese is \NF{1}{2} and the same for German:
so she will guess one of those. If however she sees an old man,
\review{%which happens
with probability \NF{1}{4}}, she will guess \review{(definitely)} English; and if she sees a young woman, she will guess German. (There are no old women in the dataset.) Her expected gain is now $\$4\times(\NF{1}{2}{\times}\NF{1}{2}+\NF{1}{4}{\times}1+\NF{1}{4}{\times}1) = \$3$.
%\cmdone[I tried to fix the table for you (and I also think some numbers were wrong, can you please check I didn't introduce any mistake)? 
%I also separated the prior from the hyper, because I thought it was confusing with them together; if you don't like it, I can revert it easily later.
%In any case, if you want, I can create the other tables to put in the appendix when I'm back from my break, I think it'll take me less time than it'll take you, since I'm used to them now.][\ms]
%\msdone[They look good. Thanks! And the numbers look fine. (What mistake did I make?)][\cm]
%\cmtodo[Well, I don't remember now :-) We'd have to recover the old files...][\ms]
%\cmdone[This is a nice narrative! But I'm not sure it covers exactly how the hyper was produced from the channel and the prior. I thought this was the purpose of this section.][\tash]
%\tashdone[You are right: it does not; and it was the purpose. There are more figures (that M{\'a}rio refers to above) that I think include more of the process; but we might avoid the space they would take by referring forward to the complete process in a larger example --- if there is one. Is there? If not, then maybe those figures could indeed go into the appendix, and be referred to from here.][\cm]
Therefore, the leak has the effect of increasing our particular adversary's expected gain from \$2 to \$3.

The example illustrates further orthogonal decomposition (beyond \Sec{s191720}) that {\QIF} enables:
%\alltodo[The ``orthogonal decomposition'' we are referring to here is not exactly the same decomposition of attacks into axes we have presented before, is it? Should we clarify further what we are decomposing, or maybe use a different choice of words?][\ms]
\begin{enumerate}
\item The dataset(s) and their structure are separated from the attacks that might be mounted: they are simply ``there''. 
The datasets used in a particular longitudinal attack are aggregated by some method:
\review{if%in case
} there is a persistent unique identifier for all individuals across all datasets, the aggregation can be done with a simple left outer join keyed on
\review{% this attribute (we discuss general alternatives for when such attribute is not available in \Sec{s191632}).
that attribute. (In \Sec{s191632} we discuss general alternatives for when such an attribute is not available.)}
% WOMBAT!!! Why is this OK, but for ARX it was not?
%\alltodo[Regarding WOMBAT: It would be ok for ARX if the datasets had a persistent unique identifier for all individuals. Then ARX would work just fine. If such an attribute doesn't exist, the linkage would be probabilistic, and ARX can't account for uncertainty in the dataset itself, so the resulting analyses would be imprecise. I rewrote the paragraph above, see if it's ok now. For more info, look for CAPYBARA.][\ms]
\item The ``\review{selection} prior''
\review{%belief 
(on records, often} uniform, as above), is separated from the actual prior (\review{induced by the attack,} here that the language spoken is twice as likely to be German as either of the other two).
%\cmdone[I'm not sure I understand this point number 2.][\ms]
\item The \review{%selection
selections} of ``what attribute(s) are sought'' and ``what attributes are leaked'' are separated from the adversary's other characteristics: they determine only what become the rows and columns of the synthesized channel matrix.
\item The posterior inferences the 
leaks might enable (revised-belief distributions over the secret) are separated from their worth to the adversary (\review{i.e.\ are captured independently in the} gain function).
\item \review{Indeed the%The
} worth to the adversary of the information a leak delivers (gain function) 
%\colorB{
is %in fact 
\review{%completely
\emph{completely}}  independent
%} 
of all other factors,
\review{%independent
in} particular of the prior, and of how many datasets were involved.
\end{enumerate}

\review{\textbf{The flexiblity of the \QIF\ framework.}
\QIF\ models are flexible by design:
once a data release is modeled as a channel,
it is easy to switch between various attack scenarios 
by changing the probability distribution 
modeling the adversary's prior knowledge, and the gain 
function modeling her goals and capabilities.
Moreover, even in scenarios where the adversary's 
prior knowledge, goals, and/or capabilities are not 
fully known, the framework provides 
quantified worst-case %but still quantified 
estimates of damage
based on the theory of ``channel capacities'' \cite{Alvim:14:CSF}.%
%%\colorB{
%quantified worst-case %but still quantified 
%estimates of damage
%%}
%based on ``channel capacities'' \cite{Alvim:14:CSF}.
%;we discuss that in \Sec{s191632}.
%\See{1522}
}

\review{\textbf{Computing leakage with a \QIF\ model.} 
\QIF\ is a
\emph{way of modeling attacks},
not an \emph{implemented tool} in itself.
%Indeed, we need to code \QIF\ models into some programming language 
%to effectively \emph{compute} leakage.
Existing general-purpose implementations of the 
\QIF\ framework
make leakage computation tractable in a range of small to 
medium-sized scenarios, without the need to write new code for new attack models: it suffices 
to simply change some parameters.\footnote{\review{An example is
		\textit{LibQIF}: \url{https://github.com/chatziko/libqif/}}}
Alternatively, \QIF\ concepts could be implemented in existing anonymization tools 
such as ARX.
%, taking advantage of their optimizations for large datasets.
However, not all \QIF\ features are native to these tools,
and capturing all attack models allowed by \QIF\ in them
may become a challenge (e.g. dealing with non-uniform priors 
on records, or adopting information measures 
not hard-coded into the tools).\footnote{\review{It has 
		been proven that \emph{every} %reasonable 
		model of inference attack (including worst-case attacks~\cite{DBLP:conf/csfw/0001FP19})
		%%(ie., 
		%satisfying a set of fundamental 
		%information-theoretic axioms %) 
		is captured in the 
		\QIF\ framework~\cite{Alvim:16:CSF,Alvim:19:TCS}. The primary limitation of \QIF\ is computational tractability rather than generality. }}%
%		as a %simple 
%		choice of gain function and prior distribution,
%		%on secrets, 
%		which are just parameters to the model.}}

As is typical of information-- and decision-theoretic frameworks, scalability to very large scenarios is a challenge
in \QIF.
In such cases it may be necessary to write and optimize 
specialized code, as we had to do for \INEP's scenario
(see \Sec{s201143} ahead).
This is in itself a contribution
of this work: to show that \QIF\ can, indeed, scale.}

\section{%\colorB{
Application of {\QIF} to a large-scale privacy problem: {\INEPallcaps}'s datasets%}
}\label{s091634a}

% !TEX root = main.tex

In this section we apply the rationalization of privacy analyses 
in the {\QIF} framework, discussed in the 
previous sections, to the large-scale scenario 
of {\INEP}'s Educational Censuses.
%mentioned in Sec.\ \ref{s103817}.
We start with the fundamentals of 
our attack models in {\QIF},  
%in \Sec{sec:inipe-case-study-fund}, 
and then provide concrete instantiations 
%of such attacks them
on a running example.
%captured in \Sec{sec:inep-case-study-examples}.
The results obtained by the application of 
these models to the full extent of {\INEP}'s 
scenarios are reserved to \Sec{s1011aa}.

%=============================
% SECTION
%=============================
\subsection{%\colorB{
Instantiating {\QIF}  
	%of Sec.\ \ref{s092134} 
to {\INEPallcaps}'s scenario
%}
}
\label{sec:inep-case-study-fund}
%\cmdraft{A brief reminder that {\INEP} was discussed in Sec.\ \ref{s103817}.
%Outline the scenario and sketch how it can be expressed in {\QIF}.}

The {\QIF} framework can be used to model 
the attacks from \Sec{s100834}. 
First, we can unify single-dataset and longitudinal attacks into a single model 
by aggregating all available datasets along a 
common axis.
We can also unify both re-identification and membership attacks with
attribute-inference attacks by considering the sensitive attribute to infer to be, respectively, each individuals' unique identifier or a special attribute indicating the individual's 
presence/absence in the dataset.
Hence all such attacks can be seen as
instances of attribute-inference attacks 
on a longitudinal collection.
%For clarity, however, here we will focus on a few
%selected attacks, and treat them as separate threats.

%Recall that 
%In the {\QIF} framework 
Using {\QIF} 
we model an adversary using a prior $\pi{:}\mathbb{D}\calx$ 
over secret values $X$, representing her prior knowledge. (We use $\mathbb{D}\calx$ for the set of distributions over the set $\calx$.)
%, 
%and a gain function $g{:}\calw{\times}\calx{\rightarrow}\mathbb{R}_{\ge 0}$, which gives the attacker's gain when taking action $w{\in}\calw$ if the secret is $x{\in}\calx$.
We assume there is a channel $C{:}\calx{\rightarrow}\mathbb{D}\caly$ which leaks information about secrets $X$ via observations $Y$.
We can then represent an adversary's prior and posterior information about the secret (i.e.\ before and after an observation from $C$) using vulnerability%\review{\footnote{\review{Note that the term \qm{vulnerability} used throughout refers to the \qm{risk} (to the secret); this is the terminology which has been adopted in the \QIF\ literature~\cite{Alvim20:Book}.}}}
functions, %as discussed earlier, 
which 
%take into account 
\review{consider}
the adversary's prior $\pi$ and gain function $g$ modeling her capabilities and preferences.
The overall privacy degradation is then computed by comparing the vulnerability of the secret 
before and after the 
attack.

\begin{figure}[tbp]
	\centering
	\includegraphics[width=\linewidth]{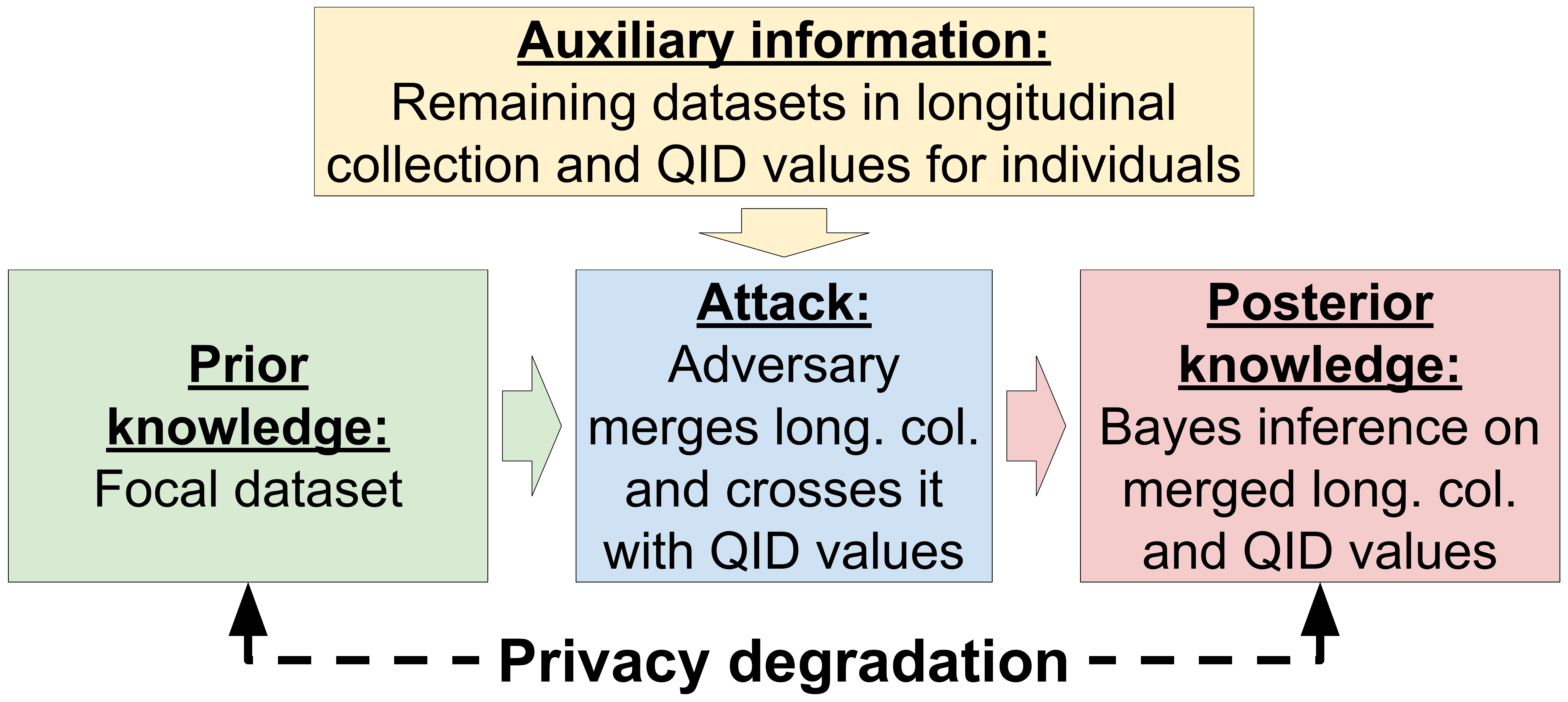}
	\caption{General schema of an attribute-inference 
	attack on a longitudinal collection, which generalizes all attacks in \Table{tab:attack-class}.
	%considered, including re-identification and membership attacks, as well as attacks on single datasets.
%	Recall that in the {\QIF} framework we model an adversary using a prior $\pi{:}\mathbb{D}X$ over secret values $X$, representing her prior knowledge.
%	%, 
%	%and a gain function $g{:}\calw{\times}\calx{\rightarrow}\mathbb{R}_{\ge 0}$, which gives the attacker's gain when taking action $w{\in}\calw$ if the secret is $x{\in}\calx$.
%	We assume there is a channel $C{:}\calx{\rightarrow}\mathbb{D}\caly$ which leaks information about secrets $X$ via observations $Y$.
%	We can then represent an adversary's prior and posterior information (i.e.\ before and after an observation from $C$) using vulnerability functions, as discussed earlier, which take into account the adversary's prior $\pi$ and gain function $g$ modeling her capabilities and preferences.
%	The overall privacy degradation is computed by comparing the vulnerability of the secret 
%	before and after the attack.
}
	\label{fig:attack-long}
\end{figure}

Fig.~\ref{fig:attack-long} 
%depicts the general schema of 
\review{schematizes}
our attack models in {\QIF}.
To accurately capture the scenario of {\INEP}'s 
Educational Censuses 
\review{from} \Sec{s102818}
%\mstodo[Has it already been discussed in the intro? If so, point to there]
--thereby constructing an appropriate 
prior, channel, and vulnerability
measure for the {\QIF} model--, 
we formalize assumptions \pg{1}--\pg{4} below.

%\mstodo[The assumptions below have changed in order and content. Make sure the formalization in the appendices reflect the new presentation.]

%\paragraph{\pg{1}: Published census data.} 
\textbf{\pg{1}: Published census data.} 
\textbf{\pg{1-A}:} There is a \emph{longitudinal collection} $\longdb{=}\{D_{1},D_{2},\ldots,D_{I}\}$ of $I$ datasets of interest. 
Each dataset $D_{i}$, with $1{\leq}i{\leq}I$, is defined over a (finite) attribute set $\attrset_{i}$.
\textbf{\pg{1-B}:} There is an \emph{attribute of unique identification} $\idattr$ common to all datasets in $\longdb$, and each individual of interest holds a persistent value for this attribute across all datasets.
%\footnote{That is, 
%there is an attribute $\idattr{\in}\attrset_{i}$, for every $1{\leq}i{\leq}I$, such that 
%whenever a given individual holds a record $x{\in}D_{i}$ and another record $x'{\in}D_{j}$, with $D_{i},D_{j}{\in}\longdb$, then $x[\idattr]{=}x'[\idattr]$.
%Throughout this paper, given a subset of attributes $\attrset'{\subseteq}\attrset$, we denote by $x[\attrset']$ the \emph{sub-record} of record $x$ consisting of the projection of $x$ onto $\attrset'$, i.e. the sub-tuple of $x$ containing only the values corresponding to the attributes in $\attrset'$.
%When $\attrset'$ is a singleton, we may overload
%notation and use $x[a]$ for $x[\{a\}]$.}%\textsuperscript{,}\footnote{Although Assumption~\pg{2} is too strong in general, it is true
%in the data we analyze (as discussed in  \Sec{sec:experimental-set-up}).

%\paragraph{\pg{2}: Adversary's prior knowledge.}
\textbf{\pg{2}: Adversary's prior knowledge.}
%Before an attack, the adversary's prior knowledge can be modeled
%as a correlation between named (identified) individuals, 
%quasi-identifiers (QIDs) and sensitive attributes. %This correlation, which we
%write as a channel, allows us to measure how much
%information about identified individuals is leaked to the adversary
%through a data release, which we can model as a joint distribution
%mapping QIDs to other sensitive data. 
%In the case of an
%attribute-inference attack, the sensitive data is a sensitive
%attribute to be learned, and in the case of a re-identification attack,
% it is the unique identifier assigned to the individual.
%Note that the adversary's prior knowledge  incorporate
% knowledge learned from querying previously available
% datasets in the longitudinal collection.
%The adversary's prior information is as follows.
\textbf{\pg{2-A}:}
%\colorB{
	In order to apply Bayesian reasoning we need to attribute a prior over secrets to the adversary. 
In this situation, the adversary has access to the \emph{focal dataset} $D_{1}{\in}\longdb$ from which she wishes to re-identify individuals. We can use the distribution of secrets in this dataset as her prior knowledge --her guess-- about which secret might belong to any particular individual. We denote by  $X{\subset}\attrset_{1}$  the set of secret attributes
to infer, and the prior distribution by  $\pi{:}\mathbb{D}\calx$. 
%}
%Before the execution of the attack,
%\colorB{the adversary's prior knowledge 
% the adversary only 
%has access to dataset $D_{1}{\in}\longdb$, called the 
%\emph{focal dataset}, and her goal is to infer sensitive 
%information about individuals in this dataset.
%Letting $X{\subset}\attrset_{1}$ 
%denote the set of secret attributes
%to infer 
%(whose domain of values we denote by $\calx$), 
%dataset $D_1$ represents the adversary's prior
%knowledge about the secrets, captured as a
%prior distribution $\pi{:}\mathbb{D}\calx$.
%\tashtodo[Should we stick with $X$ (instead of $X$) for secrets?][\ms]
%In the {\QIF} approach, her prior knowledge is modeled as
%a distribution on records of $D_{1}$, from which 
%marginal distributions on arbitrary combinations of attributes 
%can be derived in the usual way.~\footnote{Example~\ref{exa:distribution-attribute} in Appendix~\ref{sec:formal-models-datasets} exemplifies how this is done.}
%In this paper we assume a uniform prior distribution,
%representing the fact that adversary's prior knowledge
%does not allow her to favor any record over any other.
\textbf{\pg{2-B}:} The adversary assumes that 
each individual of interest holds exactly one record in $D_{1}$,
and at most one record in each other dataset in $\longdb$.~\footnote{The enforcement of assumption \pg{2-B} is discussed in \Sec{sec:inep-results-synopsis}.}

%\paragraph{\pg{3}: Channel representing the	adversary's acquisition of auxiliary information 	for the execution of the attack.} 
\textbf{\pg{3}: Channel representing the	adversary's acquisition of auxiliary information in attack execution.} 
%The general model of the adversary's collected information
%is as follows.
\textbf{\pg{3-A}:} 
%The adversary queries a dataset $D_i$
%to learn a correlation between QIDs and sensitive data. (For example, the
%adversary may apply a query to select student disability, country of origin 
%and school id from the dataset, where country of origin and school id are
%QIDs, and disability is a sensitive attribute).
%
The adversary combines
the remaining datasets $D_{2}, D_{3}, \ldots, D_{I}$, called \emph{auxiliary datasets}, with the focal dataset $D_1$ to produce an
\emph{aggregated dataset} ${\cal D}$ 
in which the records of individuals across
all datasets are linked (see for example \Table{tab:leading-example-long}).~\footnote{	
	%\colorB{
	In 
	%the case of 
	the {\INEP} Censuses analyzed,
	%}, 
	there exists a persistent unique identifier for every individual across all considered datasets, which makes the aggregation straightforward. 
In \Sec{s191632} we discuss how the {\QIF} framework can
capture more general scenarios.
%When this is not the case, the adversary may rely on QID 
%values and her prior knowledge about the population of interest
%to try to match a same individual's records across 
%different datasets. 
%The obtained aggregated dataset in this case may suffer from
%some inherent uncertainty, but this effect can be accounted for
%in the {\QIF} framework with appropriate models of an adversary's prior knowledge about the population of interest.
} 
\textbf{\pg{3-B}:} 
%\colorB{
In order to find unique mappings between named individuals and other quasi-identifiers (QIDs) in the dataset, we assume that 
%} 
the adversary mines auxiliary information derived from e.g.\ other public datasets.~\footnote{Uniqueness is not necessary for the {\QIF} model, but is used here to simplify the presentation of results.}
%\msdone[Make sure we refer to QIDs in the intro.]
The set of QIDs is defined $Y{ \subseteq}(\cup_i \attrset_{i}){\setminus}X$
(so we denote the domain of possible QID values by $\caly$).
%We assume that the adversary finds a 
%unique mapping from (some) QIDs to individuals.
%other than the longitudinal collection $\longdb$ itself, 
%the values individuals of interest have for a set of \emph{quasi-identifier} attributes $\qidset{\subseteq}\attrset_{i}$, 
%with $1{\leq}i{\leq}I$, common to all datasets in $\longdb$.~\footnote{For clarity, given a record $x{\in}\longaggr{\longdb}$, we denote by $x[(\qidset,i)]$ the sub-record of $x$ corresponding to the quasi-identifiers' values $(\qidset,i)$ of $x$ in dataset $D_{i}{\in}{\longdb}$ and by $x[\qidset]$ the sub-record corresponding to all the quasi-identifiers' values $\qidset$ in all datasets $D_{i}{\in}{\longdb}$, whose domain is $\domain{\qidset \mid \longdb}{=}\domain{\qidset}^{I}$.}
%
%
%\textbf{\pg{3-B}:} The adversary combines the data release with her
%knowledge of the correlation between the named individuals and QIDs to 
%produce a joint distribution over named individuals and sensitive attributes.
%The adversary marginalizes over the named individuals to produce a channel
%describing, for each named individual, a distribution of sensitive attributes.
\textbf{\pg{3-C}:} % A deterministic attack then succeeds whenever, for any named
%individual, there is a single sensitive attribute which occurs with probability 1.
The aggregated dataset ${\cal D}$ can be rewritten as a channel $C{:}\calx{\rightarrow}\mathbb{D}\caly$ where each entry $C_{x,y}$ is the ratio between the count of individuals with QID values $y{\in}\caly$ and secret value $x{\in}\calx$, and the total count of individuals with secret value $x{\in}\calx$. %for all $x{\in}\calx, y{\in}\caly$.
%This channel models how the adversary can use
%correlations between QID values and secrets to
%perform Bayesian reasoning and infer sensitive information.

%--i.e.\ the values of  quasi-identifiers together with the capability of aggregating information from all datasets into $\longdb$--
% to infer correlations between named individuals and sensitive information in the focal dataset $D_{1}$.
%Formally, 
%Single-dataset attacks are the particular case in
%which the longitudinal set has only one element,
%the focal dataset, and no auxiliary datasets.
%\abtodo[Is premise \pg{3} clear enough?][\ms]
%\cmtodo[Is premise \pg{3} clear enough?][\ms]

%\paragraph{\pg{4}: The attack and its privacy degradation.}
\textbf{\pg{4}: The attack and its privacy degradation.}
\textbf{\pg{4-A}:} 
The attack consists in the adversary 
combining her prior knowledge $\pi{:}\mathbb{D}\calx$ with the channel $C{:}\calx{\rightarrow}\mathbb{D}{\caly}$, and then applying Bayesian inference to produce posterior (conditional) distributions on secret values for each possible observed value of QID (i.e.\ a hyper giving a probability of inferring $x{\in}\calx$ for each $y{\in}\caly$, together with the probability of
$y$ itself occurring).
Combined with the adversary's mined knowledge from \pg{3-B}, this posterior knowledge can be used to guess the secret values corresponding to named individuals.
\textbf{\pg{4-B}:}
In a \emph{deterministic} attack,
the threat is quantified considering
the proportion of individuals whose
secret values can be inferred 
with probability 1 using the adversary's 
knowledge.
\textbf{\pg{4-C}:} In a \emph{probabilistic} attack, the threat can be quantified
using the Bayes vulnerability function~\cite{Smith09,Alvim20:Book},
%\msdone[Review the use of ``Bayes vulnerability'' as a name here.]
which gives an optimal adversary's probability
of correctly inferring the secret value in one try.
%the adversary's optimal guess of the secret
%given an observation from the channel. This is
%computed as 
%\begin{equation}\label{eqn:bv}
%V_{bv}(\pi, C) = \sum_q \max_s \pi_s C_{x, y}~.
%\end{equation} %where $p_q$ is the marginal probability of $y \in Y$ and $\delta_q$ is the posterior distribution over $x$ given $y$. 
\textbf{\pg{4-D}:} 
The leakage of information caused by the attack
can be quantified using either the ratio or the difference between the adversary's prior and posterior information about the secret (be it probabilistic or deterministic). 

\begin{table*}[!tb]
	\begin{subtable}[t]{0.28\linewidth}
		%\centering
		\renewcommand{\arraystretch}{0.9}
		\begin{small}
			$
			\begin{array}{|c|c|c|c|c|}
				\hline
				\id & \age & \textit{gend.} & \textit{grd.} & \textit{dis.} \\ \hline %\hline
				\rowcolor{sand!50!white} 1  & 25 & \female & \one   & \no  \\ %\hline
				2  & 25 & \female & \one   & \yes \\% \hline
				\rowcolor{sand!50!white} 3  & 25 & \female & \three & \yes \\% \hline
				4  & 25 & \male   & \two   & \yes \\ %\hline
				\rowcolor{sand!50!white} 5  & 25 & \male   & \two   & \no  \\ %\hline
				6  & 49 & \female & \three & \yes \\ %\hline
				\rowcolor{sand!50!white} 7  & 49 & \female & \three & \yes \\ %\hline
				8  & 49 & \female & \five  & \no  \\ %\hline
				\rowcolor{sand!50!white} 9  & 49 & \male   & \four  & \no  \\ %\hline
				10 & 60 & \male   & \four  & \no  \\ \hline
			\end{array}
			$
		\end{small}
		\caption{Focal dataset $D_1$.}
		\label{tab:leading-example-long-1}
	\end{subtable}
	%\hfill
	\begin{subtable}[t]{0.16\linewidth}
		%	%\centering
		\renewcommand{\arraystretch}{0.9}
		\begin{small}
			$
			\begin{array}{ |c|c|c| }
				\hline
				\id & \age & \textit{grd.} \\ \hline %\hline
				\rowcolor{sand!50!white} 1  & 26 & \two   \\ %\hline
				2  & 26 & \one   \\ %\hline
				\rowcolor{sand!50!white} 3  & 26 & \three \\ %\hline
				4  & 26 & \two   \\ %\hline
				\rowcolor{sand!50!white} 5  & 26 & \two   \\ %\hline
				6  & 50 & \four  \\ %\hline
				\rowcolor{sand!50!white} 7  & 50 & \three \\ %\hline
				8  & 50 & \five  \\ %\hline
				\rowcolor{sand!50!white} 9  & 50 & \four  \\ %\hline
				11 & 19 & \one   \\ \hline
			\end{array}
			$
		\end{small}
		\caption{Aux. dataset $D_2$.}
		\label{tab:leading-example-long-2}
	\end{subtable}		
	%\hfill
	\begin{subtable}[t]{0.55\linewidth}
		%\centering
		\renewcommand{\arraystretch}{0.9}
		\begin{small}
			$
			\begin{array}{|c|c|c|c|c|c|c|}
				\hline
				(\id,1) & (\age,$1$) & (\textit{gend.},$1$) & (grd.,$1$) & (\textit{dis.},$1$) & (\age,$2$) & (\textit{grd.},$2$) \\ \hline %\hline
				\rowcolor{sand!50!white} 1  & 25 & \female & \one   & \no  & 26 & \two   \\ %\hline
				2  & 25 & \female & \one   & \yes & 26 & \one   \\% \hline
				\rowcolor{sand!50!white} 3  & 25 & \female & \three & \yes & 26 & \three \\% \hline
				4  & 25 & \male   & \two   & \yes & 26 & \two   \\ %\hline
				\rowcolor{sand!50!white} 5  & 25 & \male   & \two   & \no  & 26 & \two   \\ %\hline
				6  & 49 & \female & \three & \yes & 50 & \four  \\ %\hline
				\rowcolor{sand!50!white} 7  & 49 & \female & \three & \yes & 50 & \three \\ %\hline
				8  & 49 & \female & \five  & \no  & 50 & \five  \\ %\hline
				\rowcolor{sand!50!white} 9  & 49 & \male   & \four  & \no  & 50 & \four  \\ %\hline
				10 & 60 & \male   & \four  & \no  & {-} & {-}   \\ \hline
			\end{array}
			$
		\end{small}
		\renewcommand{\arraystretch}{1}
		\caption{Aggregated dataset ${\cal D}{=}D_{1} {\leftouterjoin} D_{2}$, with each attribute tagged with its origin.}
		\label{tab:leading-example-long-1-2}
	\end{subtable}
	\caption{Example of longitudinal collection of datasets $\longdb{=}\{D_{1},D_{2}\}$ and their aggregation ${\cal D}$.
	Note that the record with \id\ $10$ is only present in $D_{1}$, so attributes $(\age,2)$ and $(\grade,2)$ have null values in the aggregated dataset $\cald$,
	whereas the record with \id\ $11$ is only present in $D_{2}$ and hence is absent from $\cald$.
}
	\label{tab:leading-example-long}	
\end{table*}
%	\msdone[Tab.~\ref{tab:leading-example-long-1-2} is a bit wider than the two columns.][\ghn]
%	\msdone[While revising the literature to solve some of the tasks you left for me, I've found this paper by Domingo-Ferrer: \qm{Assessing Disclosure Risk via Record Linkage by a Maximum-Knowledge Intruder} \cite{Domingo15:Maximum}. It may be interesting to cite it somewhere.][\ghn]
%	\ghndone[Add it to related work, or background.][\ms]
%
%\mstodo[Flesh out in a bit more detail
%how these assumptions are used in the {\QIF} model:
%how to build the prior, what's the channel, what's the vulnerability measure, etc.]

%We are now ready to flesh out %in more detail 
%our attacks of interest.

%=============================
% SECTION
%=============================
\subsection{%\colorB{
	Concrete 
		%attack 
		example: 
		col\-lec\-tive-target attribute-inference attack on a longitudinal collection (\CAL)
		%}
	}
\label{sec:inep-case-study-examples}

%\msdone[This section has been shortened a lot because of lack of space, and I don't think it'll stand on its own without a previous
%example --e.g. on \Sec{s100834} --on how {\QIF} works (an example of a prior being pushed through a channel and all that). An alternative to consider would be to move this whole section the appendix, and have it back into the main body of the paper if it's accepted.]
%\msdone[I can do that, once we've decided which example to do. Maybe it should go in \Sec{s092134}.][\cm]

%\alldone[This section has been rewritten on 2021-09-11 to provide a stand-alone example of 
%how {\QIF} works without relying on formulas or on any previous example from the background section. 
%I think it's fine now, but after everyone reads
%it we may decide that we have to add more detail.
%Also, if we need space perhaps the \CRL\ example can be moved to an appendix.][\ms]

We now illustrate the instantiation
of our general {\QIF} model to  
%concrete attacks.
a concrete \CAL\ attack.
Other attacks (as in \Table{tab:attack-class}) 
can be modeled as special cases;
Appendix~\ref{s2110071202} exemplifies a \CRL\ attack.
We consider the following scenario,
%which is assumed to satisfy 
under 
assumptions
\pg{1}--\pg{4} above.

\begin{example}[Running example  based on \Table{tab:leading-example-long}]
	\label{exa:leading-example-long}
	Consider a longitudinal collection of  two datasets $\longdb = \{D_{1},D_{2}\}$. 
	The focal dataset, $D_{1}$, is defined on the set of attributes $\attrset_{1}{=}\{\id,\age,\gender,\grade,\disability\}$ and is represented in \Table{tab:leading-example-long-1}.
	The auxiliary dataset, $D_{2}$, is defined on the set of attributes $\attrset_{2}{=}\{\id,\age,\grade\}$ and is represented in \Table{tab:leading-example-long-2}. 
	%According to Assumption~\pg{3-A}, t
	The adversary merges the datasets in $\longdb$,
	via a left outer join keyed on the persistent attribute of unique identification \id,
	to produce the aggregated dataset
	${\cal D}{=}D_{1}{\leftouterjoin}D_{2}$ in %
	%defined on the set of attributes $\attrset_{\longaggr{\longdb}}{=}\{(\id,1), \allowbreak (\age,1), \allowbreak (\gender,1), \allowbreak (\grade,1), \allowbreak (\disability,1), \allowbreak (\age,2), \allowbreak (\grade,2)\}$ and 
	%This is represented in 
	\Table{tab:leading-example-long-1-2}.
%	~\footnote{Note that the record with \id\ $10$ is only present in $D_{1}$, so attributes $(\age,2)$ and $(\grade,2)$ have null values in the aggregated dataset $\cald$,
%	whereas the record with \id\ $11$ is only present in $D_{2}$ and hence it absent from $\cald$.}	
%	%Note that the resulting dataset's cardinality is
	%$|\longaggr{\longdb}|=10$.
\end{example}

%We provide here an example of 
%%only the 
%the \CAL\ attack (as in Tab.~\ref{tab:attack-class}).
%Other attacks can be modeled as special cases;
%Appendix~\ref{s2110071202} exemplifies the instantiation of a
%\CRL\ attack.
% (excluding membership inference) 
%\tashtodo[Do we really need to exclude memberhip? We can do like people do in dp and have an attribute indicating a person's presense/absence on a dataset, can't we?][\ms]
%\mstodo[``Simpler'' might be to say that everyone is in the dataset, but that some of them have all attributes undefined. But a ``present/absent'' attribute would also work, except when both ``absent'' and some real stuff appear.][\cm]

%=============================
% SECTION
%=============================
%\subsubsection{Collective-target attribute-inference attack on a longitudinal collection (\CAL)}\label{s2110071205}

Recall that in a \emph{collective-target attribute-inference attack on a longitudinal collection (\CAL)}, %\footnote{A rigorous mathematical formalization of \CRL\ attacks is given in Definition~\ref{def:CRL} in Appendix~\ref{sec:formal-models-attacks-full}.}
the adversary's goal 
is to infer the value of a sensitive attribute
for as many individuals as possible in the focal dataset $D_{1}$,
no matter who they might be.
%this way to the re-identification \CRL\ attack from the 
%previous section.
%The sole difference is that the adversary's 
%success is measured with respect to to her ability to infer the value of some sensitive attribute of individuals 
%(rather than to re-identify these individuals).~\footnote{A rigorous mathematical formalization of \CAL\ attacks is given in Definition~\ref{def:CAL} in Appendix~\ref{sec:formal-models-attacks-full}.}
%
%\textbf{Execution of the attack.}
%\begin{example}[Execution of \CAL\ attack on running example]
%	\label{exa:CAL}
Assume that in our running example the adversary wants to infer the value of the sensitive attribute $X{=}\{\disability\}$. 

\textbf{Attack execution.}
Before the attack the adversary only has access to the focal
dataset $D_{1}$, and her prior knowledge about
\disability\ is determined by this attribute's distribution in this dataset.
Since (from \Table{tab:leading-example-long-1}) \disability\ is distributed uniformly ($50\%$ \qm{\no} and $50\%$ \qm{\yes}), the adversary's prior is uniform.
%for as many records as possible in the focal dataset $D_{1}$. 
Now consider that during the attack the adversary
gains access to the auxiliary dataset $D_{2}$ and
merges it with $D_{1}$ to obtain the aggregated
dataset $\cald$ (as in \Table{tab:leading-example-long-1-2}).
Furthermore, 
%\colorB{
we assume that
%} 
she obtains as auxiliary information (e.g. via other public datasets)
the values of the QIDs $Y {=}\{\gender, \grade\}$ for all 
individuals in $\cal D$.
%\colorB{
Using this auxiliary information,
%} 
she performs Bayesian reasoning %on the 
%collected information 
and updates her knowledge 
about the secret value from the prior to 
a set of revised conditional distributions (given the learned value of each individual's QIDs) on \disability\ s.t.\
each of these posterior distributions has its own probability of occurring --- i.e. she updates her knowledge to a hyper on the secret value.

This whole process is modeled in {\QIF} as in \Table{tbl:cal_example_1}.
First the adversary extracts from $\cald$ all co-occurrences of 
values for the secret and for QIDs (\Table{tbl:cal_eg1_correlation}),
and from that she derives a joint probability distribution on these
values (\Table{tbl:cal_eg1_joint}).
By marginalizing the joint distribution, we
get the adversary's prior $\pi$ on the secret 
value \disability, and by conditioning the joint distribution on the prior we get the channel representing the adversary's information-gathering process during the attack (\Table{tbl:cal_eg1_channel}).
The adversary's posterior knowledge is then represented by the hyper in \Table{tbl:cal_eg1_hyper}.
Finally, the overall degradation of privacy can 
be computed as follows.

% CAL attack example 
\begin{table*}[!tb]
	\begin{subtable}[t]{0.48\linewidth}
		\renewcommand{\arraystretch}{0.9}
		\centering
		\begin{small}
			$
			\begin{array}{|c|c c c c c c c c|}
				\hline
				\raisebox{-0.75mm}{QIDs $\blacktriangleright$}
   				& 
   				\multirow{3}{*}{\rotatebox{90}{\small{(\female,\one,\two)}}} 
   				&			\multirow{3}{*}{\rotatebox{90}{\small{(\female,\one,\one)}}}
   				&			\multirow{3}{*}{\rotatebox{90}{\small{(\female,\three,\three)}}}
   				&			\multirow{3}{*}{\rotatebox{90}{\small{(\male,\two,\two)}}}
   				&
   				\multirow{3}{*}{\rotatebox{90}{\small{(\female,\three,\four)}}}
   				&			\multirow{3}{*}{\rotatebox{90}{\small{(\female,\five,\five)}}} 
   				&			\multirow{3}{*}{\rotatebox{90}{\small{(\male,\four,\four)}}} 
   				&			\multirow{3}{*}{\rotatebox{90}{\small{(\male,\four,-)}}} 
   				\\[1mm] \cline{1-1}
   				& 
   				& 
   				& 
   				& 
   				& 
   				& 
   				& 
   				& 
   				\\[-2mm]
				\disability\ \blacktriangledown
				& 
				& 
				& 
				& 
				& 
				& 
				& 
				& 
				\\
				\hline
				\rowcolor{sand!50!white} \yes & 0 & 1 & 2 & 1 & 1 & 0 & 0 & 0 \\
				\no & 1 & 0 & 0 & 1 & 0 & 1 & 1 & 1 \\ \hline
			\end{array}
			$
		\end{small}
		\caption{Co-occurrence of values for secret $X{=}\{(\disability,1)\}$ and for observable QIDs $Y{=}\{(\gender,1),(\grade,1),(\grade,2)\}$,
			derived from the aggregated dataset $\cald$ from \Table{tab:leading-example-long-1-2}.
			E.g. exactly one record 
			has \disability\ status \qm{\no}
			and at the same time is a female with grade \one\ in the 
			focal dataset $D_{1}$, and grade \two\ in the auxiliary dataset $D_{2}$.}
		\label{tbl:cal_eg1_correlation}
	\end{subtable} 
	\hfill
	\begin{subtable}[t]{0.48\linewidth}
		\renewcommand{\arraystretch}{0.9}
		\centering
		\begin{small}
			$
			\begin{array}{|c|c c c c c c c c|}
				\hline
				\raisebox{-0.75mm}{QIDs $\blacktriangleright$}
				& 
				\multirow{3}{*}{\rotatebox{90}{\small{(\female,\one,\two)}}} 
				&			\multirow{3}{*}{\rotatebox{90}{\small{(\female,\one,\one)}}}
				&			\multirow{3}{*}{\rotatebox{90}{\small{(\female,\three,\three)}}}
				&			\multirow{3}{*}{\rotatebox{90}{\small{(\male,\two,\two)}}}
				&
				\multirow{3}{*}{\rotatebox{90}{\small{(\female,\three,\four)}}}
				&			\multirow{3}{*}{\rotatebox{90}{\small{(\female,\five,\five)}}} 
				&			\multirow{3}{*}{\rotatebox{90}{\small{(\male,\four,\four)}}} 
				&			\multirow{3}{*}{\rotatebox{90}{\small{(\male,\four,-)}}} 
				\\[1mm] \cline{1-1}
				& 
				& 
				& 
				& 
				& 
				& 
				& 
				& 
				\\[-2mm]
				\textit{disab.} \blacktriangledown
				& 
				& 
				& 
				& 
				& 
				& 
				& 
				& 
				\\
				\hline
				\rowcolor{sand!50!white} \yes & 0 & \nicefrac{1}{10} & \nicefrac{2}{10} & \nicefrac{1}{10} & \nicefrac{1}{10} & 0 & 0 & 0 \\
				\no & \nicefrac{1}{10} & 0 & 0 & \nicefrac{1}{10} & 0 & \nicefrac{1}{10} & \nicefrac{1}{10} & \nicefrac{1}{10} \\ \hline
			\end{array}
			$
		\end{small}
		\caption{Joint distribution of values for secret $X{=}\{(\disability,1)\}$ and for observable QIDs $Y{=}\{(\gender,1),(\grade,1),(\grade,2)\}$, derived from the
			co-occurrence matrix from \Table{tbl:cal_eg1_correlation},
			and assuming a uniform distribution on the records in $\cald$.
			E.g. there is a probability $\nicefrac{1}{10}$ that an
			individual does not present a disability and has 
			QID vaues (\female,\one,\two).}
		\label{tbl:cal_eg1_joint}
	\end{subtable} 
	\\[2mm]
	\begin{subtable}[t]{0.48\linewidth}
		\renewcommand{\arraystretch}{0.9}
		\centering
		\begin{small}
			$
			\begin{array}{|c|}
				\hline
				\pi \\ \hline
				\rowcolor{sand!50!white} \nicefrac{1}{2} \\
				\nicefrac{1}{2} \\
				\hline
			\end{array}
			\hfill
			\begin{array}{|c|c c c c c c c c|}
				\hline
				\raisebox{-0.75mm}{QIDs $\blacktriangleright$}
				& 
				\multirow{3}{*}{\rotatebox{90}{\small{(\female,\one,\two)}}} 
				&			\multirow{3}{*}{\rotatebox{90}{\small{(\female,\one,\one)}}}
				&			\multirow{3}{*}{\rotatebox{90}{\small{(\female,\three,\three)}}}
				&			\multirow{3}{*}{\rotatebox{90}{\small{(\male,\two,\two)}}}
				&
				\multirow{3}{*}{\rotatebox{90}{\small{(\female,\three,\four)}}}
				&			\multirow{3}{*}{\rotatebox{90}{\small{(\female,\five,\five)}}} 
				&			\multirow{3}{*}{\rotatebox{90}{\small{(\male,\four,\four)}}} 
				&			\multirow{3}{*}{\rotatebox{90}{\small{(\male,\four,-)}}} 
				\\[1mm] \cline{1-1}
				& 
				& 
				& 
				& 
				& 
				& 
				& 
				& 
				\\[-2mm]
				\textit{disab.} \blacktriangledown
				& 
				& 
				& 
				& 
				& 
				& 
				& 
				& 
				\\
				\hline
				\rowcolor{sand!50!white} \yes & 0 & \nicefrac{1}{5} & \nicefrac{2}{5} & \nicefrac{1}{5} & \nicefrac{1}{5} & 0 & 0 & 0 \\
				\no & \nicefrac{1}{5} & 0 & 0 & \nicefrac{1}{5} & 0 & \nicefrac{1}{5} & \nicefrac{1}{5} & \nicefrac{1}{5} \\ \hline
			\end{array}
			$
		\end{small}
		\caption{Prior distribution $\pi$ on the values 
			for secret $X{=}{(\disability,1)}$, and the channel
			for the \CAL\ attack, each derived from the joint 
			distribution from \Table{tbl:cal_eg1_joint}
			by marginalization and conditioning, respectively.
			E.g. the prior indicates that before the attack
			(i.e. without learning any QID value)
			the adversary believes that the probability of
			any individual having a disability is $\nicefrac{1}{2}$.
			On the other hand, the channel indicates that
			during the attack the adversary can use the fact that if an individual without a disability is the owner of a record, then the probability
			that that record has QID values (\female,\one,\two) is $\nicefrac{1}{5}$.}
		\label{tbl:cal_eg1_channel}
	\end{subtable} 
	\hfill
	\begin{subtable}[t]{0.48\linewidth}
		\renewcommand{\arraystretch}{0.9}
		\centering
		\begin{small}
			$
			\begin{array}{|c|c c c c c c c c|}
				%\cline{2-9}
				\multicolumn{1}{c}{\text{outers $\blacktriangleright$}}
				& \multicolumn{1}{c}{\nicefrac{1}{10}}
				& \multicolumn{1}{c}{\nicefrac{1}{10}}
				& \multicolumn{1}{c}{\nicefrac{1}{5}}
				& \multicolumn{1}{c}{\nicefrac{1}{5}}
				& \multicolumn{1}{c}{\nicefrac{1}{10}}
				& \multicolumn{1}{c}{\nicefrac{1}{10}}
				& \multicolumn{1}{c}{\nicefrac{1}{10}}
				& \multicolumn{1}{c}{\nicefrac{1}{10}} \\ \hline
				\raisebox{-0.75mm}{\text{QIDs $\blacktriangleright$}}
				& 
				\multirow{3}{*}{\rotatebox{90}{\small{(\female,\one,\two)}}} 
				&			\multirow{3}{*}{\rotatebox{90}{\small{(\female,\one,\one)}}}
				&			\multirow{3}{*}{\rotatebox{90}{\small{(\female,\three,\three)}}}
				&			\multirow{3}{*}{\rotatebox{90}{\small{(\male,\two,\two)}}}
				&
				\multirow{3}{*}{\rotatebox{90}{\small{(\female,\three,\four)}}}
				&			\multirow{3}{*}{\rotatebox{90}{\small{(\female,\five,\five)}}} 
				&			\multirow{3}{*}{\rotatebox{90}{\small{(\male,\four,\four)}}} 
				&			\multirow{3}{*}{\rotatebox{90}{\small{(\male,\four,-)}}} 
				\\[1mm] \cline{1-1}
				& 
				& 
				& 
				& 
				& 
				& 
				& 
				& 
				\\[-2mm]
				\textit{disab.} \blacktriangledown
				& 
				& 
				& 
				& 
				& 
				& 
				& 
				& 
				\\
				\hline
				\rowcolor{sand!50!white} \yes & 0 & 1 & 1 & \nicefrac{1}{2} & 1 & 0 & 0 & 0 \\
				\no & 1 & 0 & 0 & \nicefrac{1}{2} & 0 & 1 & 1 & 1 \\ \hline
			\end{array}
			$
		\end{small}
		\caption{Hyper-distribution (with column labels added for clarity) representing the adversary's knowledge after completing the \CAL\ attack.
		The top row (``outers'') gives the probability of each
		possible combination of QID values being revealed, and each column gives the 
		posterior probability distribution on secret values given that 
		the corresponding QID values were revealed.
		E.g. after the attack, 
		the adversary has a probability $\nicefrac{1}{10}$
		of learning that an individual's QID values	are 
		(\female, \one, \two), and in this case she assigns 
		probability 1 to the corresponding individual 
		having no disability.
%		 -- 
%		meaning that the value of the sensitive attribute has
%		been inferred with absolute certainty.
}
		\label{tbl:cal_eg1_hyper}
	\end{subtable}
	
	\caption{Step-by-step derivation of prior, channel, and hyper-distribution for \CAL\ attack on the longitudinal collection $\longdb$ from \Table{tab:leading-example-long}, considering secret $X = \Set{\disability}$ and observable QIDs $Y = \Set{\gender,\grade}$.}
	\label{tbl:cal_example_1}
\end{table*}	

\textbf{Deterministic degradation of privacy.}
%Since the focal dataset $D_{1}$ presents more than one value for the sensitive
%attribute \disability\ (five records hold value \yes\ and the other five hold value \no),
%in the absence of auxiliary information
%no individual can have their sensitive value inferred with certainty.
Recall that deterministic success is concerned 
with the proportion of individuals whose value
for the sensitive attribute can be inferred  with
absolute certainty.
In this example, the adversary's deterministic prior 
success is $0\%$, since before the attack no individual's \disability\ status
can be inferred with certainty. %this is the fraction of records that can have its sensitive value inferred.
After the attack, however, the adversary's 
knowledge is updated to the hyper in \Table{tbl:cal_eg1_hyper}.
Note that in that hyper the posteriors containing only $1$ and $0$ values --i.e.\ all columns but the one labeled as (\male,\two,\two)-- have unique QIDs and therefore allow the adversary to infer with probability 1 the \disability\ status of the corresponding individuals.
The adversary's deterministic posterior success is the fraction of individuals whose attribute is inferred in this way, which is exactly
$80\%$, or 8 out of 10 
(note that some posteriors in the hyper 
represent more than one individual, 
which is reflected by the posterior's weight).
%can identify six records in $\longaggr{\longdb}$ with unique values for the quasi-identifiers $\{ (\gender,1), \allowbreak (\grade,1), (\grade,2) \}$ (i.e.\ the records with \id\ 1, 2, 6, 8, 9, and 10), so their values for the sensitive attribute can be immediately inferred. Also, even though the records with \id\ 3 and 7 share the same value for the quasi-identifiers, they also share the same value for the sensitive attribute (\yes),
%so the adversary can immediately infer its value. Since the adversary can infer the values of the sensitive attribute \disability\ for 8 out of 10 records in the focal dataset $D_{1}$, 
%Thus her deterministic posterior success is $\nicefrac{8}{10}{=}80\%$ and 
We describe the overall deterministic degradation of privacy additively, as $80\%{-}0\%=80\%$, meaning
that the execution of the attack increases the proportion of individuals with inferrable \disability\ status by an absolute value of 80\%.

\textbf{Probabilistic degradation of privacy.}
Recall that probabilistic success is concerned with the chance that randomly selected individuals can have their sensitive attributes inferred, even
if without certainty.
In this example, the prior vulnerability of the dataset is $50\%$, since before the attack
the adversary's prior on \disability\ is uniform
and therefore 50\% is the maximum chance with which she can guess the secret value for an individual. 
%Since in $D_{1}$ half of individuals hold value \yes\ for the sensitive attribute \disability\, and half hold value \no, in the absence of any auxiliary information, the adversary's prior probability of correctly inferring a random individual's sensitive value is of $50\%$.
After executing the attack and updating her knowledge
to the hyper from \Table{tbl:cal_eg1_hyper}, 
the adversary's posterior success is measured as
the expected value of Bayes vulnerability
(which, recall, is the probability of guessing the
secret correctly in one try) 
taken over all posteriors distributions.
Indeed, since 7 of the posteriors 
allow the adversary to guess the secret
with probability 1 --and 6 of these posteriors
occur themselves with probability $\nicefrac{1}{10}$,
whereas 1 occurs with probability $\nicefrac{1}{5}$--,
and 1 of the posteriors allows a correct guess with probability $\nicefrac{1}{2}$ --and this posterior occurs
itself with probability $\nicefrac{1}{10}$--,
the overall
posterior Bayes vulnerability is $6{\cdot} \nicefrac{1}{10}{\cdot}1{+}1{\cdot}\nicefrac{1}{5} {\cdot}1{+}1{\cdot}\nicefrac{1}{10}{\cdot} \nicefrac{1}{2}{=}90\%$.
We describe the overall probabilistic degradation caused by the attack multiplicatively, as $\nicefrac{90\%}{50\%}{=}1.8$,
meaning that the adversary's chance of 
inferring a randomly selected individual's 
\disability\ status in the focal dataset 
increases by a factor of $1.8$ 
%\colorB{
-- so
the completion of the \CAL\ attacks almost doubles the
adversary's success in inferring the sensitive information.
%}
%so an adversary with access to the dataset 
%almost doubles her success compared to one who does not.
%\footnote{For comparison, consider the adversary were restricted to a single-dataset attack, i.e.\ the adversary could access only the dataset $D_{1}$.
%The prior probabilistic success would still be of $50\%$, but in this case the dataset could be partitioned in five blocks by using the attributes \gender\ and \grade\ as quasi-identifiers. Therefore, the adversary's posterior, probabilistic success would be equal to $10\% \cdot 100\% + 2 \cdot 20\% \cdot 50\% + 20\% \cdot 100\% + 30\% \cdot 100\% = 80\%$ and the degradation of privacy would be an increase of $\nicefrac{80\%}{50\%}{=}1.6$ in the adversary's chance, less than the degradation of privacy achieved in a longitudinal attack.}

%\end{example}

%=============================
% SECTION
%=============================

\subsection{Outline of the developed software}
\label{s201143}

As explained in \Sec{s191650},
no existing tool 
%\colorB{
	met the needs 
%provided the capabilities needed 
for the scope of our analyses:
%}:
either they did not support 
all attack models we consider (especially attribute-inference),
did not support longitudinal analysis, or
simply could not run analyses on data as large 
as {\INEP}'s.
Hence, we implemented and optimized our own tool.

Our software is implemented on 
Python 3.9.10 using 
%the additional libraries 
\texttt{numpy} 1.22.2 and \texttt{pandas} 1.4.1
to streamline some operations.
%}
To optimize the use of hardware, we employ the Python \texttt{multiprocessing} standard library to simultaneously analyze different sets of QIDs --up to the number of available CPU threads. 
Instead of relying on the \texttt{pandas} built-in functions to partition a dataset based on QIDs, 
we perform our own sorting of the records according to a given set of QIDs and compute all the values related to that attack on a single pass through the whole dataset. 
The re-identification and sensitive attribute inference attacks are carried out simultaneously for each selection of QIDs and of 
sensitive attributes.

Under these optimizations and using 20 threads from two \emph{Intel Xeon E5-2620 v2} processors with 96 GB DDR3-1866 RDIMM, all 2,047 single-dataset attacks performed on the School Census of 2018 were conducted in 40 hours. 
Due to our choice of only one set of QIDs for the longitudinal attacks, all the 4 analyses were performed in less than one hour.
We describe the results of such analyses in the next section.
%\ghntodo[We should mention here that they are available somewhere!][\ms]
%\mstodo[I'm currently working on the code trying to make it agnostic to the input dataset, so we can publish a general-use tool.][\ghn]

%\ghntodo[Add a paragraph starting like this: ``Under these optimizations, all analyses described in the following section were conducted in a period of blah in a hardware of blah.''][\ms]

%-----------------------------

%=============================
% SECTION
%=============================
\section{%\colorB{
	Privacy analyses of the \INEPallcaps\ datasets
	%}
% Summary of vulnerabilities discovered
}\label{s1011aa}

% !TEX root = main.tex

%\colorB{
We now summarize the main results of
employing the attack models from \Sec{s091634a} 
to extensive experimental privacy analyses on 
{\INEP}'s Educational Censuses.
These results were critical information for {\INEP}'s
decision making, % process, 
and we discuss their implications
in \Sec{s091634}.
%}
%Detailed descriptions of experimental results
%can be found on Appendix~\ref{sec:appendix-experimental-results}.

%=============================
% SECTION
%=============================
\subsection{Overall synopsis}
\label{sec:inep-results-synopsis}

%\cmdraft{How comprehensive; the impact of the longitudinal and other features of the dataset}

%\alldone[This first paragraph maybe should go in the intro, and we could start here from the second paragraph on.][\ms]
As mentioned, {\INEP}'s Educational Censuses 
datasets contain microdata
% (i.e. records at individual level)
for every student at all levels of education in Brazil,
%\colorB{
	including elementary,
middle, high, professional, and college
%, and special-purpose 
education.
%}
%Since 2007, every year the data are collected at school level, 
%and then processed, organized, and treated by {\INEP} before official 
%public release.
The datasets have been published yearly since 2007, and the only privacy protection techniques employed 
are \emph{de-identification} (i.e. the removal of obvious personal identifiers, such as name or governmental-issued ID numbers) and
\emph{pseudonymization} (i.e. the substitution of such obvious personal identifiers for artificially-created ones).

Our experimental analyses focused on the \emph{School Census}.
These datasets are the largest published by {\INEP},   concerning all students in the country enrolled at all levels of education other than college. 
Each yearly dataset
contains microdata for approximately 50 million students, with about 90 attributes per student.
For this study we selected the 5 most recent datasets at the time of the analyses,
as presented in \Table{tab:datasets}.
%The analyses of attacks on single-datasets were performed on the
%Census of 2018, whereas the analyses of longitudinal attacks
%were performed on the Censuses from 2014 to 2017.
%\ghndone[It's probably best to have a small table here, with each of the 4 datasets used (rather than just the first and last), and the number of records and of attributes in each dataset. It'd be useful to add as well the number of records before and after treatment per dataset.][\ms]
We now describe the fundamentals of these analyses.

\begin{table}[tb]
	\centering
	\renewcommand{\arraystretch}{1.2}
	\begin{small}
		$
		\begin{array}{|c|c|c|>{\centering\arraybackslash\small}m{0.10\linewidth}|c|}
			\hline
			\multirow{2}{*}{\textbf{Year}} & \multicolumn{2}{c|}{\textbf{\# of records}} & \textbf{\# of} & \textbf{Attacks} \\[-0.5mm]
			\cline{2-3}
			& \textbf{Original} & \textbf{Treated} & \textbf{attrib.} & \textbf{performed} \\[-0.5mm]
			\hline \hline
			2014 & 56,064,675 & 49,491,319 & 85 & \CRL\ / \CAL \\
			\hline
			2015 & 54,851,222 & 48,536,347 & 93 & \CRL\ / \CAL \\
			\hline
			2016 & 52,356,383 & 48,561,221 & 92 & \CRL\ / \CAL \\
			\hline
			2017 & 53,900,669 & 48,377,987 & 92 & \CRL\ / \CAL \\
			\hline
			2018 & 51,829,413 & 48,176,423 & 92 & \CRS\ / \CAS \\
			\hline
		\end{array}
		$
	\end{small}
	\caption{School Census datasets used in our experiments.}
	\label{tab:datasets}
\end{table}

%\paragraph{Treatment of the datasets.}
\textbf{Treatment of the datasets.}
Since these datasets may contain duplicated entries for the same student, we
treated them to meet the uniqueness Assumption~\pg{2-B} from
\Sec{s091634a}. For that, we randomly selected only one record for each
student with a same unique pseudonymization code in each dataset \cite{nunes_gabriel_henrique_2021_6533675}.
Notice that this treatment can only
underestimate privacy risks, so 
%the resulting
%\colorB{
	our
%}
analyses provide a lower bound on the real risks.

%\footnote{Until the Census of 2017, each student would receive a unique \texttt{Student ID} code, which would stay constant for every annual Census. Such an identification number allows an adversary to easily follow a given student through different years, which increases the amount of information leaked. From the Census of 2018 onward, {\INEP} changed how the \texttt{Student ID} code is assigned to each student, continuing to be unique for a student in a given dataset release, but changing among different releases. Therefore, from the Census of 2018 onward that code could not be used to follow a given student through different years anymore.} 
%This data treatment was performed independently for each of the datasets from the School Censuses from 2014 to 2018, resulting in new datasets with number of records presented in column \qm{Treated} of Tab.~\ref{tab:datasets}.

%\paragraph{Selection of attributes.}
\textbf{Selection of attributes.}
For computational tractability, we restricted our experimental
analyses to the attributes listed in \Table{tab:attributes-attacks},
which were selected according to the criteria below.~\footnote{%
	%We emphasize that the selection of those QIDs and sensitive attributes is to an extent subjective. 
	%Nevertheless, our 
%	\colorB{
		Notice that our goal is not to 
	%ultimately 
	define whether an
	attribute should be considered as a QID or sensitive. Instead, our results just illustrate privacy
	risks of possible real-life circumstances, and they can be reproduced for any other 
	%particular 
	choice of attributes.
%}
}

\begin{itemize}
	\item \textbf{Selection of QIDs.}
	For both re-identification and attribute-inference attacks on
	single-datasets (\CRS\ and \CAS, respectively), performed on the Census of 2018,
	we considered all possible 2,047 non-empty combinations of QIDs from a
	set of 11 attributes which we presume to be easily obtainable by an adversary as auxiliary information.
%	\colorB{
		For the longitudinal attacks (\CRL\ and \CAL), %performed 
	on the Censuses from
	2014 to 2017,
%}, 
we employed a fixed set of 3 QIDs that are expected
	to vary over the years (since attributes that tend to remain constant tend not to be particularly useful in longitudinal attacks). \Table{tab:CRS-CAS-CRL-CAL-qid} lists the QIDs selected for each attack.
	
	\item \textbf{Selection of sensitive attributes.}
	For attribute-inference attacks on both single-datasets and on longitudinal collections
	(\CAS\ and \CAL, respectively), we considered as sensitive:
	(i) the flag indicating whether the student has any disability, and
	(ii) the flag indicating whether the student uses public school transport
	(which may indicate economic status).
	%\footnote{For the flag indicating the	use of public school transport, the \texttt{n/a} value only appears on	the School Census of 2018.}
	These attributes are listed in \Table{tab:CAS-CAL-sensitive}. 
%	Recall that re-identification attacks (\CRS, \CRL) are the particular case of attribute-inference attacks where the sensitive attribute is the student's pseudonimization code.
\begin{table}[tb]
	\centering
	\begin{subtable}[t]{\linewidth}
		\centering
		\begin{small}
			$
			\begin{array}{|>{\centering\arraybackslash\small}m{0.42\linewidth}|>{\centering\arraybackslash\small}m{0.19\linewidth}|>{\centering\arraybackslash\small}m{0.19\linewidth}|}
			\hline
			\textbf{Attribute} & \textbf{\CRS\ / \CAS} & \textbf{\CRL\ / \CAL} \\ \hline \hline
			\texttt{Day of birth} & \yes & -- \\ \hline
			\texttt{Month of birth} & \yes & -- \\ \hline
			\texttt{Year of birth} & \yes & -- \\ \hline
			\texttt{Gender} & \yes & -- \\ \hline
			\texttt{Ethnicity} & \yes & -- \\ \hline
			\texttt{Nationality} & \yes & -- \\ \hline
			\texttt{Country of birth} & \yes & -- \\ \hline
			\texttt{City of birth} & \yes & -- \\ \hline
			\texttt{City of residency} & \yes & \yes \\ \hline
			\texttt{School id code} & \yes & \yes \\ \hline
			\texttt{School type \quad\quad\quad (public, private, ...)} & \yes & -- \\ \hline
			\texttt{Education level \quad\quad (middle, high, ...)} & -- & \yes \\ \hline
			\end{array}
			$
		\end{small}
		\caption{Attributes selected as QIDs in each attack.}
		\label{tab:CRS-CAS-CRL-CAL-qid}
	\end{subtable}
	
	\vspace{2mm}
	\begin{subtable}[t]{\linewidth}
		\centering
		\begin{small}
			$
			\begin{array}{|>{\centering\arraybackslash\small}m{0.65\linewidth}|>{\centering\arraybackslash\small}m{0.20\linewidth}|}
			\hline
			\textbf{Attribute} & \textbf{Domain} \\ \hline \hline
			\texttt{Disability status} 
			%(\textit{whether the student has a disability}) 
			& \texttt{yes}, \texttt{no} \\ \hline
			\texttt{Uses public school transportation} 
			%(\textit{whether the student uses public school transportation}) 
			& \texttt{yes}, \texttt{no}, \texttt{n/a} \\ \hline
			\end{array}
			$
		\end{small}
		\caption{Attributes selected as sensitive in attribute-inference attacks. 
			%For the \texttt{Transportation} attribute, the \texttt{n/a} value only appears on the School Census of 2018.
		}
		\label{tab:CAS-CAL-sensitive}
	\end{subtable}
	
	\caption{Attributes selected for the attacks.}
	\label{tab:attributes-attacks}
\end{table}
\end{itemize}

%\begin{table}[tb]
%	\centering
%	\renewcommand{\arraystretch}{1.2}
%	\begin{tabular}{|>{\centering\arraybackslash\small}m{0.26\linewidth}|>{\centering\arraybackslash\small}m{0.38\linewidth}|>{\centering\arraybackslash\small}m{0.16\linewidth}|}
%		\hline
%		\textbf{Attribute} & \textbf{Description} & \textbf{Domain of values} \\ \hline \hline
%		\texttt{Disability} & Whether the student possesses a disability. & \texttt{yes}, \texttt{no} \\ \hline
%		\texttt{Transportation} & Whether the student uses public school transport. & \texttt{yes}, \texttt{no}, \texttt{n/a} \\ \hline
%	\end{tabular}
%	\caption{Variables from the School Censuses selected as sensitive attributes in evaluation of each kind of attribute-inference attack. For the \texttt{Transportation} attribute, the \texttt{n/a} value only appears on the School Census of 2018.}
%	\label{tab:CAS-CAL-sensitive}
%\end{table}

%=============================
% SECTION
%=============================
%\subsection{Experimental analyses of single-dataset attacks (\CRS, \CAS)}

\begin{figure*}[tb]
	\centering
	\begin{subfigure}[t]{0.32\linewidth}
		\centering
		\includegraphics[width=\linewidth]{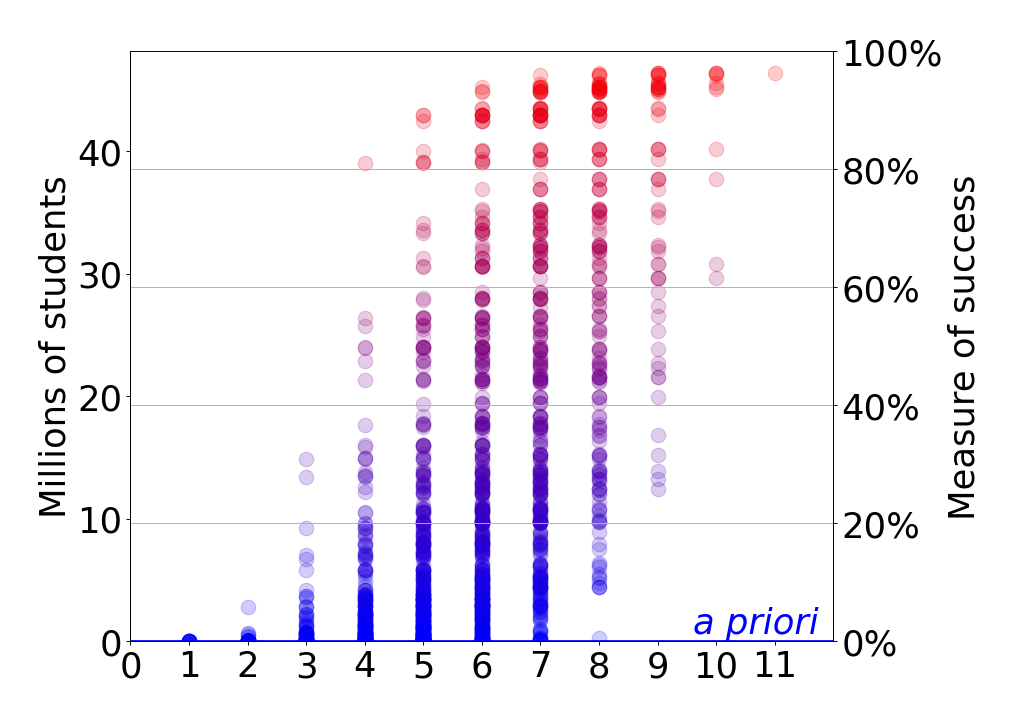}
		\caption{Deterministic success in \CRS\ attacks.}
		\label{fig:CRS-S18-det}
	\end{subfigure}
	\hfill
	\begin{subfigure}[t]{0.32\linewidth}
		\centering
		\includegraphics[width=\linewidth]{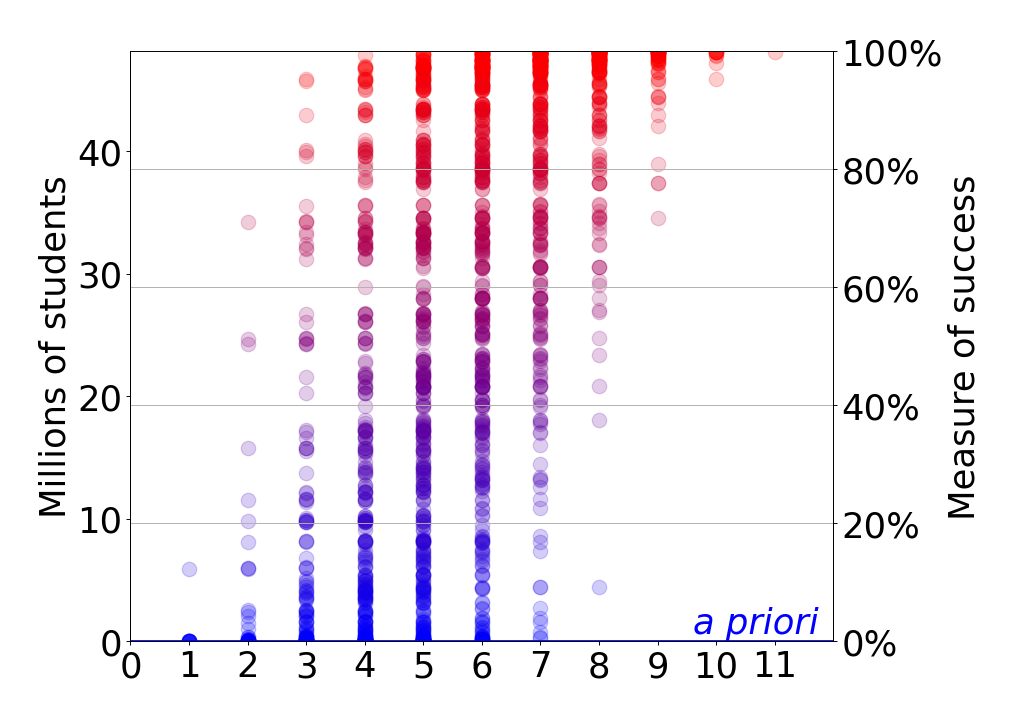}
		\caption{Det.\ success in \CAS\ attacks (disability).}
		\label{fig:CAS-S18-d-det}
	\end{subfigure}
	\hfill
	\begin{subfigure}[t]{0.32\linewidth}
		\centering
		\includegraphics[width=\linewidth]{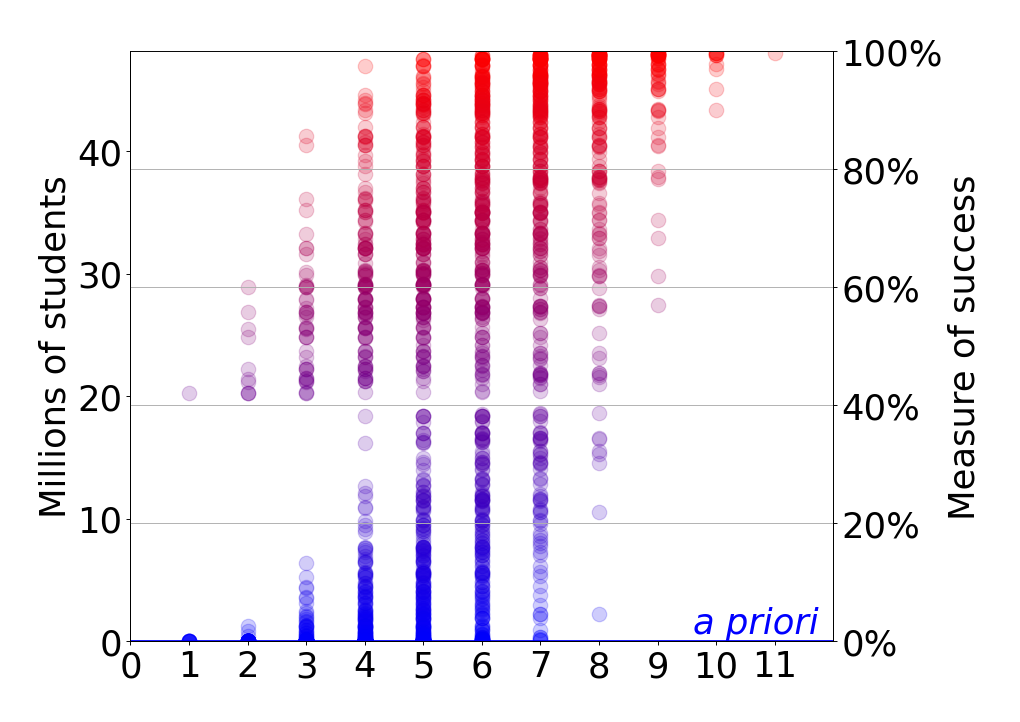}
		\caption{Det.\ success in \CAS\ attacks (transportation).}
		\label{fig:CAS-S18-f-det}
	\end{subfigure}
	
	\begin{subfigure}[t]{0.32\linewidth}
		\centering
		\includegraphics[width=\linewidth]{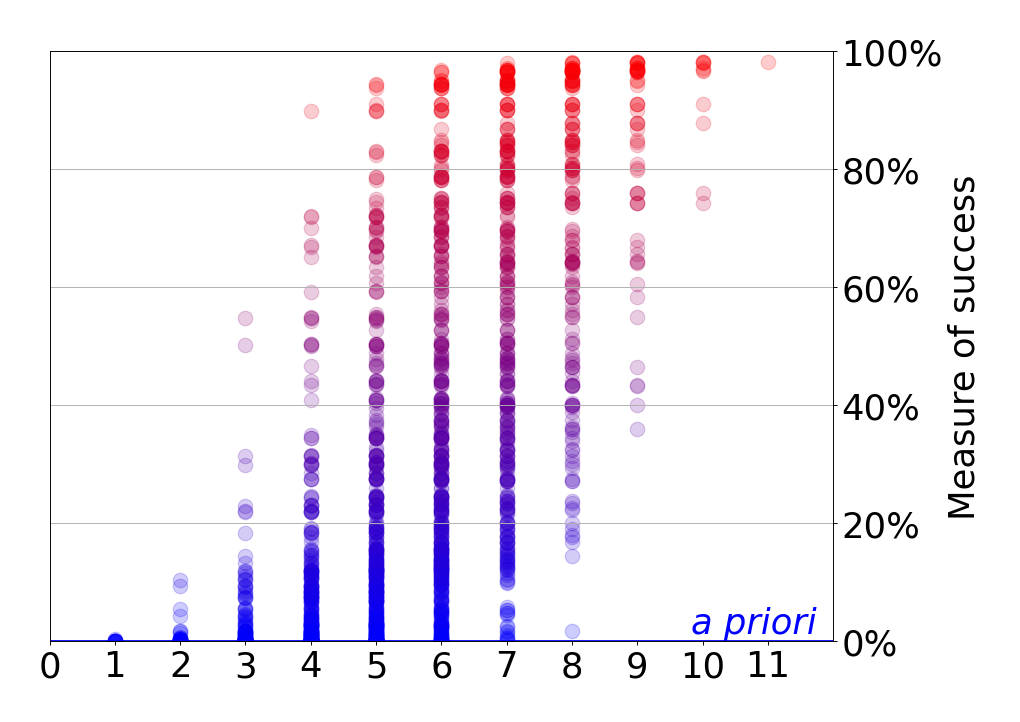}
		\caption{Probabilistic success in \CRS\ attacks.}
		\label{fig:CRS-S18-prob}
	\end{subfigure}
	\hfill
	\begin{subfigure}[t]{0.32\linewidth}
		\centering
		\includegraphics[width=\linewidth]{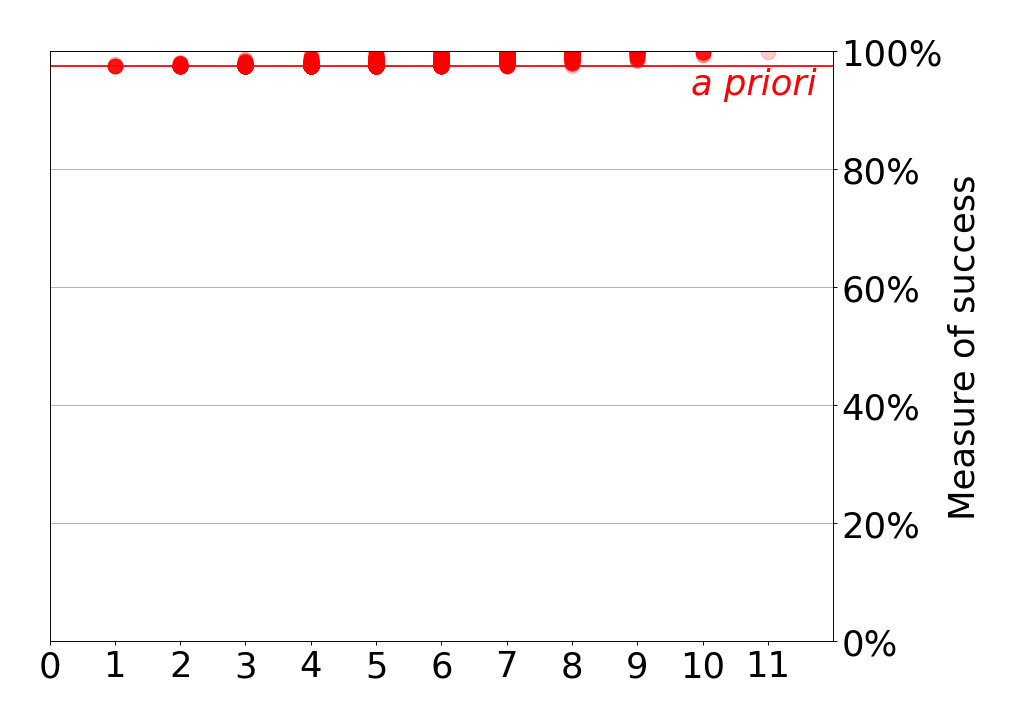}
		\caption{Prob.\ success in \CAS\ attacks (disability).}
		\label{fig:CAS-S18-d-prob}
	\end{subfigure}
	\hfill
	\begin{subfigure}[t]{0.32\textwidth}
		\centering
		\includegraphics[width=\textwidth]{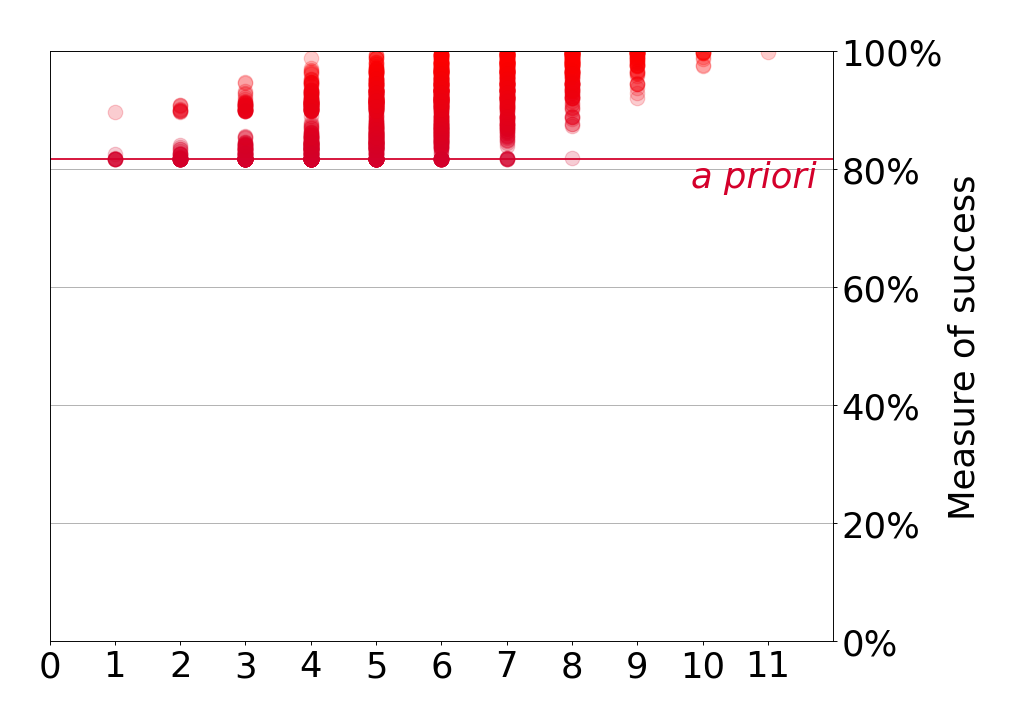}
		\caption{Prob.\ success in \CAS\ attacks (transport.).}
		\label{fig:CAS-S18-f-prob}
	\end{subfigure}
	\caption{Adversary's success in re-identification (\CRS)
		and attribute-inference (\CAS) attacks on the School Census of 2018.
		In each graph, the horizontal axis indicates the number of QIDs
		used by the adversary, and the vertical axis indicates the adversary's success.
		Each dot is the posterior success of a distinct adversary
		having as auxiliary knowledge one of the 2,047 possible combinations of QIDs.
		The horizontal \qm{\emph{a priori}} line represents the adversary's
		%deterministic or probabilistic 
		success before the attack.
%		The vertical distance between the prior success line and
%		each dot represents the privacy degradation in the corresponding attack
%		(the larger the distance, the higher the privacy degradation).
	}
	\label{fig:all-attacks-S18}
\end{figure*}

%\paragraph{Experimental analyses of single-dataset attacks.}% (\CRS, \CAS).}
\textbf{Experimental analyses of single-dataset attacks.} % (\CRS, \CAS).}
Collective-target re-identification (\CRS) and collective-target
attribute-inference (\CAS) attacks on a single-dataset were performed on the dataset of the 
School Census of 2018, described in \Table{tab:datasets}.
%In each attack we quantified both the deterministic and the probabilistic
%degradation of privacy
%, as described in \Sec{s091634a}.
Fig.~\ref{fig:all-attacks-S18} depicts both the
deterministic and the probabilistic degradation of privacy
in each of the 2,047 distinct scenarios considered for each attack, 
every one of them corresponding to an adversary obtaining as auxiliary knowledge a different
non-empty subset of the 11 possible QID attributes listed in
\Table{tab:CRS-CAS-CRL-CAL-qid}. 
%\colorB{
	Additionally, 
\Table{tab:results-comparing-sweeney} 
provides detailed numbers for some of the 2,047 scenarios from Fig.~\ref{fig:all-attacks-S18}.
%}
%We consider here that \texttt{School code} is an approximation for a 5-digit ZIP code and that \texttt{City of residency} is an approximation for County
%\msdone[Shouldn't we have kept at least how many students have flag \qm{yes} for the sensitive attributes in Tab.~\ref{tab:results-comparing-sweeney}?][\ghn]

%\begin{table*}[tb]
%	\centering
%	\renewcommand{\arraystretch}{1.2}
%	\begin{small}
%		\begin{tabular}{|>{\centering\arraybackslash\small}m{0.23\linewidth}||>{\centering\arraybackslash\small}m{0.2\linewidth}|>{\centering\arraybackslash\small}m{0.2\linewidth}|>{\centering\arraybackslash\small}m{0.2\linewidth}|}
%			\hline  
%			\textbf{QIDs used} & \textbf{\CRS} & \textbf{\CAS} (disability) & \textbf{\CAS} (trans\-por\-ta\-tion)\\
%			\hline \hline
%			\texttt{DoB} +
%			\texttt{City of residency} 
%			& 26.63\% ($\sim$12.8 million) 
%			& 65.54\% ($\sim$31.6 million) 
%			& 54.53\% ($\sim$26.3 million) \\
%			\hline
%			\texttt{DoB} +
%			\texttt{School code} 
%			& 89.23\% ($\sim$43.0 million) 
%			& 99.63\% ($\sim$48.0 million) 
%			& 98.63\% ($\sim$47.5 million)   \\
%			\hline
%		\end{tabular}
%	\end{small}
%	\caption{
%		Deterministic degradation of privacy in 
%		re-identification (\CRS) and
%		attribute-inference (\CAS) attacks on
%		the single dataset of 2018.
%		The numbers represent the fraction of
%		students whose sensitive attribute 
%		becomes inferrable with absolute certainty 
%		after the attack.
%		%		Similar QIDs to those used by Sweeney \cite{Sweeney00} and their respective degradation of privacy in \CRS\ attacks on the School Census of 2018. (Percentage of students re-identified with gender as geography and date of birth information vary. We consider here that \texttt{School code} is an approximation for a 5-digit ZIP code and that \texttt{City of residency} is an approximation for County.)
%	}
%	\label{tab:results-comparing-sweeney}
%\end{table*}

\begin{table}[tb]
	\centering
	\renewcommand{\arraystretch}{1}
	\begin{small}
		$
		\begin{array}{|>{\centering\arraybackslash\small}m{0.26\linewidth}||>{\centering\arraybackslash\small}m{0.26\linewidth}|>{\centering\arraybackslash\small}m{0.26\linewidth}|}
			\cline{2-3}
			\multicolumn{1}{c|}{}
			& \textbf{QIDs: \texttt{DoB}, \texttt{Gender}, \texttt{CoR}}
			& \textbf{QIDs: \texttt{DoB}, \texttt{Gender}, \texttt{CoR}, \texttt{SC}} \\ \cline{2-3} \noalign{\vskip 0.5mm} \hline
			\textbf{\CRS}
			& 26.63\% \qquad ($\sim$12.8 million)
			& 90.43\% \qquad ($\sim$43.6 million)
			\\ \hline
			\textbf{\CAS} \quad \text{(disability)}
			& 65.54\% \qquad ($\sim$31.6 million)
			& 99.67\% \qquad ($\sim$48.0 million) 
			\\ \hline
			\textbf{\CAS} \quad \text{(trans\-por\-ta\-tion)}
			& 54.54\% \qquad ($\sim$26.3 million)
			& 98.77\% \qquad ($\sim$47.6 million)
			\\ \hline
		\end{array}
		$
	\end{small}
	\caption{%\colorB{
			Deterministic degradation of privacy in re-identification (\CRS) and attribute-inference 
	(\CAS) attacks on the School Census of 2018
	for some of the 2,047 scenarios
	from Fig.~\ref{fig:all-attacks-S18}.
	\texttt{DoB} is day, month, and year of birth,
	\texttt{CoR} is city of residency,
	and 
	\texttt{SC} is school code.
	Percentages are the fraction of students 
	whose sensitive	attribute is inferrable 
	with certainty after the attack.
	%}		
		%		Similar QIDs to those used by Sweeney \cite{Sweeney00} and their respective degradation of privacy in \CRS\ attacks on the School Census of 2018. (Percentage of students re-identified with gender as geography and date of birth information vary. We consider here that \texttt{School code} is an approximation for a 5-digit ZIP code and that \texttt{City of residency} is an approximation for County.)
	}
	\label{tab:results-comparing-sweeney}
\end{table}
% The adversary has prior success equal to zero.
% Add the information on how many students have sensitive attributes.

%=============================
% SECTION
%=============================
%\subsection{Experimental analyses of longitudinal attacks (\CRL, \CAL)}

%\paragraph{Experimental analyses of longitudinal attacks.}% (\CRL, \CAL).}
\textbf{Experimental analyses of longitudinal attacks.} % (\CRL, \CAL).}
Collective-target re-identification (\CRL) and collective-target
attribute-inference (\CAL) attacks on longitudinal collections were applied to
the collection of School Census datasets from 2014 to 2017, described in
\Table{tab:datasets}. In all attacks the dataset of 2014 was considered
the focal one, and the 2015--2017 datasets were used as auxiliary
information.
In order to track the evolution of risks as the longitudinal collection grows,
in each scenario we assumed that the adversary begins with knowledge
of just the focal dataset, and then performs a new attack as each new
dataset is released through the years from 2015 to 2017.
In each case, the adversary performs an aggregation of the focal and the
auxiliary datasets by linking the unique, permanent pseudonymization code
provided by {\INEP} in all datasets considered.
%\footnote{Here we have chosen to
%	harness the availability of the persistent unique identifier published by
%	{\INEP}, the \texttt{Student ID} code. In the absence of such identifier, the
%	adversary could also use QID values instead, provided the model of attack
%	is updated accordingly to account for the extra uncertainty introduced by
%	this less precise linkage method.} 
%Moreover, recall that the aggregated dataset has the same number of records as the focal dataset.
Finally, we selected the attributes \texttt{City of residency},
\texttt{School code}, and \texttt{Educational stage} as QIDs for
both experiments, as specified in \Table{tab:CRS-CAS-CRL-CAL-qid}.
We then measured both the deterministic and the probabilistic degradation of
privacy for the set composed of those three attributes, as summarized in \Table{tab:longitudinal}.

\begin{table*}[tb]
	\centering
	\begin{subtable}[t]{0.99\textwidth}
		\centering
		\renewcommand{\arraystretch}{1.1}
		\begin{small}
			$
			\begin{array}{|>{\centering\arraybackslash\small}m{0.15\linewidth}|>{\centering\arraybackslash\small}m{0.22\linewidth}|>{\centering\arraybackslash\small}m{0.22\linewidth}|>{\centering\arraybackslash\small}m{0.22\linewidth}|}
				\cline{2-4}
				\multicolumn{1}{c|}{} 
				& \multicolumn{1}{>{\centering\arraybackslash\small}m{0.22 \linewidth}|}{\textbf{\CRL}} 
				& \multicolumn{1}{>{\centering\arraybackslash\small}m{0.22\linewidth}|}{\textbf{\CAL} (disability)}
				& \multicolumn{1}{>{\centering\arraybackslash\small}m{0.22\linewidth}|}{\textbf{\CAL} (transportation)} \\ \hline
				\textbf{\small Datasets in the} 
				& \textbf{\small prior success:} 0.00\%
				& \textbf{\small prior success:} 0.00\% 
				& \textbf{\small prior success:} 0.00\% 
				\\ \cline{2-4} 
				\textbf{\small long. collection} 
				& \textbf{\small posterior success} 
				& \textbf{\small posterior success} 
				& \textbf{\small posterior success}
				\\ \hline \hline
				\text{2014} 
				& $1.44\%$   ($\sim$0.7 million) 
				& $57.17\%$  ($\sim$28.3 million) 
				& $58.07\%$  ($\sim$28.7 million) 
				\\ \hline
				\text{2014 to 2015} 
				& $12.88\%$  ($\sim$6.4 million) 
				& $79.21\%$  ($\sim$39.2 million) 
				& $68.60\%$  ($\sim$34.0 million) 
				\\ \hline
				\text{2014 to 2016} 
				& $25.26\%$  ($\sim$12.5 million) 
				& $87.59\%$  ($\sim$43.4 million) 
				& $75.32\%$  ($\sim$37.3 million) 
				\\ \hline
				\text{2014 to 2017} 
				& $36.31\%$  ($\sim$18.0 million)  
				& $91.28\%$  ($\sim$45.2 million) 
				& $79.92\%$  ($\sim$39.6 million) 
				\\ \hline
			\end{array}
			$
		\end{small}
		\caption{Deterministic measure of privacy degradation (i.e. proportion of students whose sensitive attribute is inferred with certainty).}
		\label{tab:longitudinal-deterministic}
	\end{subtable}
	\\[2mm]
	\begin{subtable}[t]{0.99\textwidth}
		\centering
		\renewcommand{\arraystretch}{1.1}
		\begin{small}
			$
			\begin{array}{|>{\centering\arraybackslash\small}m{0.15\linewidth}|>{\centering\arraybackslash\small}m{0.22\linewidth}|>{\centering\arraybackslash\small}m{0.22\linewidth}|>{\centering\arraybackslash\small}m{0.22\linewidth}|}
				\cline{2-4}
				\multicolumn{1}{c|}{} 
				& \multicolumn{1}{>{\centering\arraybackslash\small}m{0.22 \linewidth}|}{\textbf{\CRL}} 
				& \multicolumn{1}{>{\centering\arraybackslash\small}m{0.22\linewidth}|}{\textbf{\CAL} (disability)}
				& \multicolumn{1}{>{\centering\arraybackslash\small}m{0.22\linewidth}|}{\textbf{\CAL} (transportation)} \\ \hline
				\textbf{\small Datasets in the} 
				& \textbf{\small prior success:} 0.000002\%
				& \textbf{\small prior success:} 98.21\% 
				& \textbf{\small prior success:} 82.50\% 
				\\ \cline{2-4} 
				\textbf{\small long. collection} 
				& \textbf{\small posterior success} 
				& \textbf{\small posterior success} 
				& \textbf{\small posterior success}
				\\ \hline \hline
				\text{2014} 
				& $4.25\%$ 
				& $98.71\%$ 
				& $91.64\%$ 
				\\ \hline
				\text{2014 to 2015} 
				& $20.08\%$ 
				& $99.03\%$ 
				& $93.03\%$ 
				\\ \hline
				\text{2014 to 2016} 
				& $34.37\%$ 
				& $99.30\%$ 
				& $94.17\%$ 
				\\ \hline
				\text{2014 to 2017} 
				& $45.60\%$ 
				& $99.49\%$ 
				& $95.07\%$ 
				\\ \hline
			\end{array}
			$
		\end{small}
		\caption{Probabilistic measure of privacy degradation (i.e. probability of successful inference of the sensitive attribute in one try).}
		\label{tab:longitudinal-probabilistic}
	\end{subtable}
	
	\caption{Privacy degradation in re-identification (\CRL) and
		attribute-inference (\CAL) attacks on the longitudinal collection
		containing the School Census datasets from 2014 to 2017.
		In all attacks the focal dataset is that of 2014 (and all others are used as auxiliary datasets), and the
		QIDs employed are \texttt{City of residency}, \texttt{School code},
		and \texttt{Educational stage}. 
%		The Census of 2014 was used as the
%		focal dataset, and the adversary performs a new attack as each new
%		auxiliary dataset from the years 2015 to 2017 is added to the
%		collection.
%		The deterministic degradation of privacy is the difference between the
%		posterior and prior successes, while the probabilistic
%		degradation of privacy is their ratio.
	}
	\label{tab:longitudinal}
\end{table*}

%\ghndone[Can you please check the numbers in Tab.~\ref{tab:results-comparing-sweeney} and Tab.~\ref{tab:longitudinal}?
%Also, the graphics in Fig.~\ref{fig:all-attacks-S18}
%seem wrong (wrong labels, and perhaps even wrong graphics).
%In any case, we'll need to rethink the presentation of such
%graphics, because they are taking upt too much space.
%In the deterministic graphics we can change the left-hand
%side $y$ label to ``millions of students'' only, and have it in one line. On the right-hand side of the $y$ axes we can just write ``Deterministic success'' or ``Probabilistic success'', and that will also save us one line (as the caption now already has the information on the type of success).
%Then we could try to present the graphs in a 2x3 matrix, rather than in a 3x2 one (just transpose the position of all graphs).][\ms]
%\msdone[The graphics in Fig.~\ref{fig:all-attacks-S18} are correct, but I've updated them as proposed. I still have to check the numbers in Tables~\ref{tab:results-comparing-sweeney} and \ref{tab:longitudinal}, which I'll do later while checking all the numbers on the paper.][\ghn]

%\clearpage % Forces all figures and tables that haven't been printed to be printed now, even if it costs a page break

%=============================
% SECTION
%=============================
\subsection{%\colorB{
	Vulnerabilities identified
		% by the analyses
%}
}

%\cmdraft{List them here as subsubsections or itemized list
%The highlighted stories eg disabilities etc.}
%
\review{Next we highlight key risks uncovered by our analyses.
An extensive list of all results is provided in \cite{Nunes:21,nunes_gabriel_henrique_2021_6533704}.}

%\msdone[Once paper is accepted, maybe cite my Master's thesis \cite{Nunes:21} here as a reference for additional analyses?][\ghn]
%\ghndone[Sure!][\ms]

%\subsubsection{Most data holders are at a considerable privacy risk}
%\paragraph*{A vast number of the approximately 50 million students in each of {\INEP}'s datasets are at considerable risk  even against modest adversaries.}
\textbf{A vast number of the approximately 50 million students in each of {\INEPallcaps}'s datasets are at considerable risk  even against modest adversaries.}
As an example, in the School Census of 2018,
an adversary starting with prior knowledge of only the released dataset
itself would not achieve absolute certainty in any of the
re-identification or attribute-inference attacks considered.
%\ghndone[I couldn't understand the double-negative in that sentence. Help!][\tash]
%\tashtodo[Is it understandable now?][\ms]
%
However, after acquiring as auxiliary information only 3 QIDs --day and month
of birth, and school code-- the adversary becomes able to re-identify up to
30.92\% of the records ($\sim$14.9 million students) and infer the disability
status and transportation method of 95.35\% ($\sim$45.9 million students) and
85.63\% ($\sim$41.3 million students), respectively. By adding year of birth
as a fourth QID, those numbers increase to 81.13\% ($\sim$39.1 million
students), 99.31\% ($\sim$47.8 million students), and 97.42\% ($\sim$46.9
million students), respectively.
As for probabilistic attacks, the adversary's expected prior success in
re-identifying any 
%randomly selected 
individual is only $0.000002\%$,
but with the use of the same four QIDs as before, the posterior expected
success increases to 89.93\%. On the other hand, the expected success
in inferring a random individual's disability or transportation method is
already very high even a priori, due to the highly skewed distribution of such attributes in
the population:
%~\footnote{In the School Census of 2014, 98.21\% of the records ($\sim$48.6 million students) do not have a disability and 82.50\% of the records do not use public school transportation ($\sim$40.8 million students). 
%	The corresponding fractions in the School Census of 2018 are 97.56\%  ($\sim$47.0 million students) and 81.75\% ($\sim$39.4 million students), respectively.}
 $97.56\%$ and $81.75\%$, respectively. By using again the same
four QIDs, the adversary's posterior expected success increases to 99.69\% and
98.82\%, respectively.

\textbf{There are plenty of unexpectedly powerful combinations of QIDs.}
The use of just city of birth and city of residency as QIDs, for instance,
allows for the unique re-identification of approximately 430,000 students.
The addition of ethnicity to that combination increases that number to around
800,000 uniquely re-identifiable students.
Another remarkable example is the use of School Code alone (which is a unique
identifier for each school in the country) as a QID in the School Census of
2018. This attribute alone allows the adversary to re-identify with absolute
certainty 99 students.~\footnote{More precisely, 57 of these 99 students were
	already uniquely re-identifiable in the original, pre-treated dataset
	(i.e. before the removal of duplicate entries for each  student), and the
	other 42 students became uniquely re-identifiable only after such treatment. 
%	\colorB{
	On the other
	hand, 26 
	%other
	students that were unique based on School Code
%} 
in the
	original dataset had those records removed by the treatment. 
	In any case,
	it is remarkable that there are dozens of such unique students, even in a dataset as large as the one analyzed.
	Interestingly, our analysis shows that the majority of these unique school codes
	refer to institutes in rural, indigenous, or \textit{``quilombola''} (i.e.
	traditional communities formed by descendants of slaves who escaped
	captivity) areas of the country, or to institutes of specialized education.
	This suggests that, although such re-identification cases may be relatively rare in the country, 
	they disproportionately affect protected minorities.
}
Perhaps even more impressively, the use of the same School Code on its own allows for the
inference, with absolute certainty, of the disability status of 5.9 million
students in the same dataset.

%\subsubsection{Longitudinal attacks can are extremely powerful}
%\paragraph*{Even modest longitudinal attacks can be highly damaging.}
\textbf{Even modest longitudinal attacks can be highly damaging.}
As an example, an adversary starting with prior knowledge of only the released School Census of 2014 would not achieve %absolute 
certainty either in re-identifying or in inferring any
individual's disability status or transportation method.
However, %even when 
by knowing only some seemingly innocuous QIDs
--city of residency, school, and educational stage--, 
%an adversary
% may be able
%is able to severely degrade %data holders' 
%privacy over the course of only a few years.
%Indeed, with knowledge of only these three QIDs 
and having access to three
auxiliary datasets, from years 2015--2017, the adversary can
re-identify 
%\colorB{
with 
%absolute 
certainty
%} 
up to 36.31\% of the records ($\sim$18.0
million students), and infer the disability status and transportation method
of, respectively, 91.28\% ($\sim$45.2 million students) and 79.92\% ($\sim$39.6
million students).
When considering probabilistic measures, again with prior knowledge of only the
%dataset for the 
School Census of 2014, the adversary's probability of
re-identifying a randomly selected individual is $0.000002\%$, whereas the
probability of inferring that random individual's disability status or
transportation method is, respectively, $98.21\%$ and $82.50\%$. After
acquiring access to the same three QIDs and the same 3 auxiliary datasets, the
adversary's probability of correctly re-identifying a random individual
increases to 45.60\%, whereas those of inferring disability status or
transportation method increase to 99.49\% and 95.07\%, respectively.

\section{Lessons learned}\label{s091634}

% !TEX root = main.tex

%\cmdraft{Say there are three, and name them, before moving onto their subsections.}

%\msdraft{After our re-organization of the paper on 2021-08-25/26,
%	I wonder if the scientific and technical lessons could be merged into a single topic, as they are shorter now.
%	In fact, some of the stuff that used to be here has been moved to other parts of the paper (e.g., the scientific discussion of unexpected vulnerabilities is now on the section of experiments, and the technical discussion of the tool built for our endeavor is now in the section of the {\INEP} case-study).}

%\colorB{
	The rigorous evaluation of privacy 
%risks 
in 
\INEP's Educational Censuses raised challenges 
in various fronts: scientific, technical, and of
communication.
Here we discuss the main lessons 
learned while overcoming these challenges.

%=============================
% SECTION
%=============================
\subsection{Communication and social aspects}
\label{s201148}

%\cmdraft{I suggest we put this first as social impact is probably of most interest.}

The main \emph{communication} challenge in our project was 
to identify --and even develop-- 
ways to effectively	transmit the results of our 
formal analyses to {\INEP}'s agents acting at the 
technical, managerial, and political levels.
For as mathematically sound and experimentally thorough 
any formal analysis might have been, it could only foster 
real change if it persuaded {\INEP}'s agents that the results applied specifically to
their datasets so that they could report these findings to the 
people empowered to make decisions.
%privacy issues at a concrete level, and then empowered
%t%hem to make well informed decisions.
%While presenting scientific and technical challenges
%we have already discussed some important communication lessons learned.
%We now will briefly cover some additional ones.
Next we summarize some additional lessons not yet covered in this paper.

%\paragraph{Even very well-known risks in the literature needed to be reproduced in {\INEP}'s own data.}
%\paragraph*{Results from the academic literature needed to be specifically interpreted and reproduced in the {\INEP} setting.}
\textbf{Results from the academic literature needed to be specifically interpreted and reproduced in the {\INEPallcaps} setting.}
As academics, one of the important inputs 
to a research effort is to learn from the findings of other researchers 
working on similar problems. However, 
in our early interactions with {\INEP}'s staff it became
clear that, despite the abundant evidence in the literature
pointing to the contrary, some influential (although not all) agents remained skeptical that
the literature could be relevant to {\INEP}'s own dataset.
%\abdone[There were people that believed the literature from the start, but not all. Perhaps we can say we had people on our side from the beginning too, but they were not all (specially because these people will read this paper!) I have changed the wording a bit. I suppose that the skeptics were influential o/wise why would their opinion matter?][\ms]
Indeed, these few skeptical agents hewed firmly to their belief that the
large number of individuals in each data-release
under scrutiny would automatically ensure some 
reasonable level of privacy --- and remained unmoved even when presented with
our comprehensive literature
review pointing out vulnerabilities in other datasets together with some anecdotal %evidence showing that 
examples of relatives of members of 
our research team that could be easily re-identified in the Censuses. In fact some of {\INEP}'s agents reaffirmed their belief in an intuition of \qm{safety in a crowd,} and 
brushed off our initial findings as a fluke.

As well as convincing {\INEP}'s agents of the need for change we %were also aware that we had to
also had to 
anticipate the possible impact of any modification 
in their current data-release policies on the civil society's perception of the agency's commitment to transparency.
%~\footnote{As an example, the US Census Bureau has faced serious resistance from stakeholders when discussing changes on the current balance between transparency and privacy in their data-publishing methods~\cite{Garfinkel18,Mervis:19:Science}.}

For these reasons we found that it was not enough to expect all of our {\INEP} counterparts to be able to recognize how well-known 
privacy issues described in the literature could apply to their own datasets, even if, as researchers, we were able to explain the underlying principles.
We had to reproduce those attacks and demonstrate specifically the potential for future harm. This turned out to be the only way 
%even
%even very well-known privacy risks from 
%the literature needed to be reproduced in \emph{{\INEP}'s own data} 
to convince both the agency (and public) of the relevance of prior %academic 
research.
%In that sense, we needed to provide {\INEP} with substantial
%empirical evidence of the damage attacks could cause 
%\emph{on the agency's own data} --rather than evidence 
%based only on, say, a review on the abundant literature 
%on the matter.

As described, the production of this concrete evidence posed
a serious challenge; 
but it 
turned out to be 
%was 
critical in convincing the agency 
--and, hopefully, in the near future, also a public accustomed to having access to highly useful data releases--
%for many years-- 
of the  necessity of changes in {\INEP}'s management of finding the right balance
 between transparency and privacy.

%\paragraph*{Irrational adversaries and unrealistic but simple scenarios acted as a stepping stone to explaining how realistic privacy risks in the {\INEP} datasets could potentially harm many citizens.}
\textbf{Irrational adversaries and unrealistic but simple scenarios acted as a stepping stone to explaining how realistic privacy risks in the {\INEPallcaps} datasets could potentially harm many citizens.}
As described in \Sec{s1011aa}, great effort went into analyzing ``deterministic vulnerabilities''  in spite of the sometimes misleading impression of security that they imply \cite{Alvim:16:CSF}. Such measures, for example, do not distinguish between an adversary who is $99.99\%$ accurate in her ability to identify individuals and one who is only $0.01\%$ accurate. In a deterministic assessment neither adversary is regarded as a threat because they cannot identify individuals with absolute certainty. However early on in our discussions, the deterministic measures turned out to be the most understandable for our {\INEP} counterparts; but we wanted to alert them to the potential harm that highly accurate, albeit imperfect adversaries could  pose to a large number of citizens. As well as computing the level of risk exposure, we also explained the results using scenarios such as the following.

A job agency is considering two 
candidates for a position.  However, the way that {\INEP} currently releases data allows the company to use additional information provided  in the candidates' resum\'es  to determine that the first candidate has a 30\% chance of having a disability, whereas the second candidate has only a 5\% chance. Of course now the agency has a choice whether to discriminate against the first candidate and offer the job to the second --- this would be a  decision made using a probabilistic inference but causing definite harm that, if done at scale, could reach a large number of citizens.

When these risks were explained, the {\INEP} team reported %to us 
that their perception of privacy 
%vision about the privacy problem 
changed significantly and that there were many more threats than just being able to identify individuals precisely. 
%Moreover 
This also led them to re-evaluate the simple mitigation techniques that 
they had been considering.
\subsection{Technical and scientific aspects}
\label{s201149}

The main \emph{technical} challenge in our endeavor 
consisted of adapting {\QIF} attack models --which are information-theoretic and, hence,
provide exact, rather than approximate, results-- 
to enable effective computational analyses at 
the large scale of {\INEP}'s scenario:
the datasets to be analyzed covered a period of 13 years, 
with each year's data containing approximately 50 million
records, each with up to around 90 attributes
(see \Table{tab:datasets}).
In particular, for each attack it was necessary to identify 
instantiations of the adversary's prior knowledge 
and of vulnerability measures
that were not only meaningful and persuasive 
to the decision makers at {\INEP},
but also computationally tractable.
Here we describe the main lessons learned while overcoming 
these challenges.

%\paragraph*{Applying academic research to real-world privacy problems requires consolidation and explainability.}
\textbf{Applying academic research to real-world privacy problems requires consolidation and explainability.}
One of our main scientific contributions outlined in \Sec{s102913}
%\Sec{s102913} and \ref{s100834} 
was to consolidate and systematize the body of knowledge on privacy threats.
Unfortunately, 
%we discovered that
%\colorB{
as it turns out
%}
we were unable to apply those findings directly to the {\INEP} datasets without a laborious consolidation step.
This was because 
our task was to provide a comprehensive assessment of known privacy risks; but whilst the literature provided an important input, each of the documented vulnerabilities were often performed by different teams on different datasets with their own special features and unique experimental set up. Not only were their overarching lessons not accessible to the {\INEP} agents, but it 
%\colorB{
	was not 
	%at all 
clear
%} 
which experiments were
essentially doing the same thing and which were related to a genuinely different adversarial setting. 

Our systematic treatment (\Sec{s102913}) together with its implementation in {\QIF} terms  (Sections~\ref{s100834} and \ref{s091634a}) ensured the broad coverage required for us to be confident in the advice we provided to {\INEP}. 

%\paragraph*{Engagement with real-world problems benefits basic scientific research.}
\textbf{Engagement with real-world problems benefits basic scientific research.}
The main \emph{scientific} challenge consisted of
rigorously 
formalizing both known and novel attacks to obtain a comprehensive evaluation of privacy
vulnerabilities in a real-world setting: our 
coherent {\QIF} framework enables
rigorous quantification and
comparison of privacy risks. %(\Sec{sec:formal-models}).

One of the outcomes of our project with {\INEP} was to discover new ways to use our {\QIF} framework, and a new understanding of its agility for  representing and explaining complex scenarios, \emph{and} that the hyper-distribution approach is not only useful as an abstract concept but  leads to a compact representation important for scalability.
As a positive side effect, we identified many ways in which
our framework can be used to model further, more sophisticated threats; we discuss these prospects in the next section.
\section{Conclusions and prospects}\label{s191632}

% !TEX root = main.tex

In this work we rationalized
a myriad of known and novel attack models 
in the rigorous
framework of \emph{Quantitative Information Flow},
showing how it can express concrete attacks
and quantify 
%the corresponding 
privacy risks %of very large data releases.
at very large scale using
%We used 
as an example the case of {\INEP}'s Educational Censuses datasets.
To the best of our knowledge, this is the largest privacy 
analysis ever performed on official governmental microdata --with rich records (over 90 attributes) from around 
50 million individuals across many years.
%\msdone[I changed `data publishing' to `microdata publishing' since there has been some
%analysis on the US 2020 Census statistical data publishing.][\tash]
%\tashtodo[Sure, thanks!][\ms]
Our results were crucial in enabling {\INEP} to reach
well-informed decisions on the balance between privacy and transparency,
which directly impacts the 25\% of the Brazilian
population represented in these data.
\mstodo[Discuss impact of our study and recent developments.][\ms]
%\colorB{
Indeed, the agency
	is currently considering our suggestions for coping with the problem,
	including the publication of only aggregated data protected by some 
	form of differential privacy or its variants,
	and allowing access to microdata only via safe rooms.
%\footnote{\colorB{{\INEP} 
%		is currently considering our suggestions for coping with the problem,
%		including the publication of only aggregated data protected by some 
%		form of differential privacy or its variants,
%		and allowing access to microdata only via safe rooms.}}
But, beyond \INEP's context, we hope that the lessons %we 
learned 
in our endeavor 
can help other
practitioners to communicate more effectively with decision makers and the public.
%}

%%=============================
%% SECTION
%%=============================
%\paragraph{Other, further kinds of analyses}\label{s090534}

We now consider some meaningful extensions of the analyses
performed on {\INEP}'s scenario.

%\mstodo[QIF tells you how to write code that will scale even if unique identifiers are not available. Use ``scalability'' term alone, and not ``longitudinal scalability'', specially in titles. Just say that we can do amounts of data that are bigger than before, for reasons like treating longitudinal datasets.]

%\paragraph*{Generalizing the model of longitudinal aggregation.} 
%\textbf{Generalizing the model of longitudinal aggregation.} 
\review{\textbf{Scalability in general longitudinal collections.}
In the analyzed {\INEP} Censuses, there exists a persistent 
%unique 
identifier for every individual across all datasets, 
which renders dataset aggregation straightforward. 
%(since, as discussed, the linkage between a same individual's 
%records on different datasets can be performed with a simple 
%left outer join keyed on this identifying attribute).
%% CAPYBARA !!!
Without such an identifier,
%When that is not the case, 
an adversary may rely on QID 
values and prior knowledge about the population of 
interest to try to match the same individual's records 
across different datasets. 
For instance, she may link a record for a student aged \qm{12} in one year
with one aged \qm{13} in the following year, and with the same nationality over
both years.
%(For instance, she may use the fact that 
%a person's age increases by 1 every year to try to 
%link a record aged \qm{12} in one year with a record aged 
%\qm{13} in the following year, and the fact that a person's
%nationality is typically very stable to link a record
%with ``Brazilian'' nationality in one year with a record
%with the same nationality in the following year.)
However, the aggregated dataset so obtained will
present some inherent uncertainty (e.g.\ because some people indeed \emph{do} change nationality from one year to the other, 
which may lead to a wrong record-linkage), 
and any leakage analysis performed on such a dataset needs to 
account for that uncertainty.
%While other tools (such as ARX) are not equipped to deal with this problem,\footnote{ARX can only take one dataset as input at a time, and does not account for uncertainty in the dataset's construction.}
This effect can be naturally accounted for in the {\QIF} framework with appropriate models of an adversary's prior knowledge about the population of interest, guaranteeing an accurate overall leakage
assessment.
Indeed, starting from 2018 {\INEP} has discontinued
the use of a unique individual identifier across datasets,
and we intend to perform a formal privacy analysis on the agency's new policy.
}
%% CAPYBARA !!!

% WOMBAT!!! \footnote{Is this explained somewhere?}
%\alldone[Regarding WOMBAT: I rewrote the paragraph above. Please take a look to see if it's ok now. Look for CAPYBARA and see the explanation.][\ms]

%\alldone[The above functionalities, although not present in ARX, were not needed for \emph{this particular study}.
%But by using QIF and our tool we can easily use these functionalities in future studies. Should we make that clearer, e.g.\ in \Sec{s191632}?][\ms]

%\paragraph*{Robust analysis of privacy risks using capacity.}
\textbf{Robust analysis of privacy risks using capacity.}
%\colorB{
Our analyses considered reasonable
%} 
--and modest--
adversaries, and showed that even those posed significant privacy
risks to data owners in {\INEP}'s datasets.
More precisely, such adversaries had limited prior knowledge,
and their intention was mostly to guess the secret value correctly in one try.
The {\QIF} framework allows for many more adversarial models, and, in a precise mathematical sense, it can cover all \qm{reasonable} adversarial models according to a set of fundamental information-theoretic 
axioms~\cite{Alvim:16:CSF,Alvim:19:TCS}.
Furthermore, we can use a theory of channel \emph{capacity}~\cite{Alvim:14:CSF} to
estimate the maximum risk {\INEP}'s data publishing can cause over
\emph{all} these reasonable adversarial models, providing 
%\colorB{%an effective 
a robust
%}
upper bound on the corresponding privacy risks.

%\paragraph*{Analyses of publications other than unmodified microdata.} 	
\textbf{Analyses of publications other than unmodified microdata.} 	
%\colorB{
	We performed %all of 
our %database 
attacks on the unmodified microdata released by {\INEP},
and protected only by de-identification and pseudonymization. 
As already mentioned, the institute is now considering applying
mitigation techniques that will change the published
data's format (e.g. to sanitized microdata, or to aggregated data protected by some form of differential privacy~\cite{Dwork06}).
%}
Because our {\QIF} model is agnostic to the particular form of published data
(since the adversary's posterior knowledge is represented by a hyper), it can
be easily extended to analyze these scenarios, whereas other tools (such as ARX) cannot.
\review{As an example, Appendix~\ref{s03060736} shows 
	how the \QIF\ framework can be used to assess privacy under
	the popular syntactic anonymization techniques of $k$-anonimity and $t$-closeness.}
%
% flexible and can be easily modified to deal with 
%
%We intend to apply our {\QIF} privacy analyses to these treated databases.
%
%In addition to inducing the rationalization of privacy attack models for microdata as discussed in Sec.~\ref{s100834}, \QIF\ can also be used on the analysis of microdata samples, aggregated data, and anonymized datasets. For instance, \QIF\ was used to study the relationship between accuracy and privacy in data releases and to demonstrate how correlations between attributes pose challenges for protecting privacy of individuals \cite{Alvim20:CONCUR}.

%%=============================
%% SECTION
%%=============================
%\subsection{Further features of {\QIF}}
%
%\cmdraft{IE that enable the above Sec.\ \ref{s090534}.
%Recap what is special and unique about this study. Make some recommendations for further collaboration with government organisations.}
%
%\cmtodo[I don't know what Carroll has in mind for this section, but we can discuss it in our next meeting.][\ms]

\review{\textbf{Availability.} %\label{s000000}
	The software developed for this work,
	together with our privacy analyses, is available at 
	%following 
	%GitHub repository 
	\href{https://github.com/nunesgh/bvm-library}{\texttt{nunesgh/bvm-lib}} \cite{nunes_gabriel_henrique_2021_6533704}.
	The repository includes a 
	%demonstration using 
	demo with ProPublica's data from the COMPAS tool.~\footnote{\url{https://github.com/propublica/compas-analysis}}}

%-----------------------------

%=============================
% SECTION
%=============================
\section{Further related work}\label{s095512}

Dalenius initiated a rigorous approach to statistical
disclosure control in 1977 \cite{Dalenius77:sT}.
De-identification was already 
%considered to be a necessary but insufficient measure
known to be an insufficient measure \cite{Dalenius86},
and several disclosure control methods 
have been proposed  considering various attack models~\cite{Samarati98,Machanavajjhala07,Li07,Dwork06,Raskhodnikova08,Dwork11}.
%As discussed in \Sec{sec:background-attacks}, those include the syntactic methods known as \emph{k}-anonymity \cite{Samarati98}, \emph{l}-diversity \cite{Machanavajjhala07}, and \emph{t}-closeness \cite{Li07}. But those methods were developed to be applied on single-datasets, not on longitudinal ones.
Fung et al. \cite{Fung10:Book} and Divanis et al. \cite{Divanis15:Book} provide a thorough review
of available methods. 
%Particularly, Fung et al. discuss different approaches to the problem of continued release of data, including \emph{incrementally updated data records} and its \emph{dynamic data republishing} case \cite{Byun06,Xiao07,Bu08}.
%Nevertheless, most methods for single-datasets and all for longitudinal ones are vulnerable to composition attacks \cite{Ganta08}.
%\ghndone[Cite the Handibook on data privacy][\ms]
%Byun et al. were the first to identify this problem in 2006 \cite{Byun06} and proposed a new approach for updating incremental datasets that would prevent an adversary's inferences, but only for the insertion of new records. This limitation was not present in Xiao and Tao's \emph{m}-\emph{invariance}, proposed in 2007 and which allowed both the addition and removal of records \cite{Xiao07}, but not the update of quasi-identifying or sensitive values over time. Hence, Bu et al. proposed the \qm{HD-composition} \cite{Bu08} in 2008, which considers not only that quasi-identifying and sensitive attributes may change over time, but also that some sensitive attribute values, once linked to a record, will never change, e.g. some diseases. But since all those methods are syntactic in nature, they are all vulnerable to composition attacks \cite{Ganta08}.
%
Notable re-identification attacks include 
Sweeney's seminal analysis of the US 1990 Census~\cite{Sweeney00} and 
Narayanan and Shmatikov's attacks on the Netflix dataset~\cite{Narayanan08}. 
%\colorB{
	Our work, however, covers significantly larger datasets
and is, to the best of our knowledge,
the most comprehensive on longitudinal governmental microdata.
More specifically, Sweeney's results were limited to estimates of how many people could be re-identified with certainty in the US 1990 Census, using 
only 
a few combinations of QIDs as auxiliary knowledge,
% whereas 
\review{whilst}
we considered both 
probabilistic and deterministic measures of success under thousands of combinations
of QIDs, %and 
%also 
%developed techniques to analyze threats on longitudinal collections.
\review{and also included longitudinal collections.}
%~\footnote{The US 1990 Census was published in tables according to a geographical specification (5-digit ZIP code, place, or county) and reported for each area as a count by age subdivisions, e.g. how many people under the age of 12, or how many people between 45 and 54 years of age.} 
On the other hand, Narayanan and Shmatikov's attacks on the Netflix dataset did not involve governmental data % or longitudinal collections, 
being limited to the Netflix Prize dataset and publicly available data from the Internet Movie Database (IMDb), used as auxiliary information.
%~\footnote{Their analyses empirically demonstrated serious and widespread risks of re-identification using data from 100\,480\,507 movie ratings of 480\,189 Netflix subscribers from December 1999 to December 2005.}
%~\footnote{The Netflix dataset was de-identified and anonymized, but the company did not provide details on the anonymization methods applied.}
%
% USCB.
%}
The US Census Bureau has identified that
their published tables are vulnerable to database reconstruction attacks \cite{Nissim03}, but
% as first proposed by Dinur and Nissim \cite{Nissim03}. 
%In such attack, an adversary is able to reconstruct the original microdata by performing a small set of queries on the released tables, which implies the re-identification of data holders. 
%However,
%Even though researchers from the US Census Bureau have performed experimental attacks on their datasets, which have empirically confirmed the vulnerabilities, 
they do not publish their results because of legal reasons \cite{Abowd18,USCB:Legacy}.
%\cite{Abowd18,USCB:Hist,USCB:Legacy}
%
%Another related work would be the maximum-knowledge intruder by Domingo-Ferrer et al. \cite{Domingo15:Maximum} for anonymized datasets. Here, the adversary performs linkage attacks and knows all the attributes in the dataset, if interested in identity disclosure, or all the attributes except for one, if interested in the disclosure of that attribute. 
%The measure used for disclosure risk is the distance between each pair of linked records, i.e. a record from the original dataset and another from the anonymized dataset, exemplified by the permutation distance.
%
Alternative tools to ARX (discussed in \Sec{s191650})
include the open source and continuously supported \texttt{sdcMicro} package \cite{sdcMicro,Templ15} for the R programming language. This package focuses on measuring the disclosure risk in microdata and provides some well-known anonymization methods, but it suffers from similar lack
of flexibility as ARX.
%
% Malacaria
\QIF\ was pioneered by 
Clark, Hunt, and Malacaria \cite{Clark:01:ENTCS}, %, and we believe the name and acronym were coined by Sebastian Hunt. 
followed by a growing community (see e.g. \cite{Malacaria:07:POPL,Chatzikokolakis:08,Smith09}), and its principles have been organized in~\cite{Alvim20:Book}. 
\review{\paragraph*{Acknowledgments.}
M\'{a}rio S.\ Alvim and Gabriel H.\ Nunes were supported 
by CNPq and CAPES. %, and FAPEMIG.
Carroll Morgan was supported by Trustworthy Systems.
The authors are grateful to {\INEP} for the partnership and constructive interactions, as well as to the other members of the UFMG PRICE project team:
Ramon G.\ Gonze, Jeroen van de Graaf, Igor W.\ Lemes, Lucas Lopes, and Jos\'{e} C.\ Oliveira Jr.
}
\alltodo[Complete Acknowledgments.]
%\alltodo[I think we should dedicate this paper to Geoffrey's memory. Do you all agree?][\ghn]
\mstodo[Once paper is accepted, remember to acknowledge the PRICE team.][\ghn]
\mstodo[Natasha has no acknowledgment to add.][\ghn]
\bibliographystyle{plain}
%\newpage
\bibliography{bibliography}
%-----------------------------

\appendix

%=============================
% APPENDIX
%=============================
\review{
\section{\review{Overview of Brazilian privacy and transparency laws}}\label{s2202280000}
	
% !TEX root = main.tex
% FROM introduction.tex:

%Here we provide more detail on the Brazilian legislation
%that regulates the issues of privacy and transparency 
%in the release of official statistical data about individuals.

\begin{comment}
The Brazilian Constitution from 1988 \cite{BR88} in itself already establishes some 
guiding principles.
In its Article 5, guarantees to all Brazilians and foreigners 
residing in the country some expectation of privacy:
\begin{quote}
	\begin{enumerate}
		\item[X.] the privacy, private life, honour and image of persons are inviolable, and the right to compensation for property or moral damages resulting from their violation is ensured;
	\end{enumerate}
\end{quote}
as well as of transparency:
\begin{quote}
	\begin{enumerate}
		\item[XXXIII.] all persons have the right to receive, from the public agencies, information of private interest to such persons, or of collective or general interest, which shall be provided within the period established by law, subject to liability, except for the information whose secrecy is essential to the security of society and of the State;
	\end{enumerate}
\end{quote}
\end{comment}

Article 5 of the Brazilian Constitution from 1988 \cite{BR88} establishes  
guiding principles on the right to both privacy (for individuals) and transparency (on matters of public concern). However, the Constitution does not provide guidance on
how to balance those two principles; these 
% or on what would be the legal limits of each one of them. 
%Therefore, to regulate the rights to privacy and transparency determined in those entrenched clauses of the Constitution, 
%two pieces of legislation were enacted.
 are detailed in the following pieces of legislation.

\textbf{The transparency law} -- this is 
%On one hand, there is the 
Law 12\,527 of 2011, known as {\LAI} 
(\emph{Lei de Acesso \`{a} Informa\c{c}\~{a}o}, 
or \emph{Access to Information Act}).
%, particularly that %considered to be 
%of collective or general interest, which must be made available via the Internet regardless of requirement. 
%Denial of access to information may only be granted as an exception
%to a philosophy of \qm{transparency by default}, and properly justified.
It requires that public authorities guarantee broad access to information, particularly that considered to be of collective or general interest, which must be made available via the Internet (Articles 6 and 8) or otherwise (Article 7) except where such information is considered confidential (Article 22). % regardless of requirement. 
%In addition, Article 7 guarantees the right of access to other information not immediately available via the Internet. 
%However, according to Article 22, access to information may be denied in whole or in part if the information is considered to be confidential. 
Regarding the treatment of personal information, Article 31 establishes that it must be done in a transparent manner and with respect to individual freedoms and guarantees; %, according to item X of Article 5 of the Constitution. 
however, provisions on the handling of personal information are open to subsequent regulation.

%Law 12\,527 of 2011 was sanctioned to regulate access to information and Law 13\,709 of 2018 was sanctioned to regulate the protection of personal data.

\textbf{The privacy law} -- this is 
%On the other hand, there is the 
Law 13\,709 of 2018, known as {\LGPD}~\cite{BR18}
(\emph{Lei Geral de Proteção de Dados Pessoais}, or
\emph{General Data Protection Act}), which
is based on the European 
General Data Protection Regulation ({\GDPR}).
%It establishes restrictions 
%on governmental agencies that publicly release data on individuals, 
%and prescribes sanctions in case of non-compliance. 
%In particular, data releases should not allow for the re-identification
%of data owners by \qm{reasonable means}, and should also protect 
%data owners' \textit{sensitive personal data} from disclosure.
%Law 13\,709 of 2018 \cite{BR18}, known as the General Personal Data Protection Law (\emph{Lei Geral de Prote\c{c}\~{a}o de Dados Pessoais}, or {\LGPD}, in Portuguese), 
%As per its Article 1, 
The law aims to protect the fundamental rights of freedom and privacy (Article 1). It determines that the processing of sensitive personal data is permitted with consent of the data subject or their legal guardian (Articles 7, 11). 
%Article 7 determines in which cases the processing of personal data is allowed, and Article 11 does the same specifically for sensitive personal data. In both cases, processing is permitted with the consent of the data subject or of the legal guardian.% for sensitive data. 
In its Article 5, {\LGPD} defines 
\emph{sensitive personal data} as
%\begin{itemize}
	%\item[II] Sensitive personal data: 
	\qm{personal data on racial or ethnic origin, religious belief, political opinion, union membership or affiliation to organizations of a religious, philosophical, or political nature, data relating to health or sexual life, genetic or biometric data, when linked to a natural person};
	and \emph{anonymous data} as
%	\item[III] Anonymous data: 
	\qm{data relating to an unidentifiable holder, considering the use of reasonable technical means available at the time of processing.}
%\end{itemize}
Article 12 defines that \qm{anonymous data} is not to be considered personal data, except when the anonymization process can be reversed with reasonable efforts. 
Therefore, objective factors such as the cost and time needed to reverse the anonymization process should be considered given the available technologies and disregarding the use of third party means. But again, the proper definition of what would be considered a reasonable effort, or which anonymization methods should be used, were left to subsequent regulation.
Finally, {\LGPD} is to be regulated by the National Data Protection Authority (\emph{Autoridade Nacional de Prote\c{c}\~{a}o de Dados}, or ANPD),
which is expected to face several challenges in harmonizing {\LAI} with {\LGPD}. 
%In this context, academic work such as the one we present here can help both society and the ANPD to understand the existent trade-offs in the case of INEP's Educational Censuses.

}
%-----------------------------

%=============================
% APPENDIX
%=============================
\section{Full procedure for Ex.\ in \Sec{s100834}}\label{s2110071153}

% !TEX root = main.tex

Here we present the full \QIF\ procedure 
to obtain the hyper-distribution (\Table{t104949-f}) from 
the original dataset (\Table{t104949-a}) in the example from \Sec{s100834}.
%Before the attack the adversary only has access to the the original
%dataset from Tab.~\ref{t104949-a}, and her prior knowledge about
%\textit{language} is determined by this attribute's distribution in this dataset.
%Since, out of 4 people, 1 speaks English, 1 speaks Portuguese, and 2 speak German,
%and assuming a uniform distribution on people in the dataset,
%the adversary's prior is that the probability of a randomly selected individual
%speaking each of these languages is, respectively, $\nicefrac{1}{4}$, $\nicefrac{1}{4}$, $\nicefrac{1}{2}$.
Recall that %we consider that 
when meeting a randomly selected individual,
the adversary is able to identify this person's age and gender.
She then performs Bayesian reasoning on the 
collected information and updates her knowledge 
about the language from the prior to 
%a set of revised conditional distributions 
%(given the learned value of each individual's \gender\ and \age) on \textit{language} s.t.\
%each of these posterior distributions has its own probability of occurring --- i.e. she updates her knowledge to 
a hyper on the secret value.

This whole process occurs as in \Table{t104949-full}.
First the adversary extracts from the original dataset all co-occurrences of 
values for language, gender, and age (\Table{t104949-full_correlation}),
and from that she derives a joint probability distribution on these
values (\Table{t104949-full_joint}).
By marginalizing the joint distribution, we
get the adversary's prior on language, and by conditioning the joint distribution on the prior we get the channel representing the adversary's information-gathering process during the attack (\Table{t104949-full_channel}).
The adversary's posterior knowledge is then represented by the hyper in \Table{t104949-full_hyper}, which is exactly the same as that in \Table{t104949-f}.

% Introduction attack example 
\begin{table*}[!tb]
	\begin{subtable}[t]{0.48\linewidth}
		\renewcommand{\arraystretch}{0.9}
		\centering
		\begin{small}
			$
			\begin{array}{|c|c c c c|}
				\hline
				\raisebox{-0.75mm}{\textit{gender, age} $\blacktriangleright$}
				& 
				\multirow{3}{*}{\rotatebox{90}{\small{\male\  $\leq30$~}}} 
				&			\multirow{3}{*}{\rotatebox{90}{\small{\male\  $>30$~}}}
				&			\multirow{3}{*}{\rotatebox{90}{\small{\female\  $\leq30$~}}} 
				&			\multirow{3}{*}{\rotatebox{90}{\small{\female\  $>30$~}}}
				\\[1mm] \cline{1-1}
				& 
				& 
				& 
				& 
				\\[-2mm]
				\raisebox{1.2mm}{\textit{language}~$\blacktriangledown$}
				& 
				& 
				& 
				& 
				\\
				\hline
				\rowcolor{sand!50!white} \texttt{English} & 0 & 1 & 0 & 0 \\
				\texttt{Portuguese} & 1 & 0 & 0 & 0 \\ 
				\rowcolor{sand!50!white} \texttt{German} & 1 & 0 & 1 & 0 \\ \hline
			\end{array}
		$	
		\end{small}
		\caption{Co-occurrence of values for \textit{language}, \gender, and \age,
			derived from the original dataset
			from \Table{t104949-a}.
			E.g. exactly one record (that of \id\ 1)
			represents an English-speaking male over 30.}
		\label{t104949-full_correlation}
	\end{subtable} 
	\hfill
	\begin{subtable}[t]{0.48\linewidth}
		\renewcommand{\arraystretch}{0.9}
		\centering
		\begin{small}
			$
			\begin{array}{|c|c c c c|}
				\hline
				\raisebox{-0.75mm}{\textit{gender, age} $\blacktriangleright$}
				& 
				\multirow{3}{*}{\rotatebox{90}{\small{\male\  $\leq30$~}}} 
				&			\multirow{3}{*}{\rotatebox{90}{\small{\male\  $>30$~}}}
				&			\multirow{3}{*}{\rotatebox{90}{\small{\female\  $\leq30$~}}} 
				&			\multirow{3}{*}{\rotatebox{90}{\small{\female\  $>30$~}}}
				\\[1mm] \cline{1-1}
				& 
				& 
				& 
				& 
				\\[-2mm]
				\raisebox{1.2mm}{\textit{language}~$\blacktriangledown$}
				& 
				& 
				& 
				& 
				\\
				\hline
				\rowcolor{sand!50!white} \texttt{English} & 0 & \nicefrac{1}{4} & 0 & 0 \\
				\texttt{Portuguese} & \nicefrac{1}{4} & 0 & 0 & 0 \\ 
				\rowcolor{sand!50!white} \texttt{German} & \nicefrac{1}{4} & 0 & \nicefrac{1}{4} & 0 \\ \hline
			\end{array}	
		$
		\end{small}
		\caption{Joint distribution for \textit{language}, \gender, and \age, derived from the
			co-occurrence matrix from \Table{t104949-full_correlation},
			and assuming a uniform distribution on the records in the
			original dataset.
			E.g. the probability that an individual
			is an English-speaking male over 30 is $\nicefrac{1}{4}$.}
		\label{t104949-full_joint}
	\end{subtable} 
	\\[2mm]
	\begin{subtable}[t]{0.48\linewidth}
		\renewcommand{\arraystretch}{0.9}
		\centering
		\begin{small}
			$
			\begin{array}{|c|}
				\hline
				prior \\ \hline
				\rowcolor{sand!50!white} \nicefrac{1}{4} \\
				\nicefrac{1}{4} \\
				\rowcolor{sand!50!white} \nicefrac{1}{2} \\
				\hline
			\end{array}
			\quad
			\begin{array}{|c|c c c c|}
				\hline
				\raisebox{-0.75mm}{\textit{gender, age} $\blacktriangleright$}
				& 
				\multirow{3}{*}{\rotatebox{90}{\small{\male\  $\leq30$~}}} 
				&			\multirow{3}{*}{\rotatebox{90}{\small{\male\  $>30$~}}}
				&			\multirow{3}{*}{\rotatebox{90}{\small{\female\  $\leq30$~}}} 
				&			\multirow{3}{*}{\rotatebox{90}{\small{\female\  $>30$~}}}
				\\[1mm] \cline{1-1}
				& 
				& 
				& 
				& 
				\\[-2mm]
				\raisebox{1.2mm}{\textit{language}~$\blacktriangledown$}
				& 
				& 
				& 
				& 
				\\
				\hline
				\rowcolor{sand!50!white} \texttt{English} & 0 & 1 & 0 & 0 \\
				\texttt{Portuguese} & 1 & 0 & 0 & 0 \\ 
				\rowcolor{sand!50!white} \texttt{German} & \nicefrac{1}{2} & 0 & \nicefrac{1}{2} & 0 \\ \hline
			\end{array}
		$	
		\end{small}
		\caption{Adversary's prior knowledge about
			a randomly selected individual's \textit{language},
			and the channel that probabilistically maps
			\textit{language}, \gender, and \age,
			each derived from the joint 
			distribution from \Table{t104949-full_joint}
			by marginalization and conditioning, respectively.}
			%E.g. the prior indicates that before meeting the individual and observing gender and age,
			%the adversary believes that the probabilities of
			%any individual speaking English, Portuguese, or German are,
			%respectively, $\nicefrac{1}{4}$, $\nicefrac{1}{4}$, and $\nicefrac{1}{2}$.
			%On the one hand, the channel indicates that
			%the adversary knows that if an 
			%English-speaking individual is the owner of a record, then 
			%with probability 1 that record indicates a male over 30.
			%On the other hand, if a German-speaking individual owns a record,
			%with probability $\nicefrac{1}{2}$ that record indicates a 
			%male under 30, and with probability $\nicefrac{1}{2}$ it indicates a female under 30.}
		\label{t104949-full_channel}
	\end{subtable} 
	\hfill
	\begin{subtable}[t]{0.48\linewidth}
		\renewcommand{\arraystretch}{0.9}
		\centering
		\begin{small}
			$
			\begin{array}{|c|c c c c|}
				\multicolumn{1}{c}{\text{outers~$\blacktriangleright$}}
				& \multicolumn{1}{c}{$\nicefrac{1}{2}$}
				& \multicolumn{1}{c}{$\nicefrac{1}{4}$}
				& \multicolumn{1}{c}{$\nicefrac{1}{4}$}
				& \multicolumn{1}{c}{$0$}
				\\[1mm]
				\hline
				\raisebox{-0.75mm}{\textit{gender, age} $\blacktriangleright$}
				& 
				\multirow{3}{*}{\rotatebox{90}{\small{\male\  $\leq30$~}}} 
				&			\multirow{3}{*}{\rotatebox{90}{\small{\male\  $>30$~}}}
				&			\multirow{3}{*}{\rotatebox{90}{\small{\female\  $\leq30$~}}} 
				&			\multirow{3}{*}{\rotatebox{90}{\small{\female\  $>30$~}}}
				\\[1mm] \cline{1-1}
				& 
				& 
				& 
				& 
				\\[-2mm]
				\raisebox{1.2mm}{\textit{language}~$\blacktriangledown$}
				& 
				& 
				& 
				& 
				\\
				\hline
				\rowcolor{sand!50!white} \texttt{English} & 0 & 1 & 0 & 0 \\
				\texttt{Portuguese} & \nicefrac{1}{2} & 0 & 0 & 0 \\ 
				\rowcolor{sand!50!white} \texttt{German} & \nicefrac{1}{2} & 0 & 1 & 0 \\ \hline
			\end{array}	
			$	
		\end{small}
		\caption{Hyper-distribution (with column labels added for clarity) representing the adversary's knowledge about \textit{language} after meeting the person and learning
		\gender\ and \age.
		This is identical to the final result of \Table{t104949-f}.
		%The top row (``outers'') gives the probability of each
		%possible combination of \gender\ and \age\ being revealed, 
		%and each column gives the posterior probability distribution 
		%on \textit{language} given that the corresponding attributes
		%were revealed.
		%E.g. the adversary has a probability $\nicefrac{1}{4}$
		%of meeting a male over 30,
		%and in this case she assigns probability 1 to the 
		%corresponding individual being an English-speaker.
		%On the other hand, the adversary has a probability $\nicefrac{1}{2}$
		%of meeting a male under 30,
		%and in this case she assigns probability $\nicefrac{1}{2}$ 
		%to the individual either speaking Portuguese or German.
			%		 -- 
			%		meaning that the value of the sensitive attribute has
			%		been inferred with absolute certainty.
		}
		\label{t104949-full_hyper}
	\end{subtable}
	
	\caption{Step-by-step derivation of prior, channel, and hyper-distribution for the native-language example from \Table{t104949}.}
	\label{t104949-full}
\end{table*}	
%-----------------------------

%=============================
% APPENDIX
%=============================
\section{Collective-target re-identification attack on a longitudinal collection (\CRL)}\label{s2110071202}

% !TEX root = main.tex

%\alldone[If we are out of space, this example could go into the appendix.][\ms]

\review{Here we revisit our claim from \Sec{sec:inep-case-study-examples}
that \CAL\ attacks can be seen as generalizations
of all others (as in \Table{tab:attack-class}).

The key idea is that in the \QIF\ framework we can consider
that the secret (i.e. the values the adversary is trying to 
make inferences about) 
consists in the whole collection of real, un-sanitized records 
for all individuals of interest, 
some of which may be treated and published, 
and some of which may not be published at all.
In this model, each real record contains 
the accurate value for all attributes of an individual, 
including: 
(i) those which are typically removed in any private
	microdata release, such as personal identifiers like 
	name or, in our case, a uniquely identifying \emph{id}
	attribute; 
(ii) those which are published in a possibly sanitized form, such as QIDs or sensitive attributes; and 
(iii) a special \emph{membership attribute} indicating
	whether or not the record in question is published
	at all in the data release.

In \QIF, the adversary's goals and capabilities 
in an attack are modeled as a gain function selecting the part
of the secret she is interested in.
In an attribute-inference attack, such as that from
\Sec{sec:inep-case-study-examples}, the gain function 
represents the adversary's goal of inferring the 
mapping  from individuals' unique {\id}s\ to their 
sensitive attribute.
Note that, in that example, all individuals of interest 
were known to be represented in the published dataset, 
but that is not required in general.
Indeed, \QIF\ allows for the assessment of leakage of information
even about individuals \emph{not present} in the data release.

Now, notice that re-identifying individuals is the same
as finding a mapping from published records to the real 
{\id}s\ of their owners.
Hence, a re-identification attack is just an instance 
of attribute-inference attacks in which the attribute to 
be inferred is the unique {\id} of individuals associated with 
published records. 
Notice that, although the {\id} values to be inferred are not 
present in the published dataset, they are part of the secret 
collection of real records, and \QIF\ allows us to measure
the information leaked about them.
Similarly, we can model a membership attack
as an attribute-inference attack in which the attribute to 
be inferred is the attribute in the complete, secret collection 
of records indicating which individuals are part of the 
published data.

Now we provide a concrete example of how to model
a \emph{collective-target re-identification attack on a longitudinal collection} (\CRL) attack as an instance of a \CAL\ attack.
Recall in such an attack, %\footnote{A rigorous mathematical formalization of \CRL\ attacks is given in Definition~\ref{def:CRL} in Appendix~\ref{sec:formal-models-attacks-full}.}
the adversary's goal is to re-identify as many individuals as possible in the focal dataset $D_{1}$, no matter who they might be.
Hence we consider a \CAL\ attack in which the
attribute to be inferred is the individual's identification itself, 
so
$X=\{\id\}$.

\textbf{Attack execution.}
%\textbf{Execution of the attack.}
%\tashtodo[Can't the adversary launch an attack on $D_1$ without requiring $D_2$? (e.g.\ record 8).]
%	
%\begin{example}[Execution of a \CRL\ attack]
%	\label{exa:CRL}
In the absence of %any 
further prior knowledge, before the attack 
the adversary considers that all individuals 
of interest have the same probability of being the owner 
of any record in the focal dataset $D_{1}$,
meaning that her prior $\pi$ on $\id$ is uniform.
Consider again that during the attack the adversary obtains the values of the QIDs $Y {=}\{\gender, \grade\}$ 
for every individual in $\cal D$
as auxiliary information.
She then performs Bayesian reasoning to update her knowledge
about the secret value from the prior $\pi$ to a hyper.
This whole process is analogous to that
presented for the \CAL\ attack, and 
is modeled in {\QIF} as presented in \Table{tbl:crl_example_1}.
%First the adversary extracts from $\cald$ all co-occurrences of 
%values for the secret and for QIDs (Tab.~\ref{tbl:crl_eg1_correlation}),
%and from that she obtains a joint probability distribution on these
%values (Tab.~\ref{tbl:crl_eg1_joint}).
%By marginalization, from the joint distribution we 
%obtain the prior $\pi$ on the secret value \id, 
%and by conditioning we obtain the channel representing the information-gathering (from the part of the adversary) 
%of QID values during the attack (Tab.~\ref{tbl:crl_eg1_channel}).
%The adversary's state of knowledge after the attack is 
%represented by the hyper in Tab.~\ref{tbl:crl_eg1_hyper},
%which is the result of applying Bayesian update to the prior 
%and the channel.
The degradation of privacy can be computed as follows.}

% CRL attack example
\begin{table*}[!tb]
	\begin{subtable}[t]{0.48\linewidth}
		\renewcommand{\arraystretch}{0.9}
		\centering
		\begin{small}
			$
			\begin{array}{|c|c c c c c c c c|}
				\hline
				\raisebox{-0.75mm}{QIDs $\blacktriangleright$}
				& 
				\multirow{3}{*}{\rotatebox{90}{\small{(\female,\one,\two)}}} 
				&			\multirow{3}{*}{\rotatebox{90}{\small{(\female,\one,\one)}}}
				&			\multirow{3}{*}{\rotatebox{90}{\small{(\female,\three,\three)}}}
				&			\multirow{3}{*}{\rotatebox{90}{\small{(\male,\two,\two)}}}
				&
				\multirow{3}{*}{\rotatebox{90}{\small{(\female,\three,\four)}}}
				&			\multirow{3}{*}{\rotatebox{90}{\small{(\female,\five,\five)}}} 
				&			\multirow{3}{*}{\rotatebox{90}{\small{(\male,\four,\four)}}} 
				&			\multirow{3}{*}{\rotatebox{90}{\small{(\male,\four,-)}}} 
				\\[1mm] \cline{1-1}
				& 
				& 
				& 
				& 
				& 
				& 
				& 
				& 
				\\[-2mm]
				\id\ \blacktriangledown
				& 
				& 
				& 
				& 
				& 
				& 
				& 
				& 
				\\
				\hline
				\rowcolor{sand!50!white} 1 & 1 & 0 & 0 & 0 & 0 & 0 & 0 & 0 \\
				2 & 0 & 1 & 0 & 0 & 0 & 0 & 0 & 0 \\
				\rowcolor{sand!50!white} 3 & 0 & 0 & 1 & 0 & 0 & 0 & 0 & 0 \\
				4 & 0 & 0 & 0 & 1 & 0 & 0 & 0 & 0 \\
				\rowcolor{sand!50!white} 5 & 0 & 0 & 0 & 1 & 0 & 0 & 0 & 0 \\
				6 & 0 & 0 & 0 & 0 & 1 & 0 & 0 & 0 \\
				\rowcolor{sand!50!white} 7 & 0 & 0 & 1 & 0 & 0 & 0 & 0 & 0 \\
				8 & 0 & 0 & 0 & 0 & 0 & 1 & 0 & 0 \\
				\rowcolor{sand!50!white} 9 & 0 & 0 & 0 & 0 & 0 & 0 & 1 & 0 \\
				10 & 0 & 0 & 0 & 0 & 0 & 0 & 0 & 1 \\
				\hline
			\end{array}
			$
		\end{small}
		\caption{Co-occurrence of values for secret $X{=}\{(\id,1)\}$ and for observable QIDs $Y{=}\{(\gender,1),(\grade,1),(\grade,2)\}$,
			derived from the aggregated dataset $\cald$ from \Table{tab:leading-example-long-1-2}.
			E.g. exactly one record 
			has \id\ 1
			and at the same time is a female with grade \one\ in the 
			focal dataset $D_{1}$, and grade \two\ in the auxiliary dataset $D_{2}$.}
		\label{tbl:crl_eg1_correlation}
	\end{subtable} 
	\hfill
	\begin{subtable}[t]{0.48\linewidth}
		\renewcommand{\arraystretch}{0.9}
		\centering
		\begin{small}
			$
			\begin{array}{|c|c c c c c c c c|}
				\hline
				\raisebox{-0.75mm}{QIDs $\blacktriangleright$}
				& 
				\multirow{3}{*}{\rotatebox{90}{\small{(\female,\one,\two)}}} 
				&			\multirow{3}{*}{\rotatebox{90}{\small{(\female,\one,\one)}}}
				&			\multirow{3}{*}{\rotatebox{90}{\small{(\female,\three,\three)}}}
				&			\multirow{3}{*}{\rotatebox{90}{\small{(\male,\two,\two)}}}
				&
				\multirow{3}{*}{\rotatebox{90}{\small{(\female,\three,\four)}}}
				&			\multirow{3}{*}{\rotatebox{90}{\small{(\female,\five,\five)}}} 
				&			\multirow{3}{*}{\rotatebox{90}{\small{(\male,\four,\four)}}} 
				&			\multirow{3}{*}{\rotatebox{90}{\small{(\male,\four,-)}}} 
				\\[1mm] \cline{1-1}
				& 
				& 
				& 
				& 
				& 
				& 
				& 
				& 
				\\[-2mm]
				\id\ \blacktriangledown
				& 
				& 
				& 
				& 
				& 
				& 
				& 
				& 
				\\
				\hline
				\rowcolor{sand!50!white} 1 & \nicefrac{1}{10} & 0 & 0 & 0 & 0 & 0 & 0 & 0 \\
				2 & 0 & \nicefrac{1}{10} & 0 & 0 & 0 & 0 & 0 & 0 \\
				\rowcolor{sand!50!white} 3 & 0 & 0 & \nicefrac{1}{10} & 0 & 0 & 0 & 0 & 0 \\
				4 & 0 & 0 & 0 & \nicefrac{1}{10} & 0 & 0 & 0 & 0 \\
				\rowcolor{sand!50!white} 5 & 0 & 0 & 0 & \nicefrac{1}{10} & 0 & 0 & 0 & 0 \\
				6 & 0 & 0 & 0 & 0 & \nicefrac{1}{10} & 0 & 0 & 0 \\
				\rowcolor{sand!50!white} 7 & 0 & 0 & \nicefrac{1}{10} & 0 & 0 & 0 & 0 & 0 \\
				8 & 0 & 0 & 0 & 0 & 0 & \nicefrac{1}{10} & 0 & 0 \\
				\rowcolor{sand!50!white} 9 & 0 & 0 & 0 & 0 & 0 & 0 & \nicefrac{1}{10} & 0 \\
				10 & 0 & 0 & 0 & 0 & 0 & 0 & 0 & $\nicefrac{1}{10}$ \\
				\hline
			\end{array}
			$
		\end{small}
		\caption{Joint distribution of values for secret $X{=}\{(\disability,1)\}$ and for observable QIDs $Y{=}\{(\gender,1),(\grade,1),(\grade,2)\}$, derived from the
			co-occurrence matrix from \Table{tbl:cal_eg1_correlation},
			and assuming a uniform distribution on the records in $\cald$.
			E.g. there is a probability $\nicefrac{1}{10}$ that an
			individual has \id\ 1 and has 
			QID values (\female,\one,\two).}
		\label{tbl:crl_eg1_joint}
	\end{subtable} 
	\\[2mm]
	\begin{subtable}[t]{0.48\linewidth}
		\renewcommand{\arraystretch}{0.9}
		\centering
		\begin{small}
			$
			\begin{array}{|c|}
				\hline
				\pi \\ \hline
				\rowcolor{sand!50!white} \nicefrac{1}{10} \\
				\nicefrac{1}{10} \\
				\rowcolor{sand!50!white} \nicefrac{1}{10} \\
				\nicefrac{1}{10} \\
				\rowcolor{sand!50!white} \nicefrac{1}{10} \\
				\nicefrac{1}{10} \\
				\rowcolor{sand!50!white} \nicefrac{1}{10} \\
				\nicefrac{1}{10} \\
				\rowcolor{sand!50!white} \nicefrac{1}{10} \\
				\nicefrac{1}{10} \\
				\hline
			\end{array}
			\quad
			\begin{array}{|c|c c c c c c c c|}
				\hline
				\raisebox{-0.75mm}{QIDs $\blacktriangleright$}
				& 
				\multirow{3}{*}{\rotatebox{90}{\small{(\female,\one,\two)}}} 
				&			\multirow{3}{*}{\rotatebox{90}{\small{(\female,\one,\one)}}}
				&			\multirow{3}{*}{\rotatebox{90}{\small{(\female,\three,\three)}}}
				&			\multirow{3}{*}{\rotatebox{90}{\small{(\male,\two,\two)}}}
				&
				\multirow{3}{*}{\rotatebox{90}{\small{(\female,\three,\four)}}}
				&			\multirow{3}{*}{\rotatebox{90}{\small{(\female,\five,\five)}}} 
				&			\multirow{3}{*}{\rotatebox{90}{\small{(\male,\four,\four)}}} 
				&			\multirow{3}{*}{\rotatebox{90}{\small{(\male,\four,-)}}} 
				\\[1mm] \cline{1-1}
				& 
				& 
				& 
				& 
				& 
				& 
				& 
				& 
				\\[-2mm]
				\id\ \blacktriangledown
				& 
				& 
				& 
				& 
				& 
				& 
				& 
				& 
				\\
				\hline
				\rowcolor{sand!50!white} 1 & 1 & 0 & 0 & 0 & 0 & 0 & 0 & 0 \\
				2 & 0 & 1 & 0 & 0 & 0 & 0 & 0 & 0 \\
				\rowcolor{sand!50!white} 3 & 0 & 0 & 1 & 0 & 0 & 0 & 0 & 0 \\
				4 & 0 & 0 & 0 & 1 & 0 & 0 & 0 & 0 \\
				\rowcolor{sand!50!white} 5 & 0 & 0 & 0 & 1 & 0 & 0 & 0 & 0 \\
				6 & 0 & 0 & 0 & 0 & 1 & 0 & 0 & 0 \\
				\rowcolor{sand!50!white} 7 & 0 & 0 & 1 & 0 & 0 & 0 & 0 & 0 \\
				8 & 0 & 0 & 0 & 0 & 0 & 1 & 0 & 0 \\
				\rowcolor{sand!50!white} 9 & 0 & 0 & 0 & 0 & 0 & 0 & 1 & 0 \\
				10 & 0 & 0 & 0 & 0 & 0 & 0 & 0 & 1 \\
				\hline
			\end{array}
			$
		\end{small}
		\caption{Prior distribution $\pi$ on the values 
			for secret $X{=}{(\id,1)}$, and the channel
			for the \CRL\ attack, each derived from the joint 
			distribution from \Table{tbl:cal_eg1_joint}
			by marginalization and conditioning, respectively.}
			%			The prior models the adversary's knowledge
			%			about the secret before the attack is executed, 
			%			whereas the channel
			%			models the information the adversary learns about the QID values for each individual during the attack.
			%E.g. the prior indicates that before the attack
			%(i.e. without knowing any QID value)
			%the adversary believes that the probability of
			%any 
			%randomly selected
			%record belonging
			%to the individual of \id\ 1 is $\nicefrac{1}{10}$.
			%On the other hand, the channel indicates that
			%during the attack the adversary can use the fact that if a record has \id\ 1, then
			%the probability that this record's QID values are (\female,\one,\two) is $1$.}
		\label{tbl:crl_eg1_channel}
	\end{subtable} 
	\hfill
	\begin{subtable}[t]{0.48\linewidth}
		\renewcommand{\arraystretch}{0.9}
		\centering
		\begin{small}
			$
			\begin{array}{|c|c c c c c c c c|}
				%				\cline{2-9}
				\multicolumn{1}{c}{\text{outers $\blacktriangleright$}}
				& \multicolumn{1}{c}{\nicefrac{1}{10}}
				& \multicolumn{1}{c}{\nicefrac{1}{10}}
				& \multicolumn{1}{c}{\nicefrac{1}{5}}
				& \multicolumn{1}{c}{\nicefrac{1}{5}}
				& \multicolumn{1}{c}{\nicefrac{1}{10}}
				& \multicolumn{1}{c}{\nicefrac{1}{10}}
				& \multicolumn{1}{c}{\nicefrac{1}{10}}
				& \multicolumn{1}{c}{\nicefrac{1}{10}} \\ \hline
				\raisebox{-0.75mm}{QIDs $\blacktriangleright$}
				& 
				\multirow{3}{*}{\rotatebox{90}{\small{(\female,\one,\two)}}} 
				&			\multirow{3}{*}{\rotatebox{90}{\small{(\female,\one,\one)}}}
				&			\multirow{3}{*}{\rotatebox{90}{\small{(\female,\three,\three)}}}
				&			\multirow{3}{*}{\rotatebox{90}{\small{(\male,\two,\two)}}}
				&
				\multirow{3}{*}{\rotatebox{90}{\small{(\female,\three,\four)}}}
				&			\multirow{3}{*}{\rotatebox{90}{\small{(\female,\five,\five)}}} 
				&			\multirow{3}{*}{\rotatebox{90}{\small{(\male,\four,\four)}}} 
				&			\multirow{3}{*}{\rotatebox{90}{\small{(\male,\four,-)}}} 
				\\[1mm] \cline{1-1}
				& 
				& 
				& 
				& 
				& 
				& 
				& 
				& 
				\\[-2mm]
				\id\ \blacktriangledown
				& 
				& 
				& 
				& 
				& 
				& 
				& 
				& 
				\\
				\hline
				\rowcolor{sand!50!white} 1 & 1 & 0 & 0 & 0 & 0 & 0 & 0 & 0 \\
				2 & 0 & 1 & 0 & 0 & 0 & 0 & 0 & 0 \\
				\rowcolor{sand!50!white} 3 & 0 & 0 & \nicefrac{1}{2} & 0 & 0 & 0 & 0 & 0 \\
				4 & 0 & 0 & 0 & \nicefrac{1}{2} & 0 & 0 & 0 & 0 \\
				\rowcolor{sand!50!white} 5 & 0 & 0 & 0 & \nicefrac{1}{2} & 0 & 0 & 0 & 0 \\
				6 & 0 & 0 & 0 & 0 & 1 & 0 & 0 & 0 \\
				\rowcolor{sand!50!white} 7 & 0 & 0 & \nicefrac{1}{2} & 0 & 0 & 0 & 0 & 0 \\
				8 & 0 & 0 & 0 & 0 & 0 & 1 & 0 & 0 \\
				\rowcolor{sand!50!white} 9 & 0 & 0 & 0 & 0 & 0 & 0 & 1 & 0 \\
				10 & 0 & 0 & 0 & 0 & 0 & 0 & 0 & 1 \\
				\hline
			\end{array}
			$
		\end{small}
		\caption{Hyper-distribution (with column labels added for clarity) representing the adversary's knowledge after completing the \CRL\ attack.
			%The top row (``outers'') gives the probability of each
			%possible combination of QID values being revealed, and each column gives the 
			%posterior probability distribution on secret values given that 
			%the corresponding QID values were revealed.
			%E.g. after the attack, 
			%the adversary has a probability $\nicefrac{1}{10}$
			%of learning that an individual's QID values	are 
			%(\female, \one, \two), and in this case she assigns 
			%probability 1 to the corresponding individual 
			%being that of \id\ 1.
			%, meaning that a re-identification has occurred with absolute certainty.
		}
		\label{tbl:crl_eg1_hyper}
	\end{subtable}
	
	\caption{Step-by-step derivation of prior, channel, and hyper-distribution for \CRL\ attack on the longitudinal collection $\longdb$ from \Table{tab:leading-example-long}, considering secret $X = \Set{\disability}$ and observable QIDs $Y = \Set{\gender,\grade}$.}
	\label{tbl:crl_example_1} 
\end{table*}	

\textbf{Deterministic degradation of privacy.} 
%Deterministic success is concerned with the fraction of
%individuals who can be re-identified with absolute certainty.
The deterministic prior vulnerability of the dataset is $0\%$, since before the attack no individual can be re-identified with certainty.
After the attack, 
the adversary's posterior knowledge is given by
%can be computed by pushing her (uniform) prior through the channel 
%$C$ to obtain the hyper depicted in Tab.~\ref{tbl:crl_eg1_hyper}.		
the hyper of \Table{tbl:crl_eg1_hyper}.
Note that in that hyper posteriors containing only $1$ and $0$ 
values -- i.e.\ records with {\id}s\ 1, 2, 6, 8, 9, and 10 -- have unique QIDs and can therefore be re-identified with certainty.
The adversary's posterior success is the proportion of 
individuals identified in this way, which is exactly 6 
out of 10, i.e. $\nicefrac{1}{10}{\cdot}6{=}60\%$. %			
The overall deterministic degradation of
privacy is $60\%{-}0\%{=}60\%$, meaning
that the execution of the attack increases the proportion of re-identifiable individuals
by an absolute value of 60\%.
%\footnote{For comparison, consider the adversary were restricted to a single dataset scenario, i.e.\ she could access only the dataset $D_{1}$.
%	Prior success would still be $0\%$.
%	However, in this case there would be only one record in $D_{1}$ with unique values for the attributes \gender\ and \grade\ (the record with \id\ 8), giving a posterior success of $\nicefrac{1}{10}{=}10\%$. 
%	Therefore, the overall deterministic privacy degradation would be $10\%$, significantly less than that achieved in the longitudinal attack.}

\textbf{Probabilistic degradation of privacy.}
The prior vulnerability of the dataset is given by its Bayes vulnerability (i.e. the maximum probability of guessing any secret value), which is $\nicefrac{1}{10}{=}10\%$ (given the prior is uniform).
%Without any auxiliary information, the adversary's best strategy prior is to blindly guess which individual is the owner of any record in the focal dataset $D_{1}$.
%\footnote{According to the \emph{maximum entropy principle}, given the known information, the probability distribution that best represents a state of knowledge is that in which the entropy has its maximum value. Therefore, in the absence of any information, i.e. a state in which nothing is assumed from the adversary's knowledge, the best possible distribution would be the uniform distribution and the best possible action for the adversary would be to choose a random value for the secret \cite{Jaynes03}.} Since there are 10 records in $D_{1}$, the adversary's probabilistic prior success is of $10\%$.
%\ghndone[Cite the maximum entropy principle in a footnote.][\ms]
After the attack, the adversary's knowledge
is given by the hyper in \Table{tbl:crl_eg1_hyper}.
The posterior Bayes vulnerability %(recall Eqn \ref{eqn:bv}) 
is the expected maximum probability of success over
all posteriors.
More precisely, since in 6 of the posteriors the probability of
a correct guess is $1$ --and each of these posteriors occur themselves with probability \nicefrac{1}{10}--, and in 2 of the posteriors the probability of success is  $\nicefrac{1}{2}$ --and each of them occurs with probability $\nicefrac{1}{5}$--, the overall
posterior Bayes vulnerability is $6 {\cdot} \nicefrac{1}{10} {\cdot} 1 {+} 2 {\cdot} \nicefrac{1}{5} {\cdot}
\nicefrac{1}{2}{=}80\%$.
%We describe
The overall probabilistic degradation of privacy caused by the attack is $\nicefrac{80\%}{10\%}{=}8$,
meaning that the adversary's chance of re-identifying a randomly selected record in the focal dataset $D_{1}$ increases by a factor of eight after the \CRL\ attack.%\footnote{For comparison, consider the adversary were restricted to a single-dataset attack, i.e. the adversary could access only the dataset $D_{1}$. In this case, the adversarys
%prior success would remain at $10\%$, but a posteriori she would be able to partition $D_{1}$ into five blocks according to QIDs \gender\ and \grade. 
%Therefore, her probabilistic posterior success would be of $60\% \cdot 50\% + 30\% \cdot 33\% + 10\% \cdot 100\% \approx 50\%$, and the degradation of privacy would be approximately 5, a value significantly smaller than that of the longitudinal attack.}
%\end{example}

%-----------------------------

%=============================
% APPENDIX
%=============================
\review{
\section{\review{Using the \QIF\ framework in other large-scale scenarios}}
\label{s03060736}

% !TEX root = main.tex

\review{We now exemplify how the \QIF\ framework, which
grounded the \INEP\ privacy analyses, can be 
generalized to other other large-scale scenarios.
More precisely, we show how the \QIF\ framework can 
be used to assess privacy under popular syntactic 
anonymization techniques.}

\review{The techniques considered partition the set of records 
into blocks of records with the same values for QIDs, 
and then perform \emph{generalization}, 
\emph{suppression}, or \emph{swapping}. % of records.
Here we considered as QIDs the 11 attributes
in \Table{tab:CRS-CAS-CRL-CAL-qid},
and as sensitive the attribute \disability.
We then employed the ARX tool 
(extended with our update to treat datasets 
larger than $2^{31}{-}1$ cells)
to anonymize the School Census of 2018 
(see \Table{tab:datasets}), 
using the following techniques \cite{nunes_gabriel_henrique_2022_6533684}.\footnote{\review{We also initially considered \emph{$\ell$-diversity}~\cite{Machanavajjhala07}, 
	which ensures that each block with the same values for QIDs 
	has \qm{well-represented} values for the sensitive attribute 
	according to some suitable metric and threshold $\ell$.
	%	By \qm{well-represented} values, they suggested some different interpretations, 
	%	including \emph{entropy} \emph{l}-\emph{diversity} and \emph{recursive} (\emph{c,l})-\emph{diversity}.
	%We considered both entropy $\ell$-diversity and
	%recursive $(c,\ell)$-diversity for values 
	%of $\ell\approx2$ and of $c\approx 40$.
	However, due to the skewness}
\review{of the distribution, all solutions found by ARX suppressed 
	all QID values, and we discarded them from our experiments.}}
}
\review{%
%\begin{itemize}
%	\item 
First, \emph{$k$-anonymity}~\cite{Samarati98}, 
	which ensures that each block with the}
\review{same} 
\review{values for QIDs has at least $k$ records.}
	%We adopted $k\in \{4, 8, 12, 16, 20, 24\}$.
%	
%	\item \emph{$\ell$-diversity}~\cite{Machanavajjhala07}, 
%	which ensures that each block with the same values for QIDs 
%	has \qm{well-represented} values for the sensitive attribute 
%	according to some suitable metric and threshold $\ell$.
%	%	By \qm{well-represented} values, they suggested some different interpretations, 
%	%	including \emph{entropy} \emph{l}-\emph{diversity} and \emph{recursive} (\emph{c,l})-\emph{diversity}.
%	We considered both entropy $\ell$-diversity and
%	recursive $(c,\ell)$-diversity for values 
%	of $\ell\approx2$ and of $c\approx 40$.
%	However, due to the skewness of the distribution on
%	disability, all solutions found by ARX suppressed 
%	all QID values, and we did not further 
%	consider them.% in our experiments.
%	
%	\item 
\review{Second, \emph{$t$-closeness}~\cite{Li07}, 
	which ensures that the distance (according to some suitable metric, e.g. Earth Mover's Distance) between the distribution 
	on the sensitive attribute in each block with the same QID values
	and the overall distribution on the sensitive attribute 
	is bounded by a threshold $t$.%
	%We adopted $t \in \{0.1, 0.2, 0.3, 0.4, 0.5\}$.
%\end{itemize}
}

\review{
On each anonymized dataset we then performed the same 
privacy analyses of deterministic and probabilistic
privacy degradation for both collective-target re-identification 
(\CRS) and collective-target attribute-inference attacks (\CAS) 
on \disability.
\Table{tab:anonymization} 
%and \Fig{fig:anonymization} 
summarizes our results, and confirms
the intuitions that larger values of $k$ and smaller
values of $t$ lead to more private data releases
(at a usually increasing cost on utility).
\review{\footnote{\review{We have used ARX in a standard configuration (which looks for an optimal solution to the $k$-anonymity or $t$-closeness problem wrt.\ the tool's default utility metric) to compute anonymized datasets for varying values of the parameters. Because ARX does not keep the same anonymity groupings across different values of the parameters, the resulting privacy guarantees (in terms of inferences) do not necessarily increase monotonically. For instance, note on \Table{tab:anonymization-deterministic} the \CAS\ values for $k{=}4$ and $k{=}12$, and on \Table{tab:anonymization-probabilistic} the \CRS\ values for $k{=}4$ and $k{=}12$ and for $t{=}0.1$ and $t{=}0.3$. This is a strength of our \QIF\ analysis, as it uncovers unexpected consequences of anonymization techniques not tailored towards inference attacks.}}}}

\begin{table*}[tb]
	\centering
	\begin{subtable}[b]{0.48\linewidth}
		\centering
		\renewcommand{\arraystretch}{1.1}
		\begin{small}
			\review{
				$
				\begin{array}[t]{|c|>{\centering\arraybackslash\small}m{0.34\linewidth}|>{\centering\arraybackslash\small}m{0.34\linewidth}|}
				\cline{2-3}
				\multicolumn{1}{c|}{} 
				& \multicolumn{1}{>{\centering\arraybackslash\small}m{0.34 \linewidth}|}{\textbf{\CRS}} 
				& \multicolumn{1}{>{\centering\arraybackslash\small}m{0.34\linewidth}|}{\textbf{\CAS} (disability)}
				\\ \hline
				\multirow{2}{*}{\textbf{\small Dataset}}
				& \textbf{\small prior success:} 0.000000\%
				& \textbf{\small prior success:} 0.000000\% 
				\\ \cline{2-3} 
				%			\textbf{\small dataset} 
				& \textbf{\small posterior success} 
				& \textbf{\small posterior success} 
				\\ \hline \hline
				\text{Original} 
				& $96.342560\%$ \qquad ($\sim$46.4 mi.)
				& $99.890918\%$ \qquad ($\sim$48.1 mi.)
				\\ \hline \hline
				$k=4$
				& $0.000000\%$ ($0$)
				& $0.048254\%$ ($\sim$23,200)
				\\ \hline
				%			$k=8$
				%			& $0.000000\%$ ($0$)
				%			& $0.039663\%$ ($\sim$19,100)
				%			\\ \hline
				$k=12$
				& $0.000000\%$ ($0$)
				& $0.079773\%$ ($\sim$38,400)
				\\ \hline
				%			$k=16$
				%			& $0.000000\%$ ($0$)
				%			& $0.008977\%$ ($\sim$4,300)
				%			\\ \hline
				$k=20$
				& $0.000000\%$ ($0$)
				& $0.008977\%$ ($\sim$4,300)
				\\ \hline
				%			$k=24$
				%			& $0.000000\%$ ($0$)
				%			& $0.008977\%$ ($\sim$4,300)
				%			\\ \hline \hline
				$t=0.1$
				& $0.000000\%$ ($0$)
				& $0.008977\%$ ($\sim$4,300)
				\\ \hline
				%			$t=0.2$
				%			& $0.000000\%$ ($0$)
				%			& $0.048254\%$ ($\sim$23,200)
				%			\\ \hline
				$t=0.3$
				& $0.000023\%$ ($\sim$11)
				& $0.013258\%$ ($\sim$6,300)
				\\ \hline
				%			$t=0.4$
				%			& $0.000073\%$ ($\sim$35)
				%			& $0.026102\%$ ($\sim$12,500)
				%			\\ \hline
				$t=0.5$
				& $0.000166\%$ ($\sim$79)
				& $0.041118\%$ ($\sim$19,800)
				\\ \hline
				\end{array}
				$
			}
		\end{small}
		\caption{\review{Deterministic measure of privacy degradation (i.e. proportion of students whose sensitive attribute is inferred with certainty).}}
		\label{tab:anonymization-deterministic}
	\end{subtable}
%
%	\vspace{2mm}
%	
\hfill
	\begin{subtable}[b]{0.48\linewidth}
		\centering
		\renewcommand{\arraystretch}{1.1}
		\begin{small}
			\review{
				$
				\begin{array}[t]{|c|>{\centering\arraybackslash\small}m{0.34\linewidth}|>{\centering\arraybackslash\small}m{0.34\linewidth}|}
				\cline{2-3}
				\multicolumn{1}{c|}{} 
				& \multicolumn{1}{>{\centering\arraybackslash\small}m{0.34 \linewidth}|}{\textbf{\CRS}} 
				& \multicolumn{1}{>{\centering\arraybackslash\small}m{0.34\linewidth}|}{\textbf{\CAS} (disability)}
				\\ \hline
				\multirow{2}{*}{\textbf{\small Dataset}}
				& \textbf{\small prior success:} 0.000002\%
				& \textbf{\small prior success:} 97.556444\% 
				\\ \cline{2-3} 
				%\textbf{\small dataset} 
				& \textbf{\small posterior success} 
				& \textbf{\small posterior success} 
				\\ \hline \hline
				\text{Original} 
				& $98.138799\%$
				& $99.946785\%$
				\\ \hline \hline
				$k=4$
				& $0.008369\%$
				& $97.556444\%$
				\\ \hline
				%			$k=8$
				%			& $0.006974\%$
				%			& $97.556444\%$
				%			\\ \hline
				$k=12$
				& $0.029857\%$
				& $97.556444\%$
				\\ \hline
				%			$k=16$
				%			& $0.002790\%$
				%			& $97.556444\%$
				%			\\ \hline
				$k=20$
				& $0.002790\%$
				& $97.556444\%$
				\\ \hline
				%			$k=24$
				%			& $0.002790\%$
				%			& $97.556444\%$
				%			\\ \hline \hline
				$t=0.1$
				& $0.002790\%$
				& $97.556444\%$
				\\ \hline
				%			$t=0.2$
				%			& $0.008369\%$
				%			& $97.556444\%$
				%			\\ \hline
				$t=0.3$
				& $0.002750\%$
				& $97.556444\%$
				\\ \hline
				%			$t=0.4$
				%			& $0.005467\%$
				%			& $97.556444\%$
				%			\\ \hline
				$t=0.5$
				& $0.006615\%$
				& $97.556444\%$
				\\ \hline
				\end{array}
				$
			}
		\end{small}
		\caption{\review{Probabilistic measure of privacy degradation (i.e. probability of successful inference of the sensitive attribute in one try).}}
		\label{tab:anonymization-probabilistic}
	\end{subtable}
	
	\caption{\review{Comparison of 
			privacy degradation in re-identification (\CRS) and
			attribute-inference (\CAS) attacks on the School Census of 2018
			before and after sanitization by $k$-anonymity and by $t$-closeness. In all of the attacks, the QIDs employed are the 11 listed on Table~\ref{tab:CRS-CAS-CRL-CAL-qid} for \CRS\ / \CAS\ attacks.}}
	\label{tab:anonymization}
\end{table*}

}
%-----------------------------

%\DraftOnly{
%%=============================
%% APPENDIX
%%=============================
%\section{Anonymization Experiments}\label{s2202280001}
%
%\input{appendix-anon-experiments}
%%-----------------------------
%}

%\DraftOnly{
%=============================
% SARGASSO
%=============================
%\newpage
%\input{sargasso.tex}
%-----------------------------
%}

\end{document}